\documentclass[11pt]{article}
\usepackage{amsmath, amsthm, amssymb}
\usepackage{amsmath}
\usepackage{comment}
\usepackage{color}
\usepackage{ragged2e}
\usepackage{booktabs,caption}
\usepackage[flushleft]{threeparttable}
\usepackage[cm]{fullpage}
\usepackage{amsfonts}
\usepackage{graphicx}
\usepackage{adjustbox}
\usepackage{amsfonts}
\usepackage{setspace}
\usepackage{booktabs}
\usepackage[english]{babel}
\usepackage{array}
\usepackage{cancel}
\usepackage{rotating}
\usepackage{dcolumn}
\usepackage{rotfloat}
\usepackage{graphicx}
\usepackage{pdflscape}
\usepackage{textcomp}
\usepackage{eurosym}
\usepackage{tabularx}
\usepackage{float}
\usepackage[authoryear,sort]{natbib}
\usepackage{geometry}
\usepackage{hyperref}
\hypersetup{pdftex,colorlinks=true,allcolors=blue}
\emergencystretch=\maxdimen
\newgeometry{left=2.81cm,right=2.81cm,bottom=2cm,top=2cm}

\usepackage[symbol*]{footmisc}
\DefineFNsymbolsTM{myfnsymbols}{
	\textdagger    \dagger
	\textdaggerdbl \ddagger
	\textsection   \mathsection
	\textasteriskcentered *
	\textparagraph \mathparagraph
	\textbardbl    \|%
}%
\setfnsymbol{myfnsymbols}

\newcommand{\sym}[1]{\rlap{#1}}
\makeatletter
\let\estinput=\@@input

\newtheorem{res}{Result}

\usepackage{siunitx} 
\usepackage{subfigure}
\begin{document} 
	
	\title{Identity and Cooperation in Multicultural Societies}
	\author{Natalia Montinari\footnote{University of Bologna, Piazza A. Scaravilli 2, 40126 Bologna. e-mail: natalia.montinari2@unibo.it}, Matteo Ploner\footnote{University of Trento, Via Vigilio Inama, 5, 38122 Trento. e-mail: matteo.ploner@unitn.it}, Veronica Rattini\footnote{University of Bologna, Piazza A. Scaravilli 2, 40126 Bologna. e-mail: veronica.rattini2@unibo.it
			\newline
			Contact author: Natalia Montinari, natalia.montinari2@unibo.it
\newline  
The study received the approval of the Bioethics Committee of Bologna University, N. 0111100 on 22/05/2019. The study was not pre-registered; however, the hypotheses associated with both awarded grants were clearly specified prior to data collection. At the time the project was initiated, pre-registration was not yet a standard practice in experimental economics—except for field experiments—so no formal pre-analysis plan was filed.
\newline
N.Montinari gratefully acknowledges financial support from the Unicredit Foundation through the Modigliani Research Grant (“The Power of Identity in Multicultural Societies”, June 2018), as well as from the Almaidea Junior 2018 program of the University of Bologna.
\newline  
V. Rattini gratefully acknowledges funding by the European Union - NextGenerationEU, Mission 4, Component 2, in the framework of the GRINS -Growing Resilient, INclusive and Sustainable project (GRINS PE00000018 – CUP J33C22002910001).
   \newline
   We acknowledge with deep gratitude Marco Tecilla for his valuable contribution to the development of the experimental software. His premature passing represents a significant loss for our community.
   \newline
			We would like to thank the participants of the ESA 2022-Bologna and of the 2019 Rome BEEN meeting. We thank Pol Campos, Ginevra Marandola, and Erik Wengstrom for their comments on an early draft of this project. We also thank all the students and teachers participating in the study.}}
	
	\date{January 2026}
	
	\maketitle

	\begin{abstract}
		\linespread{1}
		\selectfont
This paper tests whether low-cost identity priming can raise cooperation in diverse groups. In a lab-in-the-field experiment with 390 adolescents in ethnically mixed classrooms in Italy, students were individually randomized to a neutral priming condition, a common-identity priming condition that emphasized shared school belonging, or a multicultural-identity priming condition that prompted reflection on family origins and local cultural diversity, and then played a repeated public goods game first without and then with punishment. In the neutral treatment, students with an immigrant background contributed around 13 percent more than natives. Making multicultural identity salient increased natives’ contributions by about 3 percentage points, closing this baseline gap, whereas common-identity priming had no detectable effect. When punishment was available, priming had limited effects on contributions but continued to influence sanctioning: both natives and immigrants exposed to the multicultural prime were more likely to punish free riders. These findings indicate that promoting cooperation in diverse groups requires engaging host-society members as well as immigrants and that highlighting multicultural identity, rather than generic shared belonging, can foster cooperative norms among natives.

        
		
		\medskip 
		
		
	\end{abstract}

	\newpage
	\newgeometry{left=2.81cm,right=2.81cm,bottom=3.81cm,top=3.81cm}
	
	\DefineFNsymbolsTM{myfnsymbols}{
	}%
	\setfnsymbol{myfnsymbols}

\newpage


	\newpage
	\newgeometry{left=2.81cm,right=2.81cm,bottom=3.81cm,top=3.81cm}
	
	\DefineFNsymbolsTM{myfnsymbols}{
	}%
	\setfnsymbol{myfnsymbols}

\newpage

\section{Introduction}

Over the past decade, both international migration flows and the share of foreign-born residents have risen markedly. In 2023, about 146 million people lived outside their country of birth across OECD countries—over 25\% more than ten years earlier. Foreign-born populations now represent 14\% of the U.S. population, 17\% in Germany, and 10.6\% in Italy \citep{oecd2024}. Looking ahead, conflicts and climate change are expected to further accelerate displacement and make host societies increasingly diverse \citep{unhcr2024,ipcc2023}.
This growing diversity raises the question of how institutions and policies can foster cooperation and social cohesion in increasingly heterogeneous societies. Without adequate economic, political, and social inclusion, immigrants may face marginalization and limited opportunities in the host country. At the same time, perceived threats to jobs, housing, or cultural identity among natives can fuel concerns about fiscal costs, segregation, and social tensions \citep{gathmann2023citizenship}. Unsurprisingly, European citizens consistently identify immigration and integration as among the most pressing issues for the European Union \citep{eurobarometer2022}. Yet relatively little is known about which concrete, low-cost interventions can strengthen cooperative behavior across native–immigrant lines in everyday settings such as schools.

Integration is a multifaceted phenomenon that cannot be fully captured by a single measure \citep{hainmueller2015naturalization}. Although economists have traditionally focused on comparing the outcomes of immigrants with those of natives—such as educational attainment \citep{schuller2015parental}, labor market performance \citep{rienzo2022performance,sweetman2015immigration}, and health status \citep{qureshi2022health}—an equally important dimension of integration concerns social interactions, namely, social integration. These interactions—ranging from casual daily encounters to collaborative tasks and service provision—frequently require cooperation to achieve socially optimal outcomes. However, disparities in opportunities \citep{hoffman1996social}, as well as differences in socioeconomic background, ethnicity, and cultural identity between immigrants and natives, can undermine cooperation \citep{akerlof2000economics}. As such, when assessing integration policies or interventions, it is not sufficient to measure success solely in terms of reducing disparities in education or employment. Rather, it is also crucial to evaluate their potential to foster cooperation between diverse social groups. However, there is limited causal evidence on whether simple, scalable interventions can durably affect cooperation between natives and immigrants, especially in naturally diverse youth populations

In this paper, we experimentally investigate whether cooperation within mixed groups—composed of natives and migrants—can be influenced by low-cost identity priming that exogenously alters the salience of \textit{naturally existing }social identities. To our knowledge, this study is among the first to examine how identity priming affects public-good cooperation and punishment in naturally diverse groups of natives and migrants.\footnote{Previous studies on cooperation and trust between migrants and natives have primarily examined discrimination \citep{fershtman2001discrimination, guillen2011trust, chen2014hat, cox2015trust, cettolin2019return}. A smaller set of papers investigates how such biases change exogenously, for instance, after refugee inflows \citep{albrecht2022social} or reforms to birthright citizenship \citep{felfe2021more}. While our study does not focus on trust or altruism per se, it contributes by testing whether cooperation can be shaped through priming interventions that alter the salience of social identities. Our approach builds on the priming methodology introduced by \cite{shih1999stereotype} and applied to migration contexts by \cite{chen2014hat}. Relatedly, \cite{alan2021building} examine how curriculum changes in Turkish schools receiving Syrian refugees affect prosocial behavior. We further review this literature in Section~\ref{sec:lit}.}
Following the framework of \cite{akerlof2000economics}, we view identity as a multidimensional construct that, for immigrants, combines ethnic components linked to their country of origin and host-country elements. This perspective is consistent with developmental work on bicultural youth, which emphasizes that ethnic and national identities often coexist and interact during adolescence (e.g. \cite{vedder201415}). Our study investigates whether cooperation between natives and migrants can be shaped by making different identity dimensions salient. We implement three identity priming treatmens: (i) a \emph{Common Identity} treatment, which highlights shared belonging to the same school community; (ii) a \emph{Multicultural Identity} treatment, which encourages students to reflect on their family origins while emphasizing the coexistence of multiple cultural backgrounds within the broader group; and (iii) a \emph{Neutral Identity} treatmen, involving no specific identity activation.
To this end, we conducted a lab-in-the-field experiment with more than 390 adolescents (aged 12–14), 33\% of whom have a migrant background, in two middle schools in Bologna, Italy. Students in each classroom were randomly assigned to one of three priming treatments and then participated in a repeated Public Good Game. This within-class randomization kept constant class-level characteristics —such as peer networks, teaching quality, and extracurricular activities— across treatment arms. Data were collected between December 2019 and February 2020. We also gathered rich individual background information, social network data, and incentivized measures of altruism, risk preferences, general trust, ethnic stereotypes, as well as an open-ended self-description task following \cite{bugental1950investigations}. This lab-in-the-field design allows us to study cooperation and sanctioning in pre-existing peer groups where ethnic diversity, social networks, and classroom composition are policy-relevant features rather than purely experimental artefacts.
The results show that immigrants are substantially more cooperative than natives at baseline, contributing about 13\% more than the baseline mean. Natives significantly increase their contributions when exposed to the \emph{Multicultural Identity} treatment, which makes the coexistence of multiple cultural backgrounds salient; this increase of roughly 3 percentage points closes the initial cooperation gap between natives and immigrants. When punishment is introduced in the second phase, the effects of priming on contributions become small and imprecisely estimated, but priming continues to shape \textit{social punishment}: immigrants remain more likely to sanction free riders, and so do natives exposed to the multicultural prime. Together, these findings point to two insights: immigrants exhibit higher baseline cooperation, and making multicultural identity salient encourages natives to align their cooperative behavior with that of immigrants. From an economic perspective, the multicultural priming effect on natives is comparable in magnitude to the baseline immigrant–native gap, suggesting that identity salience can substitute for part of the cooperative advantage associated with immigrant status in these classrooms. 
These results indicate that fostering cooperation in diverse groups cannot focus solely on immigrants. The responses of the Host-society members to diversity also matter, highlighting that integration is inherently a mutual process.
Our analysis further shows that these effects are not driven by differences in socioeconomic background or baseline generosity, as measured with an incentivized Dictator Game. Moreover, the impact of the \emph{Multicultural Identity} priming on natives is particularly strong in classrooms with fewer immigrant students, a higher share of first-generation immigrants, or less dense social networks in which immigrant students occupy more peripheral friendship positions—contexts in which natives may not clearly perceive the multicultural nature of their everyday environment. These heterogeneous effects speak directly to debates over classroom composition policies, suggesting that the returns to identity-based interventions are largest where diversity is less visible or where immigrants are more socially peripheral.
Finally, we rule out several alternative explanations: our findings are not driven by differences in comprehension of the instructions, experimenter demand effects, or variation in emotional states.

This paper contributes to the literature on the \emph{contact hypothesis}, which argues that interaction with minority groups can reduce prejudice, strengthen social bonds, and promote integration.\footnote{According to the contact hypothesis, cross-group interactions reduce bias when treatments such as shared goals, cooperation, and institutional support are present \citep{williams1947reduction,allport1954nature}. In contrast, the \emph{social comparison hypothesis} \citep{festinger1954theory} suggests that individuals evaluate themselves relative to peers in their immediate environment, and that comparisons with higher-status or higher-performing peers may induce negative emotions.} Using experimentally elicited data combined with natural variation in classroom diversity, we assess whether contextual classroom factors explain our main findings. As classrooms become more diverse, they provide a natural setting to study how intergroup contact shapes cooperation. In economic terms, they offer a field environment in which group composition, peer networks, and exposure to diversity jointly determine the production of cooperative norms (e.g. \cite{lavy2011mechanisms,patacchini2016social}). Schools, in particular, generate frequent peer interactions across backgrounds and can produce both conflict and cooperation \citep{allport1954nature,lowe2021types,mousa2020building}. We leverage the quasi-random assignment of students to classes within schools to exploit variation in peer-group composition and network structure.\footnote{This work also relates to research on the non-pecuniary returns to education, which shows that exposure to diverse peers fosters cross-cultural interaction and social cohesion \citep{goldin2008transitions,oreopoulos2011priceless}.}
We show that policy-relevant identity priming---specifically, interventions that emphasize the multicultural dimensions of society---can significantly enhance cooperation, as reflected in contributions to a public good. This suggests shifting from integration strategies focused solely on assimilating immigrants toward approaches that also engage native populations. Finally, our priming intervention provides a low-cost, easily scalable tool for enhancing cooperation in diverse contexts. Schools offer a particularly promising platform for such interventions, with potential long-term benefits for social cohesion and economic performance \citep{gradstein2002education}.

The remainder of this paper is organized as follows. Section~\ref{sec:lit} reviews the relevant literature. Section~\ref{sec: institutional setting} presents the institutional background of the study. Section~\ref{sec:design} describes the experimental design and provides details on the experimental setting and procedures. Section \ref{sec:result} reports the main findings and Section \ref{sec:mec} examines potential mechanisms. Section \ref{sec:rob} shows the robustness checks we run on the main results. Section~\ref{sec:conclusion} concludes with policy implications and directions for future research. Additional analyses and extended institutional details are provided in the Appendix \ref{sec:app}.

\section{Literature Review}\label{sec:lit}

Our study uses a lab-in-the-field experiment with naturally occurring identities to provide new evidence on the relationship between immigration, social identity, and cooperation. By integrating insights from experimental economics, social psychology, and labor economics, we contribute to a broader understanding of cooperative behavior in ethnically diverse environments.

Closest to our work, \citep{butler2024causal} studies how cultural identity affects cooperation in a public-good setting, showing that identity can significantly shape non-kin cooperation in the lab.\footnote{See, for example, \cite{butler2024causal}.} We differ by focusing on naturally occurring native–migrant identities in adolescent classrooms, by comparing multicultural and common-identity priming, and by studying punishment as well as contributions.

A large body of research on immigrant–native integration has focused on economic outcomes such as labor market performance, health, and educational attainment \citep[see][]{brell2020labor, edo2019impact, giuntella2018reason, schneeweis2011educational}. Other work examines social integration through identity formation, cultural values, intermarriage, residential patterns, and political participation \citep{laurentsyeva2017social}. Yet, integration also depends on everyday social interactions and on the ability of migrants and natives to cooperate across cultural lines—a dimension that is comparatively underexplored in the economic literature.\\
Developmental work on bicultural youth similarly emphasizes that immigrant adolescents often navigate multiple, overlapping identities, with implications for their social interactions and adjustment (see e.g. \cite{vedder201415}).

Most experimental work on native–migrant interactions studies discrimination in trust-based games. The seminal contribution by \cite{fershtman2001discrimination} documented systematic trust discrimination between Jewish subgroups in Israel. Similar patterns have been observed in Australia \citep{guillen2011trust}, the United States \citep{cox2015trust}, and the Netherlands \citep{cettolin2019return}. More recent studies demonstrate that exogenous shocks can reshape trust discrimination: \cite{albrecht2022social} show that refugee resettlement increased trust toward refugees in Australia, while \cite{felfe2021more} find that exposure to a German birthright citizenship reform increased immigrant children's trust in native peers. Relatedly, \cite{achard2025local} find that local exposure to refugees during the European crisis improved Dutch residents' attitudes toward ethnic minorities and reduced far-right voting, with effects driven by interethnic contact rather than large-scale shocks. However, this literature largely focuses on dyadic trust and discrimination, rather than on group-level public good provision and norm enforcement in naturally diverse groups

A related strand investigates how identity shapes cooperative behavior in public-good and related games. \cite{chen2014hat} study interactions between Caucasian and Asian participants in Prisoner’s Dilemma and minimum-effort games and test whether salient ethnic identity affects discriminatory behavior. They find heterogeneous effects: activating ethnic identity reduces effort in the minimum-effort game, whereas a common-identity prime improves coordination in the Prisoner’s Dilemma. These results highlight that identity salience interacts with the strategic environment. Similarly, \citet{butler2024causal} show that cultural identity can causally affect cooperation toward non-kin in a public-good experiment, underscoring that identity is an important determinant of cooperative behavior even outside migration contexts. Experimental work on identity and social exclusion also documents how minority status can affect patterns of inclusion and cooperation in school-age samples \citep{candelo2017identity}.

Our work builds on these insights, shifting the focus from dyadic interactions to group-level cooperation and punishment dynamics in naturally diverse populations.

While we do not study discrimination directly, we make a distinct contribution by testing whether cooperation in mixed groups can be shaped through identity salience. To our knowledge, this is the first study to examine the effect of identity priming on Public Goods Game behavior in groups comprising both native and migrant adolescents.

Following the priming framework introduced by \cite{shih1999stereotype} and applied in economics by \cite{chen2014hat}, we activate identity dimensions through a short pre-experimental questionnaire and visual cues shown during the game. Our design follows the broader social-priming literature in using subtle identity cues to shift the salience of particular group memberships without altering incentives (e.g. \cite{molden2014understanding}). These subtle interventions heighten the salience of specific identity components without requiring explicit reflection. We implement two priming treatments: a \emph{Multicultural Identity} treatment that highlights ethnic diversity, family origins, and cultural heritage; and a \emph{Common Identity} treatment that emphasizes shared membership in the school community. These dimensions align with \cite{akerlof2000economics}, who conceptualize immigrants’ identities as comprising both ethnic and institutional components.\footnote{Two experimental strategies dominate the literature on identity: induced identities generated in the laboratory and priming of naturally occurring identities \citep{charness2020social}. While most studies rely on the former, we prime naturally held ethnic and school-based identities that are meaningful to participants. Naturally occurring identity primes have been shown to affect economic behavior across domains \citep{peffley2007persuasion, mcleish2011social, chen2014hat, hetey2014racial, cohn2015bad, li2017common, cohn2017professional, hetey2018numbers, chang2019rhetoric}.}

Our study departs from prior work in several important ways. First, we examine group-level cooperation and punishment rather than bilateral trust or coordination, thereby capturing dynamics that more closely resemble interaction structures in real social settings. In doing so, we provide, to our knowledge, the first experimental evidence on how identity priming shapes public-good cooperation and punishment between natives and migrants in naturally diverse youth groups. Second, we vary group size (3, 4, or 5 members) to examine how cooperative behavior responds to changes in group composition. Third, unlike experiments restricted to two predefined ethnic groups, our subject pool exhibits substantial natural heterogeneity in migration backgrounds. Finally, our participants—middle school students aged 11–13—belong to the first Italian cohorts widely exposed to ethnic diversity from early schooling.\footnote{Participants were born between 2006 and 2011 and began primary school during a period of rising diversity in Italian classrooms \citep{miur2021}.}

Our work also complements \cite{alan2021building}, who study prosociality in ethnically mixed Turkish schools following a major curricular reform. Their setting involves a large policy shock and two dominant ethnic groups. In contrast, our identity-priming interventions are low-cost, scalable, and suited to environments where diversity emerges gradually through long-term migration. Moreover, many immigrant students in our sample were born in Italy but do not hold Italian citizenship, providing a unique perspective on identity salience and cooperation among second-generation youth. More broadly, our results complement field evidence that large institutional changes—such as citizenship reforms—affect trust and integration (e.g. \cite{felfe2021more}), by showing that low-cost identity priming can shift cooperative behavior and punishment norms in everyday school environments.

Together, these features allow us to causally identify how different dimensions of naturally held identity shape cooperation in heterogeneous groups, and to do so in a population where the formation of social norms around diversity is still unfolding. This positions our study as one of the first to test identity priming in a public-goods environment using a naturally diverse youth population, offering new insights into the mechanisms supporting cooperation in multicultural societies.

\section{Institutional Setting}
\label{sec: institutional setting}
In this section, we outline the demographic and educational context of our study, focusing on immigration in Italy and the structure of the school system. A more detailed description is provided in Appendix~\ref{appA}.
\subsection{Immigration in Italy} 
Since the early 2000s, Italy has experienced a steady increase in the share of students without Italian citizenship. By the 2019/2020 school year---when our experiment was conducted---non-Italian students represented about 10\% of the national student population, and more than 60\% of them were born in Italy \citep{miur2021}. This highlights a distinctive feature of the Italian context: although many students are formally classified as non-Italian due to citizenship rules, most have spent their entire lives in the country and are integrated into Italian schools from an early age.

Emilia-Romagna, the region where our study took place, records the highest share of non-Italian students nationwide (17.1\%). In Bologna, this figure is 16.4\% among middle school students, mirroring the broader demographic diversity of the city. The two districts involved in our study—Borgo Panigale–Reno and Savena—are among the most diverse in Bologna, with foreign resident shares of 15–16\% and substantial variation in the socioeconomic composition of their neighborhoods.
In line with Italy’s \emph{jus sanguinis} citizenship regime, we classify as “immigrant students’’ those whose parents were not born in Italy, including both first- and second-generation immigrants. In our sample, approximately 75\% of immigrant students were born in Italy, while the remainder came primarily from other European countries, Asia, and Africa. Table~\ref{tab:country} reports the distribution of countries of origin.

\subsection{The Educational System in Italy}

Italian schooling is publicly funded and compulsory from ages 6 to 16. Students progress through primary school, middle school, and then choose one of three high school tracks (academic, technical, or vocational). Schools enjoy considerable autonomy in curriculum design and class organization, and students typically remain with the same classmates across all subjects. This structure creates stable peer groups over several years, making schools a particularly well-suited environment for studying social interactions.

Classroom diversity has become an increasingly central policy issue, and the Ministry of Education has introduced a series of national guidelines to promote inclusion and support students with non-Italian citizenship. A key recommendation is that the proportion of students with limited Italian proficiency should generally not exceed 30\% per class, although schools may adjust this threshold based on the linguistic skills of their student population. Emilia-Romagna—the region of our study—has the highest share of classes meeting or exceeding this limit.
Table~\ref{tab:school_char} summarizes the characteristics of the two schools involved in our study, where the average share of immigrant students per class is approximately 33\%, reflecting the high levels of diversity in the area.

\section{Experimental Framework}\label{sec:design} 

\subsection{Schools}

The experiment was implemented as a within-class randomized controlled trial. The pool consists of 390 students from two middle schools in Bologna, a municipality in the Emilia-Romagna region of Italy. The average number of students per class is around 19, and approximately 33\% of students in each class are immigrants.\footnote{Class sizes are slightly larger in School 2 than in School 1. However, because treatment is randomized within each class, these differences do not affect identification or treatment effect estimates.}

The timeline of our study is as follows. In spring 2019, we first examined the distribution of immigrant students across classes using administrative data from the Ministry of Education and class numbers from school websites.\footnote{We collected information on the number of classes in each school from the schools’ websites and focused on those with an immigrant-student share close to the 30\% threshold, computed using publicly available Ministry of Education data: https://dati.istruzione.it/opendata/progetto/.}
 This allowed us to prioritize schools with larger student bodies and sufficient diversity. At the start of the 2019/2020 academic year, we recruited two schools—one in the “Borgo Panigale–Reno” district and one in “Savena”—as shown in Table~\ref{tab:school_char}. Each school headteacher received a description of the project and IRB approval during our initial visit. Together with the headteachers, we then scheduled class-by-class meetings during school hours to ensure a consistent research protocol and to avoid conflicts with afternoon activities.\footnote{We also recruited a pilot school in June 2019 to test the duration of the experiment, device functionality, and the clarity of instructions. The procedure is described in Section~\ref{subsec:procedures}.}

\subsection{Experimental Design}

The experiment took place over two meetings of about two hours each, held one week apart. In the first meeting, students were randomly assigned to one of the priming treatments and played the Public Good Game on tablets. Randomization, priming, and game-play were implemented using oTree \citep{chen2016otree}. 
The second meeting was devoted to collecting the additional measures described in Section~\ref{subsec:other_measures}, including incentivized and non-incentivized tasks—administered mostly with pen and paper—as well as a questionnaire on students’ demographic characteristics.

\subsubsection{Priming Treatment}\label{subsec:priming}

The priming intervention was designed to manipulate the salience of shared or distinct identity elements and examine how these influence cooperation in the Public Good Game. Within each class, students were randomly assigned to one of three treatments, ensuring that class-specific factors (peer networks, teaching styles, curricula, etc.) remained constant across treatments. The treatments were:

\begin{itemize}
\item \textbf{Common Identity Treatment}: activates a shared school identity (e.g., attending the same middle school);
\item \textbf{Multicultural Identity Treatment}: primes students’ ethnic background while explicitly conveying the coexistence of multiple identities within the group—for example, by displaying the flags of the various communities represented in the city;
\item \textbf{Control Treatment}: provides a neutral baseline with no identity content.
\end{itemize}

\begin{figure}[H]
    \centering 
    \subfigure{\includegraphics[width=0.60\textwidth]{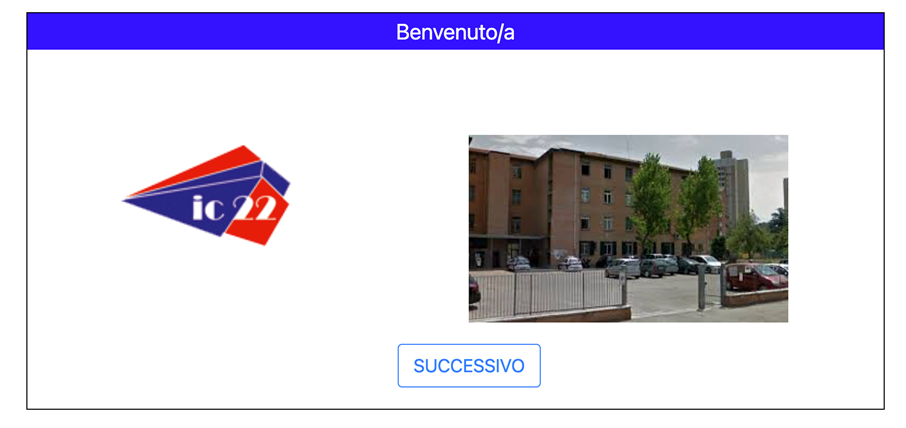}} 
    \subfigure{\includegraphics[width=0.60\textwidth]{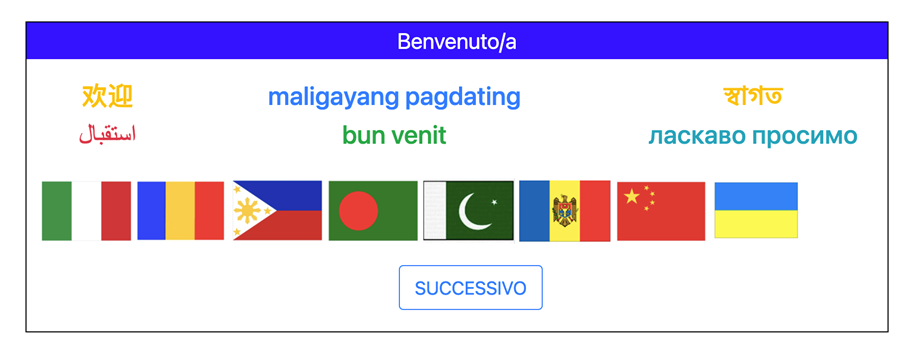}}
    \subfigure{\includegraphics[width=0.60\textwidth]{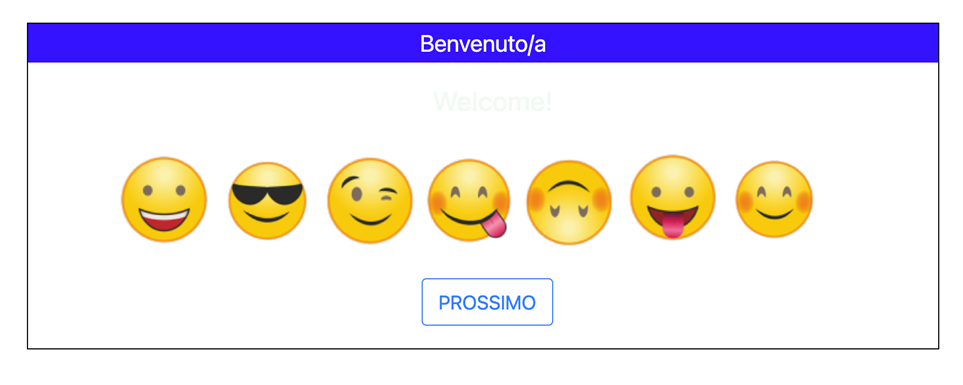}}
   \caption{Welcome screens displayed at the beginning of the session and during all transition phases between tasks, by treatment. \\
(a) \emph{Common Identity}: a screen emphasizing shared school belonging through the school's name and visual elements. 
(b) \emph{Multicultural Identity}: a multilingual welcome screen showing the flags of the largest migrant communities in Bologna, highlighting local linguistic and cultural diversity. 
(c) \emph{Control}: a neutral screen without identity-related content.}
    \label{fig:Screens1}
\end{figure}

We used the priming technique first introduced by \cite{shih1999stereotype}, according to which a specific identity can be activated unconsciously by exposing participants to certain stimuli (e.g., images, videos, and pre-experimental questionnaires) and its adaptation to migration contexts in \cite{chen2014hat}. Table~\ref{tab:questionnaires} reports the questionnaires used in each treatment, and Figure~\ref{fig:Screens1} shows the corresponding welcome and waiting screens.

In the Common Identity treatment, we primed a shared school identity. Participants answered questions about the middle school they attended—why they chose it and what they liked most (Table~\ref{tab:questionnaires}). During the experiment, they also viewed “welcome’’ and “waiting’’ screens displaying pictures and the logo of their school (see the first panel of Figure~\ref{fig:Screens1}).
In the Multicultural Identity treatment, we highlight the presence of different ethnic and cultural backgrounds by asking participants questions such as: “What is the country of origin of your family?” or “Which language is spoken in your family?”, as shown in Table~\ref{tab:questionnaires}. Unlike \cite{chen2014hat}, who study a single minority group (Chinese) within a native Australian context, our setting—as is common in Italy—includes several minority communities. Consequently, the treatment not only activates students’ own ethnic origins but also conveys a broader message of multiculturalism. To reinforce this, the “welcome’’ and “waiting’’ screens display a multilingual message and the flags of the main ethnic communities represented in the municipality of Bologna (Figure~\ref{fig:Screens1}).\footnote{Information on ethnic communities is retrieved from the statistical registry of residents in the municipality of Bologna (ISTAT): http://inumeridibolognametropolitana.it/dati-statistici/popolazione.}
For the Control group, we administered neutral questions (e.g., extracurricular activities, TV habits) and used neutral screens throughout the experiment, ensuring that participants experienced the same procedure but without activating any particular dimension of identity (third panel of Figure~\ref{fig:Screens1}).

After the pre-experimental priming questionnaires, participants played the Public Good Game, during which they repeatedly viewed the treatment-specific “welcome’’ and ``waiting'' screens. At the end of the game, all students -regardless of treatment assignment- answered the priming questions used in the other two treatment arms, which allows us to compare identity salience across treatments in a consistent way. They then completed an open-ended self-description task, framed as a letter to a student in Sweden of the same grade, following the approach in \cite{bugental1950investigations}. We use this task to validate the effectiveness of the priming manipulation (see Section \ref{subsec:effectiveness}).

In addition, we collected individual background information, social network data, and a series of incentivized and non-incentivized measures of altruism, risk-taking, trust, and ethnic stereotypes. While a few of these measures were administered during the first meeting, most were collected in the second meeting (details in Section~\ref{subsec:other_measures}).

\subsubsection{Public Good Game}\label{subsec:public_good}

To elicit cooperative preferences, we used a standard Public Good Game implemented in the take frame \citep{fosgaard2019cooperation} under a partner-matching protocol within each class. We adopted this frame because Fosgaard et al. (2019) show that it is the most intuitive and easiest to understand for non-standard (i.e., non-university) participants, particularly in repeated decision settings. 
Since the number of students present on the day of data collection could not be determined ex ante, each session began with a headcount. We then used an automated algorithm to randomly assign students to one of the three priming treatments and to groups of size 3, 4, or 5.\footnote{Although consent forms were collected about a week before each session, some students were occasionally absent due to illness or other reasons.}
After the first part (10 rounds), participants received the instructions for the second part, which was identical except for the introduction of punishment opportunities. Importantly, participants knew from the outset that the experiment consisted of several parts, but they only learned the content of each subsequent part after completing the previous one. In each round of the Public Good Game, groups started with a common pool containing 150, 200, or 250 points depending on group size (i.e., 50×N points). Each participant began with zero points and chose either to leave points in the common pool or to take between 0 and 50 points (in increments of 5). Points taken represented private earnings, while the remaining amount in the common pool was doubled and then evenly redistributed among all group members.\footnote{The experiment was conducted in Italian. An English translation of the instructions appears in Online Appendix A, while the original Italian version is available at the following \href{https://www.dropbox.com/scl/fi/iswrdcr4dv21y72got86o/Istruzioni_Take_Frame_Ita_PRINT-version.docx?rlkey=d4x6zr55rilqaac6s5cr6liik&dl=0}{link}.}
All instructions were read aloud, and participants completed comprehension questions before proceeding. The written instructions included an example for a group of four participants, and we also solved the corresponding cases for groups of three and five on the blackboard to ensure full understanding. At the end of each round, participants received detailed feedback: how many points they left in the common pool, how many the group collectively left, their private earnings, their share of the doubled common pool, and their total earnings for that round.

\subsubsection{Other measures}\label{subsec:other_measures}
\begin{figure}[h]

\includegraphics[width=12cm]{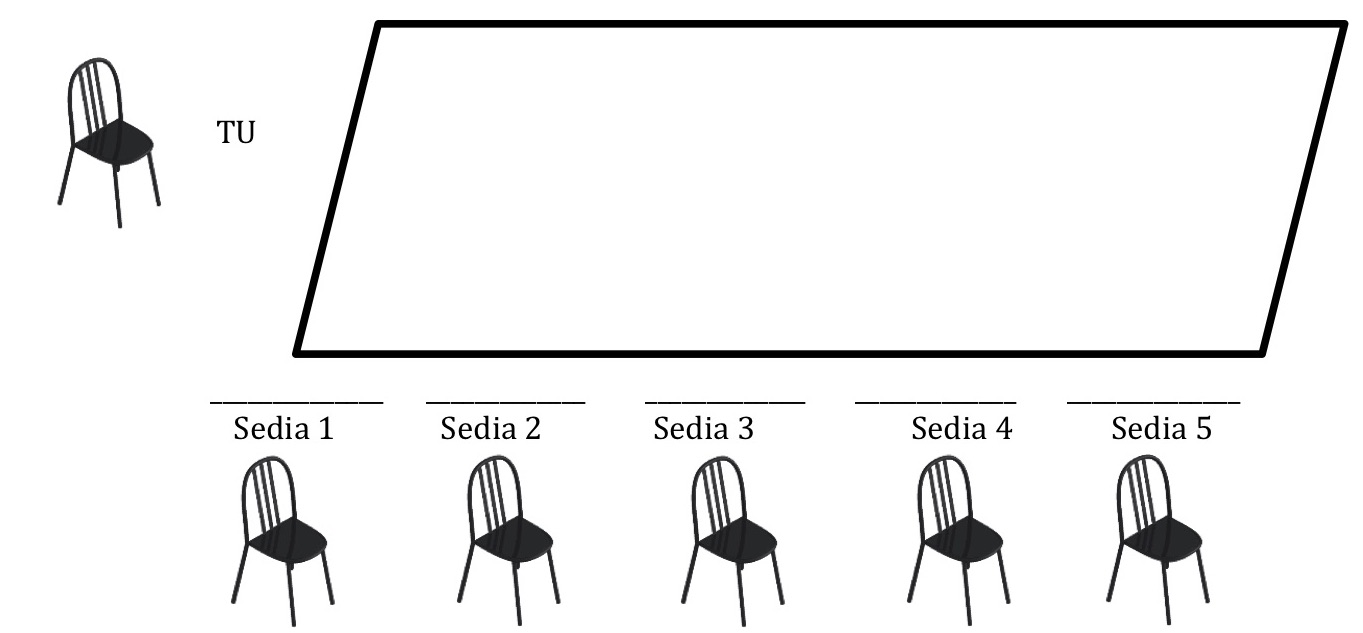}
  \caption{Friendship elicitation task. \\
Each student received a sheet depicting a table with five chairs.
They were instructed to write their own ID number on the chair at the head of the table (“TU”) and then list up to five classmates—one per remaining chair (Sedia 1-..- Sedia 5)— whom they would most like to sit next to, ordered from closest to farthest. 
These entries are used to construct directed friendship links within each classroom. \label{fig:friends}}
\end{figure}

After the Public Good Game, participants completed an open-ended self-description task, framed as a letter to a student in Sweden \citep{bugental1950investigations}. As illustrated in the priming section, this task serves to validate the manipulation, with treated students expected to refer more often to the primed identity dimension. Results are reported in Table~\ref{tab:mentions} and discussed in Section~\ref{subsec:priming}.
After this activity, we also collect each student's friends network using the methodology applied by \cite{landini2016friendship} and \cite{chen2016group}. These data enable us to construct measures of the intra- and intercultural friendship links among class peers, which can be interpreted as indicators of socialization and assimilation between students from different cultural backgrounds in the class (see, for example, \cite{patacchini2016social} and \cite{facchini2015migration}).
Specifically, as in \cite{landini2016friendship}, participants had to fill out a sheet of paper depicting a table and five chairs. Each participant had to write his/her ID number on the chair on the head of the table and (up to) five other ID numbers of his/her friends in the same class that s/he would like to have seated close to him/her on the same line (from the closer to the farther, see Figure \ref{fig:friends}). We informed participants that we would keep the IDs they reported confidential and that neither the parents, the teachers, nor other friends would know what they wrote. Children received a fixed number of points for completing this task. 

After this activity, we collected incentivized measures on generosity using the standard Dictator Game \citep{kahneman1986fairness} and on fairness views using the impartial version of the same game \citep{konow2000fair}. Participants were divided into pairs comprising one dictator and one recipient. Dictators were endowed with 10€ and had to decide how much to keep for themselves and how much to give to the recipient, who had no decision to make. A self-interested dictator should keep the entire endowment, while positive transfers are interpreted as a proxy for generosity. Here, we used role reversal; all players were asked to make decisions as dictators. Roles were assigned and revealed at the end of the experiment.

Finally, participants also answered a questionnaire that collected information on household composition, participants’ aspirations, risk preferences, general trust, and stereotypes. Notice that the questionnaire was collected in a second meeting with each class, which was scheduled one week after the first meeting; during the second meeting, we also finalized the data collection of the first meeting, in case students didn't finish all the activities in the first meeting. In the next section, we provide more details on the experimental procedures.

\subsection{Experimental Procedures}\label{subsec:procedures}

As introduced in Section~\ref{sec:design}, we coordinated with each school headteacher to schedule class-level sessions -held in December 2019 in School~1 and in February 2020 in School~2. About one week before each session, parents or legal guardians received a leaflet describing the study and were asked to sign a consent form.\footnote{Participation rates exceeded 95\%. Children who did not participate assisted the experimenters with distributing materials and received a small show-up prize to avoid absences or any stigma associated with non-participation.}

In each school, we set up a temporary laboratory in a dedicated room. At the beginning of each session, one class at a time entered the lab, accompanied by their teacher, who then left; all data collection was conducted solely by the research team. Students completed the experimental tasks individually, earning points as they progressed. Points were converted into euros at an exchange rate of 100 points = 1 euro and paid using Feltrinelli\footnote{Feltrinelli is an Italian book store, for more details see the following \href{https://www.lafeltrinelli.it/?gad_source=1&gad_campaignid=19777030988&gbraid=0AAAAAD-Pe5zoBHxqh2xKtyswUgaNSsOWK&gclid=CjwKCAiAraXJBhBJEiwAjz7MZUyiin8ZAg4EMkgG7y2RiMQ5mLvXcdau7BwvPlEDuTsONX-fOLpRdRoC61sQAvD_BwE}{link}.} or Amazon gift cards. The average payment was approximately 8 euros per participant.

For the additional measures collected in the second meeting (Section~\ref{subsec:other_measures}), we experienced partial attrition in School~2 because some class sessions were postponed to June 2020 due to the school closure during the first wave of the COVID-19 emergency. We systematically tested whether attrition was correlated with treatment assignment and found no evidence of differential attrition across treatments. Table~\ref{tab:attrition_treatment} shows no treatment differences in the likelihood of completing the activities in February versus June; Table~\ref{tab:attrition_never_treatment} confirms no differences in overall completion probabilities. Observable characteristics measured in the first meeting do not differ between early and late completers (Table~\ref{tab:attrition1_ind_char}) nor between completers and non-completers (Table~\ref{tab:attrition2_ind_char}). These checks indicate that neither midline nor endline attrition generates selection concerns.
The main analysis relies solely on outcomes collected in the first meeting
. When additional measures are used, the sample is restricted to those who completed them before the COVID-19 disruption.

\section{Theoretical Framework: Identity, Norms and Cooperation}

This section provides a conceptual framework to rationalize how identity salience affects cooperation in heterogeneous groups. The framework builds on the economics of identity \citep{akerlof2000economics} and on models of norm-based cooperation and norm enforcement \citep{fehr2000cooperation, herrmann2008antisocial}. Its key feature is that identity priming affects behavior by altering the salience of reference-group norms, rather than by changing material incentives or group composition.

Consider individual $i$, observed in round $t$. In each round, individual $i$ chooses a contribution $Y_{it} \in [0,50]$ to the public good, interpreted as the number of points left in the common pool. Monetary payoffs are given by:
\begin{equation}
\pi_{it} = \pi(Y_{it}, Y_{-i,t}),
\end{equation}
where $Y_{-i,t}$ denotes the vector of contributions by other group members.

In our experimental framework, in a group of size $N$, each point left in the common pool is doubled and evenly redistributed. The monetary payoff from contributing $Y$ points is therefore:
\begin{equation}
\pi(Y, Y_{-i}) = (50 - Y) + \frac{2}{N} \left( Y + \sum_{j \neq i} Y_j \right),
\end{equation}
where $50 - Y$ denotes points taken for private use. The marginal monetary payoff of contributing one additional point is constant and given by:
\begin{equation}
\frac{\partial \pi}{\partial Y} = -1 + \frac{2}{N},
\end{equation}
which is strictly negative for all group sizes used in the experiment ($N = 3,4,5$). Thus, in the absence of non-material motives, the unique payoff-maximizing strategy is $Y = 0$.

Following \cite{akerlof2000economics}, utility depends not only on material payoffs but also on conformity to identity-based behavioral prescriptions. Individuals experience disutility from deviating from contribution levels prescribed by the norms associated with salient identity dimensions. Each identity is associated with a reference contribution norm  $\tilde Y^\ast_i$.

In addition, individuals may exhibit other-regarding preferences, deriving utility from the overall level of cooperation in the group, as in standard models of social preferences \citep{fehr1999theory, charness2002understanding}. Utility is given by:
\begin{equation}
U_{it}
=
\pi(Y, Y_{-i})
- \lambda_i \left( Y - \tilde Y^\ast_i \right)^2
+ \phi_i W\!\left( \sum_{j} Y_j \right),
\end{equation}
where $\lambda_i \geq 0$ captures the degree of norm internalization, $\tilde Y^\ast_i$ denotes the identity-based contribution norm that is salient at the time of choice, $\phi_i \geq 0$ captures the weight placed on group welfare, and $W(\cdot)$ is an increasing function.

A key feature of the framework is that individuals may differ not only in their sensitivity to norms ($\lambda_i$) and in their other-regarding preferences ($\phi_i$), but also in whether identity-related norms are already salient in the absence of experimental manipulation. In particular, for some individuals—such as those with a migration background—identity-related norms may be chronically salient due to repeated exposure in everyday interactions, while for others these norms may be activated only situationally \citep{benjamin2016religious}.

Identity priming is modeled as affecting the salience of reference norms $\tilde Y^\ast_i$, rather than the underlying norms $Y^\ast_{i}$ themselves, norm internalization $\lambda_i$, or social preferences $\phi_i$. That is, priming alters which identity dimension is cognitively accessible at the time of decision, consistent with the social-priming literature \citep{shih1999stereotype, chen2014hat}.

The individual optimization problem is:
\begin{equation}
Y_{it}^{\text{opt}}
=
\arg\max_{Y \in [0,50]}
\left\{
(50 - Y)
+ \frac{2}{N} \left( Y + \sum_{j \neq i} Y_j \right)
- \lambda_i (Y - \tilde Y^\ast_i)^2
+ \phi_i W\!\left( \sum_j Y_j \right)
\right\}.
\end{equation}

Because the monetary payoff is linear in $Y$ and the cost of deviating from the salient identity-based norm is strictly convex and enters negatively, the objective function is strictly concave for $\lambda_i>0$, ensuring a unique interior solution. Abstracting from strategic considerations regarding expectations about others’ contributions, the first-order condition yields:
\begin{equation}
Y_{it}^{\mathrm{opt}}
=
\tilde Y^\ast_i
+ \frac{1}{2\lambda_i}
\left( -1 + \frac{2}{N} \right)
+ \Gamma(\phi_i),
\end{equation}
where $\Gamma(\phi_i)$ is a weakly increasing function capturing the contribution-enhancing effect of other-regarding preferences, under standard monotonicity assumptions.

This expression illustrates that optimal contributions increase with the salience of cooperative identity norms (captured by $\tilde Y^\ast_i$), the degree of norm internalization (captured by $\lambda_i$), and the weight placed on group welfare (captured by $\phi_i$). For expositional convenience, we summarize optimal behavior in the reduced form:
\begin{equation}
Y_{it}^{\mathrm{opt}} = \alpha + \beta \, \tilde Y^\ast_i(T_i),
\qquad \beta > 0,
\end{equation}
where $\alpha$ captures the net effect of material incentives and other-regarding preferences, and $\beta$ measures the responsiveness of contributions to shifts in the salient identity-based norm induced by the priming treatment $T_i$.

The framework developed above has the following implications for the empirical analysis:

\begin{itemize}
\item Cooperative behavior may differ systematically across social groups due to
heterogeneity in the salience or internalization of identity-based norms, even in the
absence of differences in material incentives or observable characteristics.

\item Identity priming operates by shifting the salience of reference-group norms and may therefore generate asymmetric responses across individuals for whom such norms are
already salient versus situationally activated.

\item When punishment is available, sanctioning behavior can be interpreted as a form of
norm enforcement, providing complementary evidence on the role of identity-based norms
in shaping cooperative behavior.
\end{itemize}
These considerations guide our empirical analysis of contribution and punishment behavior
under different identity priming conditions.

\section{Results}\label{sec:result}

In this section, we examine the effects of the priming interventions on cooperative behavior in the Public Good Game. We first assess whether the priming successfully activated the intended identity dimensions and then analyze contributions in the first part of the experiment (without punishment). We subsequently turn to the second part, where punishment opportunities were introduced, and study both contribution decisions and sanctioning behavior. Throughout, we compare outcomes across priming treatments and by immigration status. Because students are nested within classes, classes within schools, and schools within districts, we estimate Multilevel Mixed-Effects Models to account for dependence in the data. Section \ref{sec:rob} shows that our main findings are robust to alternative estimation strategies.

\subsection{Priming effectiveness}\label{subsec:effectiveness}

We use the open-ended self-descriptions to assess whether the priming interventions successfully activated the intended identity dimension. Table~\ref{tab:mentions} reports how the share of words referring to each identity type varies across treatments. Column~(1) shows changes in the probability of mentioning origin-related terms (e.g., ethnicity, place of birth, or residence) when exposed to the Multicultural or Common priming relative to the Neutral treatment; Column~(2) performs the analogous test for references to the school or student identity.
Across both dimensions, students in the relevant priming treatment mention the corresponding identity more frequently than those in the Neutral group. These patterns confirm that our priming interventions effectively increased the salience of the targeted identity, consistent with the validation found in previous studies using similar techniques \citep{shih1999stereotype, chen2}

\subsection{Identity Priming and Cooperative Behavior}

\begin{figure}[htpb]
\centering
    \begin{minipage}{0.75\textwidth}
        \subfigure[Overall]{
            \includegraphics[width=\textwidth]{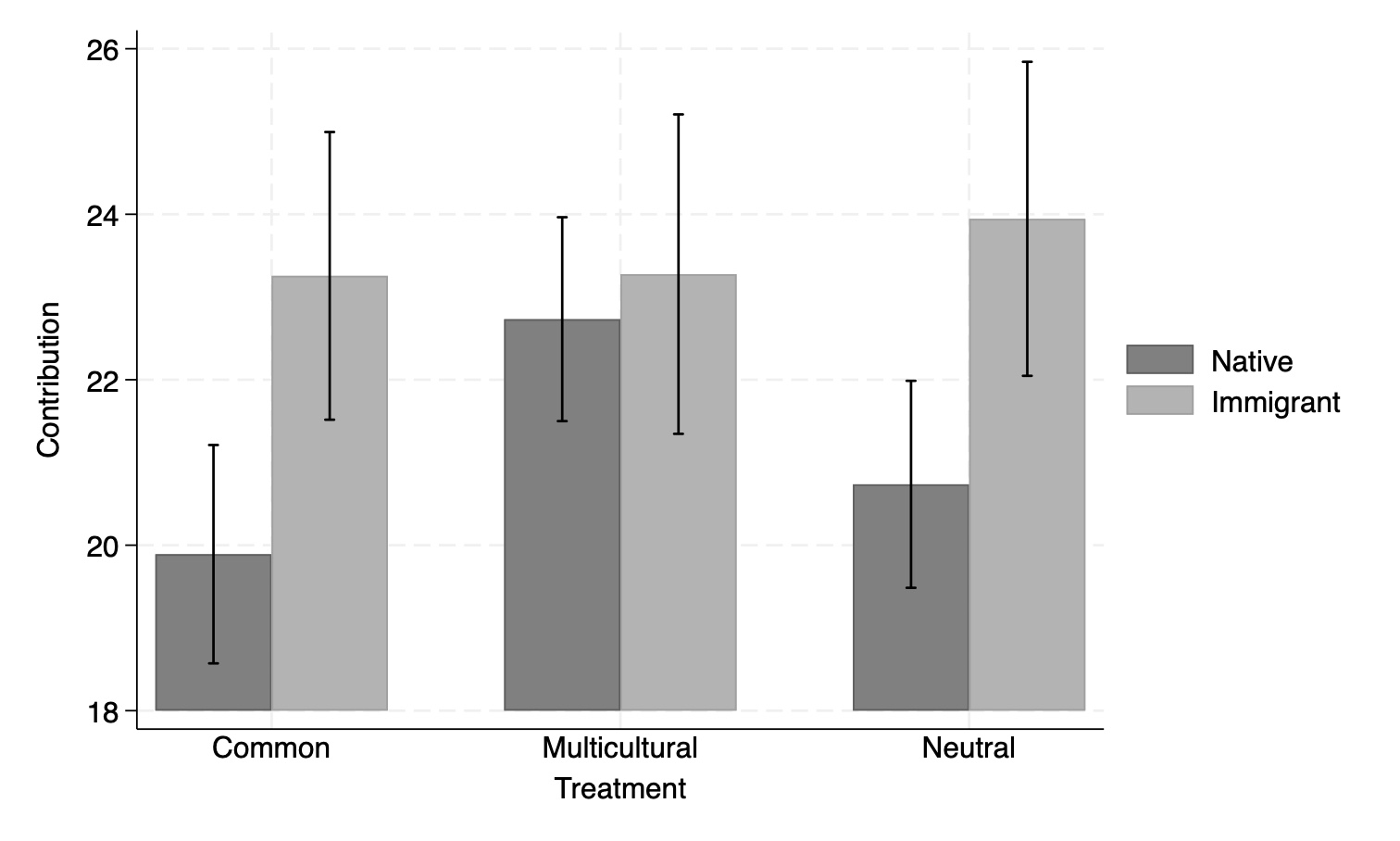}
        }
    \end{minipage}

    \vspace{0.4cm}

    \begin{minipage}{0.75\textwidth}
        \subfigure[Over rounds]{
            \includegraphics[width=\textwidth]{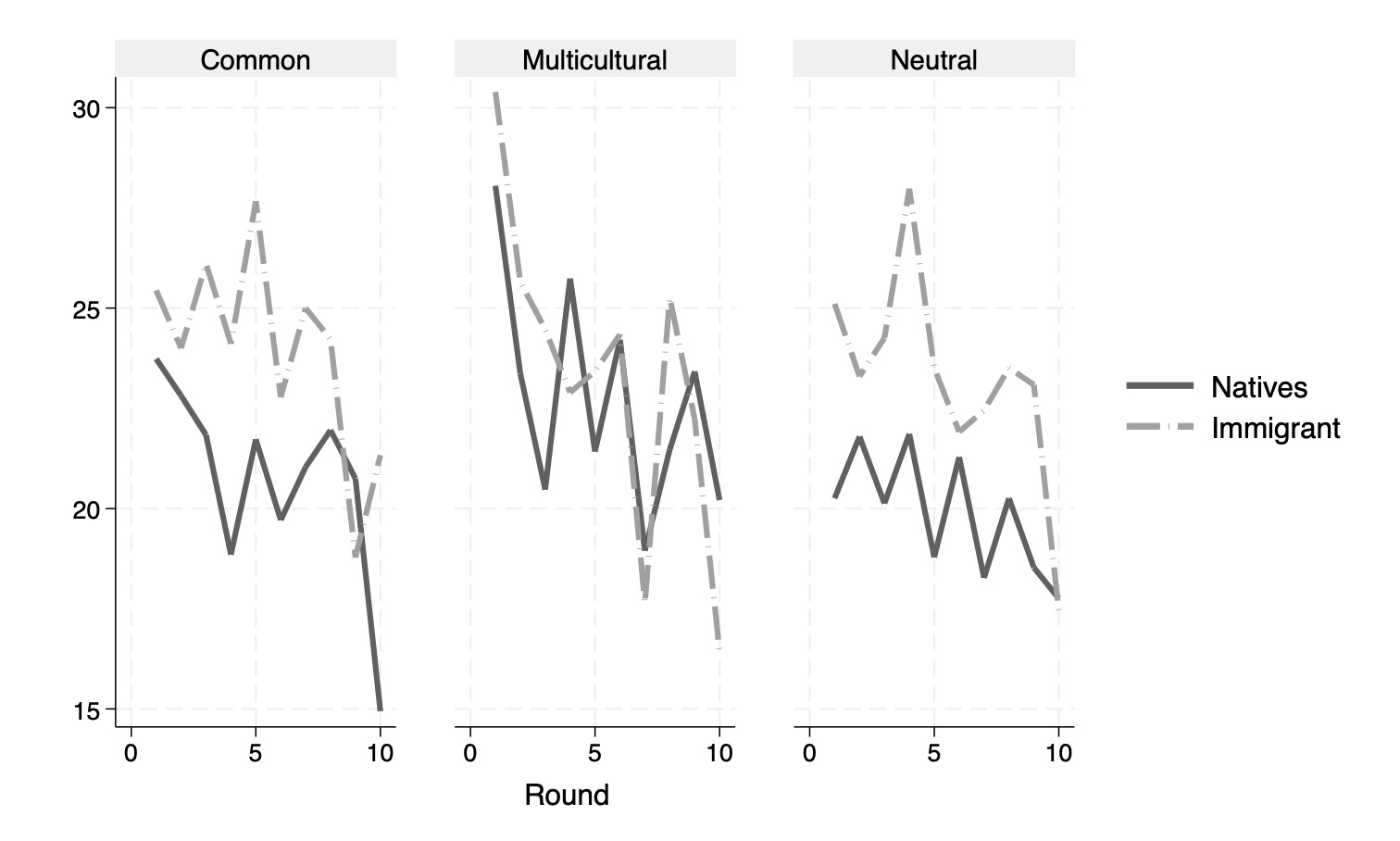}
        }
    \end{minipage}
{
\caption{Contribution to the PGG by treatment and status}
\label{fig:contrib_treatstatus}
\begin{minipage}{0.98\textwidth}\footnotesize
\textit{Notes:} Panel (a) shows average contributions over the ten rounds of the Public Good Game by treatment and immigration status. Immigrant students contribute more than natives in the Neutral and Common Identity conditions. Under the Multicultural Identity priming, natives' contributions shift upward, narrowing the baseline gap. Panel (b) displays round-by-round contributions, confirming that the Multicultural prime raises natives' cooperation across all rounds, while immigrants’ behavior remains stable across treatments.
\end{minipage}}
\end{figure}

In this subsection, we present the main results from the first part of the experiment, where participants played the Public Good Game without punishment (Section~\ref{subsec:public_good}). Figure~\ref{fig:contrib_treatstatus} plots individual contributions (i.e., points left in the common pool) by treatment and immigration status. Panel~(a) shows average contributions over all ten rounds, and Panel~(b) shows the round-by-round dynamics for natives and immigrants under each priming treatment. In the Neutral treatment, immigrants contribute substantially more than natives—both on average and in every round—indicating a clear baseline cooperation gap. This gap remains essentially unchanged under the Common Identity treatment. In contrast, under the Multicultural Identity treatment, natives consistently increase their contributions across all rounds, effectively shifting their contribution path upward. Thus, the descriptive evidence suggests that multicultural priming increases cooperation, and this effect is driven by natives rather than immigrants.

These results—the higher baseline contributions of immigrants and the positive effect of the Multicultural Identity priming on natives—are confirmed by a more formal analysis. Table~\ref{tab:xtmixed_contribution} reports estimates from a regression model using all rounds from the first part of the experiment (periods 1-10). Since contributions are observed repeatedly for each individual, and individuals are nested within groups, classes, and schools, we estimate a four-level mixed-effects model to account for dependence in the data.

		\begin{eqnarray} \label{t:eqn1}
		\begin{split}
		Y_{itgcs} & = \alpha + \beta_1 Treatment_{i} + \beta_2 Immigrant_{i} + \beta_3 Treatment_{i}* Immigrant_{i} + \\
		& +\beta_4 X_{i} +\beta_5 Z_{g} +\tau_{it} + \delta_{i} + \delta_{i}*\tau_{it} +\gamma_{g} +\lambda_{c} + \psi_{s} +\epsilon_{itgcs}
		 \end{split}
		\end{eqnarray}
        
where, \(Y_{itgcs}\) denotes individual contributions (points left in the common pool) in each round; \(Treatment_{i}\) indicates exposure to the Common, Multicultural, or Neutral priming; and \(Immigrant_{i}\) is a dummy for immigration status (equal to 1 if at least one parent is not Italian). \(X_{i}\) includes individual controls (gender and group size), while \(\tau_{t}\) captures round fixed effects. 
The model includes random intercepts and slopes at the individual level (\(\delta_{i}\), \(\delta_{i} \times \tau_{t}\)) and random intercepts at the group (\(\gamma_{g}\)), class (\(\lambda_{c}\)), and school (\(\psi_{s}\)) levels, with \(\epsilon_{itgcs}\) as the error term. This structure accounts for repeated decisions within individuals and the hierarchical clustering of participants. Appendix~\ref{tab:contribution} reports analogous results using ordinary least squares.

Columns~1--2 of Table~\ref{tab:xtmixed_contribution} report estimates of equation~\ref{t:eqn1} for the full sample (390 participants observed over 10 rounds). Columns~3–4 restrict the model to immigrants (N=130), while Columns~5–6 focus on natives (N=260). In the full sample, the coefficient on \emph{Immigrant} is positive and significant across specifications, indicating that immigrants contribute more than natives at baseline---by more than 3 points, or roughly 16\% of the average contribution in the Neutral treatment. Group size also matters: participants in groups of five contribute more than those in groups of three.
Turning to treatment effects, both \emph{Common} and \emph{Multicultural} priming are associated with higher contributions in the full sample, but the effect of the latter is substantially larger. The interactions between treatment dummies and immigration status are not significant, indicating no differential response among immigrants. Indeed, when estimating the model separately for immigrants (Columns~3–4), neither priming treatment has a significant effect, though group size remains positively related to contributions.
By contrast, results for natives (Columns~5–6) show that the \emph{Multicultural} prime significantly increases contributions at the 1\% level, whereas the \emph{Common} prime has no discernible effect. The coefficient on the female dummy is also positive and significant in both native-only models.

The results of the first part of the experiment are summarized in Result~\ref{res1}.

\begin{res}[Priming and cooperation]\label{res1}
Priming a multicultural identity increases cooperation relative to both the common-identity and the control treatments. This effect is driven by natives, whose contributions rise under the multicultural prime, while immigrants remain more cooperative than natives across all treatments.
\end{res}

In the next section, we turn to the second part of the public good (periods 11-20), where participants could punish free riders, to examine how priming affects cooperation under this alternative institutional setting.

\subsection{Identity Priming and Cooperative Behavior under Punishment}

We now turn to the second part of the public good, where the same groups played the Public Good Game with punishment. As described in Section~\ref{subsec:public_good}, after making their contribution decision in each round, participants observed the points taken by each group member, their own earnings, and the earnings of others. They then had the opportunity to reduce the earnings of any group member through costly punishment.

\subsubsection{Punishment}

Following \cite{herrmann2008antisocial}, we distinguish between two types of punishment:  
(i) \emph{social punishment}, defined as punishment directed at free riders (i.e., group members who contributed less than the punisher), and  
(ii) \emph{anti-social punishment}, defined as punishment of those who contributed more than the punisher.

Figure~\ref{fig:punishment} displays average punishment by treatment and immigration status, separating social punishment (panel~a) from anti-social punishment (panel~b). A joint reading of both panels shows that, at baseline, immigrants punish more than natives, particularly through social punishment. Panel~a further suggests that both priming treatments—and especially the Multicultural Identity treatment -increase natives' social punishment. In contrast, panel~b indicates that the Multicultural prime may reduce natives' anti-social punishment.

To formally examine the determinants of punishment behavior, we estimate the following model:

\begin{eqnarray} \label{t:eqn2}
		\begin{split}
		Y_{itgcs} & = \alpha + \beta_1 Treatment_{i} + \beta_2 Immigrant_{i} + \beta_3 Treatment_{i}* Immigrant_{i} + \\
		& + \beta_4 Punished Contribution_{it} + \beta_5 Contribution_{it} + \beta_6 Contribution Others_{it}  + \\
		&  + \beta_7 Punishment Received_{it-1}  +\beta_8 X_{i}+\sigma_5 Z_{g} +\tau_{it} + \delta_{i} + \delta_{i}*\tau_{it} +\gamma_{g} +\lambda_{c} + \psi_{s} +\epsilon_{itgcs}
		\end{split}
		\end{eqnarray}
        
where \(Y_{itgcs}\) denotes the (anti-)social punishment points assigned in each round; \(Treatment_{i}\) indicates exposure to the Common, Multicultural, or Neutral priming; and \(Immigrant_{i}\) is a dummy for immigration status (equal to 1 if at least one parent is not Italian). \(Punished\ Contribution_{it}\) is the contribution of the player being punished, \(Contribution_{it}\) is the punisher's own contribution, and \(Contribution\ Others_{it}\) is the average contribution of the remaining group members. \(Punishment\ Received_{it-1}\) captures punishment received in the previous round. \(X_{i}\) includes individual controls (gender and group size), while \(\tau_{t}\) absorbs round fixed effects. The model further incorporates random intercepts and slopes at the individual level (\(\delta_{i}\), \(\delta_{i} \times \tau_{t}\)) and random intercepts at the group (\(\gamma_{g}\)), class (\(\lambda_{c}\)), and school (\(\psi_{s}\)) levels, with \(\epsilon_{itgcs}\) as the error term.
Table~\ref{tab:xtmixed_socialp_cost} reports estimates of equation~\ref{t:eqn2} for social punishment, defined as the points deducted from free riders. Columns~1--2 present results for the full sample, while Columns~3--4 and 5--6 restrict the analysis to immigrants and natives, respectively.

In Columns~1–2, immigrants are significantly more likely to punish free riders---by roughly 12 points. Among natives (Columns~5–6), both priming treatments increase social punishment, with the \(Multicultural\) treatment producing a larger and more precisely estimated effect. For immigrants, however, neither treatment generates a statistically significant change relative to the Neutral treatment (Columns~3–4), although the point estimates are negative and non-negligible.
Consistent with \cite{herrmann2008antisocial}, social punishment is positively associated with the punisher’s own contribution (\(Contribution\)) and negatively associated with the average contribution of other group members (\(Contribution\ Others\)). Female participants generally punish less, particularly among immigrants (Columns~3–4). Finally, social punishment is stronger in larger groups: groups of four or five exhibit a higher intensive margin of punishment than groups of three.
Overall, this evidence indicates that exposure to Multicultural priming increases the intensity of social punishment among natives. Consistent with previous findings, the strongest predictors of social punishment are the punisher’s own contribution (positive correlation) and the group's average contribution (negative correlation). Immigrant students also display higher baseline levels of social punishment in the Neutral treatment. 

Turning to anti-social punishment, Table~\ref{tab:xtmixed_antisocialp_cost} in the Appendix reports estimates of the same model as in equation~\ref{t:eqn2}. We find no baseline differences between natives and immigrants: when they do punish anti-socially—i.e., when they subtract points from group members who contributed more—the two groups punish with similar intensity (about 31 points, as shown by the constant term). Among natives, both the \(Common\) and \(Multicultural\) treatments reduce anti-social punishment, with the latter producing a somewhat larger reduction (5–6 points versus about 3 points; Columns~1–2 and 5–6). For immigrants, treatment effects are larger in magnitude and opposite in sign; only the \(Common\) treatment significantly increases anti-social punishment---by roughly 7 points, or about 20\% of the baseline mean (Columns~3–4).
Other covariates behave consistently across specifications: anti-social punishment decreases with one’s own contribution and increases with the punished subject’s contribution. Moreover, it rises with lagged punishment received, indicating the presence of a ``punishment–counterpunishment'' dynamic among participants.

These patterns in both social and antisocial punishment closely parallel the contribution results from Part 1. In particular, the Multicultural treatment not only increases natives’ contributions in the no-punishment environment but also makes them more willing to sanction free riders and less prone to engage in antisocial punishment. Both behaviors reflect additional forms of cooperation in the Public Good environment, albeit operating through enforcement rather than direct contribution.

Results from the second part of the experiment can be summarized as follows:

\begin{res}[Punishment]\label{res2}
When punishment opportunities are introduced, immigrants tend to engage more in social punishment at baseline, while natives exposed to multicultural priming are more likely to punish free riders and reduce antisocial punishment.
\end{res}

\subsubsection{Contributions in Part 2 and Across All Rounds}

We now examine contribution behavior in the second part of the experiment (Rounds 11–20), where punishment was introduced, and then consider cooperation over the full 20 rounds. As before, we estimate the model in equation~\ref{t:eqn1} using a multilevel mixed-effects specification. Table~\ref{tab:xtmixed_contribution2} reports the results for Part~2: Columns 1–2 present estimates for the full sample (390 participants × 10 rounds), Columns 3–4 for immigrants, and Columns 5–6 for natives.
As in Part~1, immigrants continue to contribute more at baseline—by roughly 3 points—even when punishment opportunities are available, as shown by the positive and significant coefficient on the $Immigrant$ dummy in Columns 1–2. Among natives, the $Multicultural$ treatment has a small positive effect on contributions, but it is neither statistically significant nor meaningful in magnitude. Thus, unlike in Part~1, the effect of multicultural priming on cooperation is weaker once punishment is introduced, consistent with the finding that priming mainly operates through punishment behavior in this phase.
As before, neither priming intervention affects immigrants’ contributions, while immigrants in larger groups contribute about 4 points more than those in smaller groups. Female participants remain more cooperative in this second part of the game. We then re-estimate equation~\ref{t:eqn1} using contribution decisions from all 20 rounds.
Table~\ref{tab:xtmixed_contribution12} summarizes these results. Both priming dummies—\textit{Common} and \textit{Multicultural}—are positive and significant in the full sample, as is the $Immigrant$ dummy. In the immigrant subsample (Columns 3–4), contributions remain unaffected by priming but increase in larger groups. Among natives (Columns 5–6), the $Multicultural$ prime again raises cooperative behavior, and female participants remain more cooperative on average.

\begin{res}[Overall cooperation]\label{res3}
Across all 20 rounds of the Public Good Game, multicultural identity priming increases cooperation relative to both common-identity and neutral priming. This effect is driven entirely by native participants, while immigrants remain more cooperative than natives at baseline.
\end{res}

\section{Mechanisms}\label{sec:mec}

In this section, we examine the mechanisms underlying our two main findings: (i) immigrants display higher baseline cooperation, and (ii) natives increase their contributions under the \emph{Multicultural Identity} priming. We draw on data from the experiment, the post-experimental questionnaire, and additional incentivized tasks. As detailed in Section~\ref{sec:rob}, these results are not driven by concerns about internal validity: controlling for errors in the comprehension questions—or excluding participants who failed both—does not change the results, nor do we observe evidence of experimenter-demand effects. Reported strategies align closely with observed behavior.

\subsection{Why are immigrants more cooperative at baseline?}

We begin by testing whether migrants and natives differ along observable characteristics typically associated with cooperative behavior. Table~\ref{tab:individual_char} shows that the two groups are similar on most dimensions, except for socio-economic background (mother's education, mother's employment status, number of siblings). Prior work suggests that socio-economic background can shape voluntary cooperation \citep[e.g.,][]{gachter2004trust,andreoni2021higher,kosse2020formation,smeets2015giving,bauer2014parental,angerer2015donations,falk2018global}. To test whether these differences account for the cooperation gap, we re-estimate equation~\ref{t:eqn1} including these variables individually, jointly, and via their first principal component. As shown in Table~\ref{tab:xtmixed_mech}, the coefficient on the $Immigrant$ dummy remains large and significant across all specifications. Socio-economic background, therefore, does not explain the higher baseline cooperation of immigrant students.

A second possibility is that migrants and natives differ in unconditional generosity or norm-based prosociality. Using behavior in the Dictator Game \citep{kahneman1986fairness}—a standard measure of generosity and social norms \citep{list2007interpretation}—we find that controlling for the share of the endowment transferred does not reduce the immigrant–native gap in Public Good contributions (Column~6 of Table~\ref{tab:xtmixed_mech}). Thus, differences in unconditional generosity or social norms also cannot account for migrants’ higher baseline cooperation.

\subsection{Why does Multicultural priming increase native cooperation?}

We next explore whether classroom contextual factors shape the effectiveness of the priming treatments. Classrooms constitute a natural environment for daily cross-group interaction, which can foster both cooperation and tension \citep{allport1954nature,lowe2021types,mousa2020building}. Because students are quasi-randomly assigned to classes within schools, variation in peer composition offers a useful source of between-class diversity.

We first examine how contributions and treatment effects vary with the share of immigrant students in the class. Then, using the friendship-network data described in Section~\ref{subsec:other_measures} and in Section \ref{sec:network} of the Appendix, we construct measures of intra- and inter-cultural links, widely interpreted as indicators of socialization and cross-cultural assimilation \citep{landini2016friendship,patacchini2016social,facchini2015migration}. Results are reported in Table~\ref{tab:xtmixed_group_mech}. Column~1 shows that baseline contributions are higher in classes with a larger share of immigrant students: a 10\% increase in immigrant share is associated with roughly 10 additional points contributed. The Multicultural treatment, however, is most effective in classes with a \emph{lower} proportion of immigrants, as indicated by the positive and significant coefficient on $Multicultural$ (about 4 points). Column~2 shows that this pattern is strongest in classes with a high proportion of first-generation immigrants. Column~3 indicates that the effect is also larger in classrooms with sparser friendship networks. Columns~4 and~5 confirm that the treatment effect is amplified when immigrant students occupy more peripheral network positions or have fewer friendship links with native classmates.\footnote{For more details on how we construct these network measures, see Section \ref{sec:network} of the Appendix.}

These findings suggest that the Multicultural priming is especially powerful in environments where the multicultural nature of the classroom is less salient to natives -either because immigrants are numerically few, or socially disconnected from native peers. When natural exposure to diversity is limited, priming appears to make ethnic heterogeneity more visible and cognitively accessible. This interpretation is consistent with the ``contact hypothesis,'' which posits that intergroup contact can reduce prejudice and foster cooperation when supported by cooperative interaction and shared goals \citep{williams1947reduction,allport1954nature}. This aligns with recent field evidence that local exposure to refugees improves natives' attitudes toward ethnic minorities through interethnic contact \cite{achard2025local}, suggesting that our priming intervention effectively substitutes for or amplifies naturally occurring contact in school settings.

\section{Robustness}\label{sec:rob}

This section presents a set of robustness checks supporting the internal validity of our findings.

We first assess whether the treatment effects could be driven by emotional reactions or experimenter-demand concerns. Following \cite{benjamin2016religious}, at the end of the Public Good Game, we administered a shortened version of the Spielberger State–Trait Anxiety Inventory \citep{marteau1992development} and asked: “While you were making your choices, were you thinking about what we wanted you to do?” If participants inferred the experimenters’ intentions, treatment effects might reflect compliance rather than priming. Table~\ref{tab:individual_char} shows no baseline differences between natives and immigrants in any emotional dimension. Table~\ref{tab:feelings} further indicates that neither priming treatment affected reported emotions (Columns 1–6) or the probability of reporting demand-driven reasoning (Column 7). Thus, the results are not explained by emotional responses or experimenter-demand effects.

Second, we consider whether differential comprehension could explain the treatment effects. Table~\ref{tab:individual_char} shows that natives and immigrants performed similarly on the two control questions embedded in the instructions. In Table~\ref{tab:mistakes}, Columns 1–2 confirm that comprehension is uncorrelated with treatment assignment, as expected under randomization. Column~3 adds the number of mistakes directly to equation~\ref{t:eqn1}; treatment coefficients remain stable, indicating that comprehension does not drive the main results.

We also examine the alignment between stated and observed strategies. At the end of the experiment, participants indicated agreement with the statement: “During the game, I tried to maximize my own payoff rather than the group payoff.” While not incentivized, this provides an additional behavioral signal. Table~\ref{tab:individual_char} shows that natives are more likely to endorse this self-interested strategy, consistent with their lower baseline contributions documented earlier.

Finally, Tables~\ref{tab:contribution} and \ref{tab:contribution12} re-estimate the contribution regressions for Part~1 and for the full 20 rounds using OLS. Although OLS does not fully account for the hierarchical structure of the data, the resulting estimates closely match those from the mixed-effects models, reinforcing the robustness of our conclusions.

\section{Conclusion}\label{sec:conclusion}

The integration of immigrant-origin students and the preservation of cooperation in increasingly diverse societies are pressing policy challenges at both national and supranational levels. A central question is whether institutions can make particular dimensions of social identity salient in ways that foster cooperation between natives and immigrants.

Schools provide a natural context to explore this question: they are environments where diverse identities coexist, where students interact daily, and where norms of cooperation take shape. Leveraging this setting, we experimentally primed distinct identity dimensions and studied their causal effects on cooperative behavior in ethnically mixed classrooms.

Our results show that making a multicultural identity salient increases cooperation among native students, whereas priming a common school-based identity does not produce any behavioral change relative to the control. Immigrant-origin students, by contrast, display consistently higher cooperative behavior across all priming conditions, suggesting that they already contribute more to the public good at baseline and do not respond differentially to identity cues. Taken together, these findings highlight that integration is a two-sided process: sustaining cooperation in diverse settings requires not only supporting immigrant students — who already exhibit cooperative behavior — but also shaping how native students perceive and engage with diversity. This pattern has direct implications for policymakers designing institutions that aim to foster cohesion in multicultural societies.

A noteworthy feature of our findings is the null effect of the Common Identity priming, despite clear evidence that the prime successfully activated the intended identity dimension in the “Who am I?” task. One plausible interpretation is that middle-school identity is not a sufficiently meaningful or voluntary layer of identity for adolescents. School choice at this stage is typically constrained by residential boundaries and parental decisions, potentially limiting students’ identification with their school community. Future work should investigate which identity layers are most salient for adolescents, and under what conditions shared identities foster cooperative behavior.

Overall, our findings suggest that emphasizing multicultural belonging—rather than relying solely on generic common identities—may be a productive way to strengthen cooperative norms in diverse societies. Low-cost, scalable interventions that increase the visibility and legitimacy of multicultural identities can help native-born students engage more cooperatively with peers from different backgrounds, contributing to more cohesive and resilient communities.

\bigskip
\textbf{Declaration of generative AI and AI-assisted technologies in the writing process}\\
During the preparation of this work, the authors used ChatGPT 4 to improve language and readability, with caution. After using this tool, the authors reviewed and edited the content as needed and took full responsibility for the content of the publication.

\newpage
\bibliographystyle{chicago}
\bibliography{bib.bib}

\clearpage
\section*{Figures and Tables}

\begin{figure}[h]
\centering
\includegraphics[width=12cm]{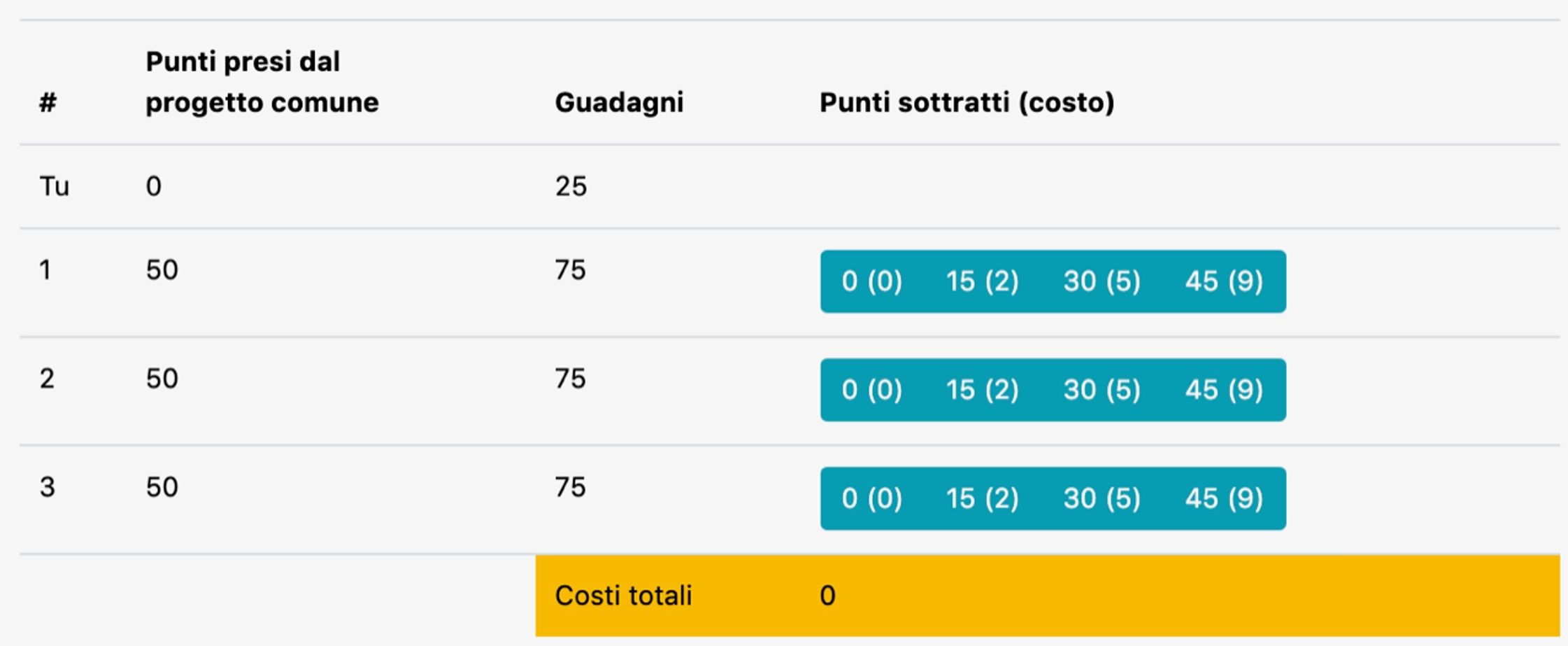}

\caption{Screenshot of the punishment stage. \\
The table shows, for each group member and each round: 
\emph{Punti presi dal progetto comune} (points taken from the common pool), 
\emph{Guadagni} (earnings), and 
\emph{Punti sottratti (costo)} (punishment points assigned, i.e.\ costly deductions). 
Rows are labeled ``TU'' (the participant) followed by the numbers assigned to the other group members. 
Participants are informed that, to preserve anonymity, player numbers are randomly reassigned and re-ordered in every round, so participants cannot track the same peer across rounds.\label{fig:punishment}}
\end{figure}

\clearpage 

\begin{figure}[htpb]
\centering
    \begin{minipage}{0.75\textwidth}
        \subfigure[Social punishment]{
            \includegraphics[width=\textwidth]{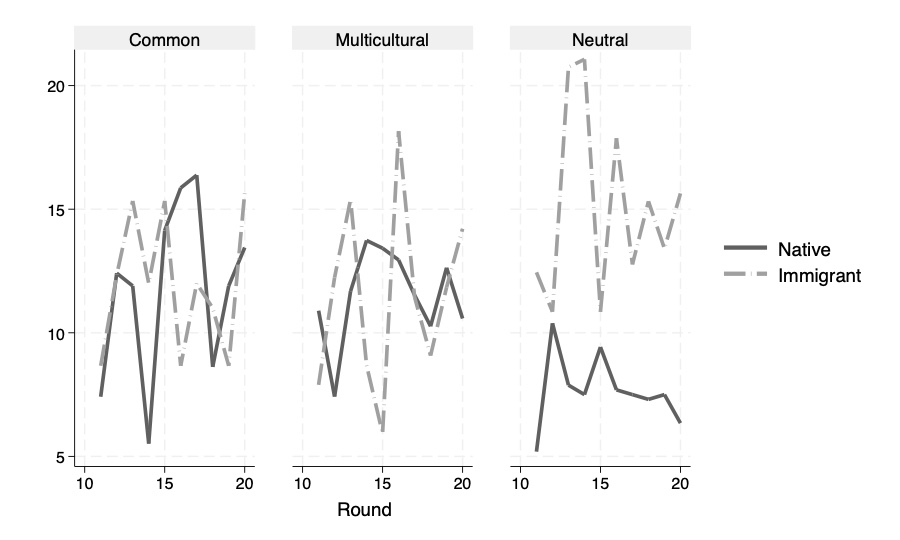}
        }
    \end{minipage}

    \vspace{0.4cm}

    \begin{minipage}{0.75\textwidth}
        \subfigure[Anti-social punishment]{
            \includegraphics[width=\textwidth]{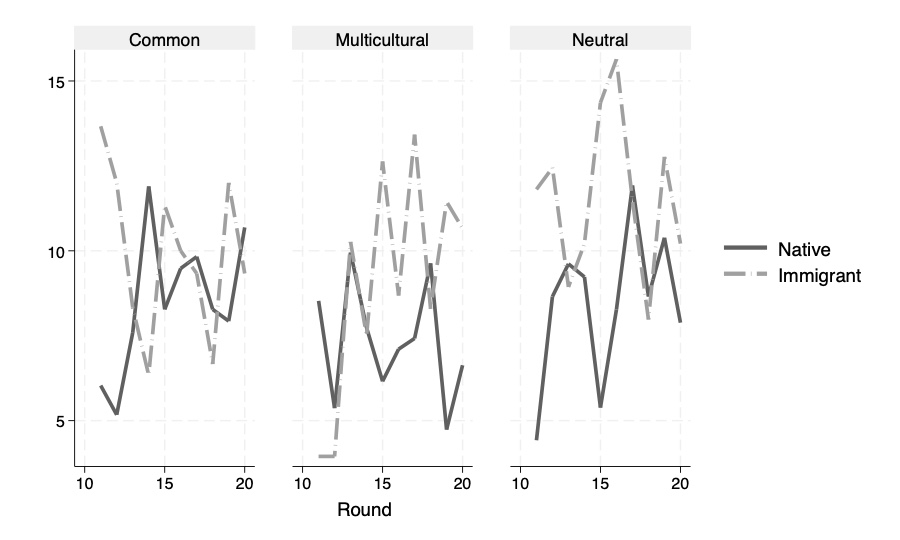}
        }
    \end{minipage}
{
\caption{Punishment by treatment and status}
\label{fig:punishment}
\begin{minipage}{0.98\textwidth}\footnotesize
\textit{Notes:} Panel (a) reports average social punishment—points subtracted from free-riders—by treatment and immigration status. Immigrant students punish more at baseline than natives, especially socially. Both identity treatments increase social punishment among natives, with the Multicultural Identity prime generating the largest rise. Panel (b) displays anti-social punishment—points subtracted from peers who contributed more than the punisher. Native students reduce anti-social punishment under the Multicultural Identity priming, whereas immigrants show stable patterns across treatments.
\end{minipage}}
\end{figure}

\clearpage
\begin{table}[ht]
\linespread{1.15}\selectfont
\centering
\begin{tabular}{p{0.18\linewidth} | p{0.75\linewidth}}
\hline
\textbf{Treatment} & \textbf{Questions} \\
\hline

\textbf{All} &
Date of Birth: DD/MM/YYYY; Gender: M F; Grade: I, II, III. \\
\hline

\textbf{Common Identity} &1) Name of the School:\\
& 2) When choosing the school you are attending, did you consider any other schools?: YES, NO. If yes, what other schools? ... (Please specify). \\
& 3) Why did you decide to choose your specific school? \\
& 4) List three things you like about your school and make it special. \\
\hline

\textbf{Multicultural Identity} &
1) Where were you born?: Italy; in a different country than Italy, that is... (Please specify). \\
& 2) Where was your mother born?: Italy; in a different country than Italy, that is... (Please specify). \\
& 3) Where was your father born?: Italy; in a different country than Italy, that is... (Please specify). \\
& (Displayed \emph{only} if one parent was not Italian) \\
& 4) Which country is the one you feel closer to? the one where my mother is born; the one where my father is born. \\
& 5) What languages do you speak at home? \\
\hline

\textbf{Control} &
1) How often do you watch television? a) every day; b) 4/5 times a week; c) 2/3 times a week; d) a few times a month; e) a few times a year; f) rarely if ever; g) never. \\
& 2) How often do you surf the web? a) never; b) about 15 minutes per day; c) about 1 hour per day; d) about 1 hour per day; e) about 1.5 hours per day; f) about 2 hours per day; g) more than 2 hours per day. \\
& 3) How often do you watch TV series? a) every day; b) 4/5 times a week; c) 2/3 times a week; d) a few times a month; e) a few times a year; f) rarely if ever; g) never. \\
\hline

\end{tabular}
\caption{Priming questionnaires. \\
Participants first answered the general questions listed under “All,” and then completed the questionnaire corresponding to their randomly assigned priming treatment. These items constituted the identity–salience manipulation: students in the \emph{Common Identity} treatment answered questions highlighting school belonging; those in the \emph{Multicultural Identity} treatment answered questions emphasizing family origins and linguistic–cultural diversity; and those in the \emph{Control} group answered questions about media habits unrelated to identity. 
At the end of the session, all participants also completed the questionnaires from the two priming treatments they had not been assigned to, which allows us to compare responses across treatments for validation and robustness.}

\label{tab:questionnaires}
\end{table}

\begin{table}[htbp]
\caption{Effectiveness of the Priming Treatments. }

	\begin{center}
		\begin{adjustbox}{width=1\textwidth}
			{
\def\sym#1{\ifmmode^{#1}\else\(^{#1}\)\fi}
\begin{tabular}{l*{6}{c}}
\toprule
                &\multicolumn{2}{c}{\begin{tabular}{@{}c@{}}Full\end{tabular}}&\multicolumn{2}{c}{\begin{tabular}{@{}c@{}}Natives\end{tabular}}&\multicolumn{2}{c}{\begin{tabular}{@{}c@{}}Immigrant\end{tabular}}\\\cmidrule(lr){2-3}\cmidrule(lr){4-5}\cmidrule(lr){6-7}
                &\multicolumn{1}{c}{(1)}         &\multicolumn{1}{c}{(2)}         &\multicolumn{1}{c}{(3)}         &\multicolumn{1}{c}{(4)}         &\multicolumn{1}{c}{(5)}         &\multicolumn{1}{c}{(6)}         \\
                &P(Cultural)         &P(School)         &P(Cultural)         &P(School)         &P(Cultural)         &P(School)         \\
\midrule
Common          &    0.111         &    0.301\sym{***}&    0.111\sym{*}  &    0.207\sym{***}&    0.194\sym{*}  &    0.518\sym{***}\\
                &  (0.087)         &  (0.022)         &  (0.058)         &  (0.014)         &  (0.100)         &  (0.009)         \\
Multicultural   &    0.165\sym{**} &    0.216         &    0.199\sym{***}&    0.201         &    0.133\sym{***}&    0.182         \\
                &  (0.075)         &  (0.227)         &  (0.062)         &  (0.203)         &  (0.047)         &  (0.213)         \\
Constant        &    0.413\sym{***}&    0.436\sym{***}&    0.406\sym{***}&    0.465\sym{***}&    0.385\sym{***}&    0.315\sym{***}\\
                &  (0.109)         &  (0.028)         &  (0.096)         &  (0.042)         &  (0.094)         &  (0.003)         \\
\midrule
N Obs.          &      160         &      160         &      113         &      113         &       47         &       47         \\
Clustering      &      Yes         &      Yes         &      Yes         &      Yes         &      Yes         &      Yes         \\
Class nesting   &      Yes         &      Yes         &      Yes         &      Yes         &      Yes         &      Yes         \\
Group nesting   &      Yes         &      Yes         &      Yes         &      Yes         &      Yes         &      Yes         \\
\bottomrule
\multicolumn{7}{l}{\footnotesize Symbols $***$, $**$, and $*$ indicate significance at the 1\%, 5\% and 10\% level, respectively.}\\
\end{tabular}
}

		\end{adjustbox}
	\end{center}
\begin{tablenotes}
\small
\item \textbf{Notes}: This table examines whether the priming interventions increased the salience of the targeted identity dimensions.
The dependent variable in Column (1) is the share of words in the open-ended self-description referring to cultural or origin-related identity.
The dependent variable in Column (2) is the share of words referring to the school-based identity.
Key independent variables are indicators for assignment to the \emph{Multicultural Identity} or \emph{Common Identity} treatments; the \emph{Neutral} treatment is the omitted category.
All models include student-level controls and class fixed effects.
Standard errors are clustered at the class level.
\item Words classified under each identity are listed in Italian, with English translations provided in parentheses. 
School-identity words include: scuol- (school), studente (student), studio (study), Rita Levi Montalcini, Dozza, medie / media (middle school). 
Cultural-identity words include: origin- (origin), Italia (Italy), African-, Rumen- (Romanian), vengo da (I come from), Perù, bengalese (Bengali), Filippine (Philippines), ghanese (Ghanaian), Pakistan, moldav- (Moldovan), svizzero (Swiss), Nigeria, Colombia, Algeria, Cina / cinese (China / Chinese), nat- (born), vivo / abito (I live), paese (country/town), spagnolo (Spanish), Svezia (Sweden).
\end{tablenotes}

	\label{tab:mentions}
\end{table}

\begin{table}[htbp]
	\caption{Attrition by Treatment -- Completion between February 2020 and June 2020} 
	\begin{center}
		\begin{adjustbox}{width=1\textwidth}
			{
\def\sym#1{\ifmmode^{#1}\else\(^{#1}\)\fi}
\begin{tabular}{l*{4}{c}}
\toprule
                &\multicolumn{1}{c}{(1)}&\multicolumn{1}{c}{(2)}&\multicolumn{1}{c}{(3)}&\multicolumn{1}{c}{(4)}\\
                &\multicolumn{1}{c}{P(Description)}&\multicolumn{1}{c}{P(Dictator)}&\multicolumn{1}{c}{P(Fairness)}&\multicolumn{1}{c}{P(Friends)}\\
\midrule
Common          &    0.038         &    0.018         &    0.029         &    0.013         \\
                &  (0.027)         &  (0.018)         &  (0.022)         &  (0.028)         \\
Multicultural   &    0.038         &    0.012         &    0.038         &    0.020         \\
                &  (0.026)         &  (0.020)         &  (0.030)         &  (0.024)         \\
Constant        &    0.179\sym{*}  &    0.705\sym{***}&    0.613\sym{***}&    0.101         \\
                &  (0.098)         &  (0.122)         &  (0.131)         &  (0.068)         \\
\midrule
N Obs.          &      390         &      390         &      390         &      390         \\
Clustering      &      Yes         &      Yes         &      Yes         &      Yes         \\
School fixed effects&      Yes         &      Yes         &      Yes         &      Yes         \\
\bottomrule
\multicolumn{5}{l}{\footnotesize Symbols $***$, $**$, and $*$ indicate significance at the 1\%, 5\% and 10\% level, respectively.}\\
\end{tabular}
}

		\end{adjustbox}
	\end{center}
	\begin{tablenotes}
		 \small \item  \textbf{Notes}: The Table show how the probability of having completed each activity before the COVID-19 first wave, rather than in June 2020, changes across treatments.
	\end{tablenotes}
	\label{tab:attrition_treatment}
\end{table}

\begin{table}[htbp]
	\caption{Attrition by Treatment -- Completion at end-line} 
	\begin{center}
		\begin{adjustbox}{width=1\textwidth}
			{
\def\sym#1{\ifmmode^{#1}\else\(^{#1}\)\fi}
\begin{tabular}{l*{4}{c}}
\toprule
                &\multicolumn{1}{c}{(1)}&\multicolumn{1}{c}{(2)}&\multicolumn{1}{c}{(3)}&\multicolumn{1}{c}{(4)}\\
                &\multicolumn{1}{c}{P(Description)}&\multicolumn{1}{c}{P(Dictator)}&\multicolumn{1}{c}{P(Fairness)}&\multicolumn{1}{c}{P(Friends)}\\
\midrule
Common          &   -0.068         &   -0.030         &   -0.042         &   -0.060         \\
                &  (0.048)         &  (0.031)         &  (0.033)         &  (0.050)         \\
Multicultural   &   -0.025         &   -0.008         &   -0.027         &   -0.009         \\
                &  (0.032)         &  (0.022)         &  (0.030)         &  (0.031)         \\
Constant        &    0.208\sym{***}&    0.100\sym{**} &    0.125\sym{**} &    0.227\sym{***}\\
                &  (0.051)         &  (0.046)         &  (0.047)         &  (0.047)         \\
\midrule
N Obs.          &      390         &      390         &      390         &      390         \\
Clustering      &      Yes         &      Yes         &      Yes         &      Yes         \\
School fixed effects&      Yes         &      Yes         &      Yes         &      Yes         \\
\bottomrule
\multicolumn{5}{l}{\footnotesize Symbols $***$, $**$, and $*$ indicate significance at the 1\%, 5\% and 10\% level, respectively.}\\
\end{tabular}
}

		\end{adjustbox}
	\end{center}
	\begin{tablenotes}
		 \small \item  \textbf{Notes}: The Table show how the probability of having completed each activity at the end-line of the study changes across treatments.
	\end{tablenotes}
	\label{tab:attrition_never_treatment}
\end{table}

\begin{table}[htbp]
	\caption{Observable characteristics and Mid-line sample attrition} 
	\begin{center}
		\begin{adjustbox}{width=1\textwidth}
			{
	\def\sym#1{\ifmmode^{#1}\else\(^{#1}\)\fi}
	\begin{tabular}{l*{5}{c}}

                    &        Full&    Sample 1&    Sample 2&       Diff.         \\
\midrule
Immigrant           &       0.333&       0.333&       0.273&       0.061         \\
                    &     (0.472)&     (0.472)&     (0.449)&                     \\
Born in Italy       &       0.913&       0.917&       0.927&      -0.011         \\
                    &     (0.282)&     (0.277)&     (0.262)&                     \\
Speak only Italian at home&       0.697&       0.696&       0.764&      -0.068         \\
                    &     (0.460)&     (0.461)&     (0.429)&                     \\
Female              &       0.413&       0.417&       0.382&       0.035         \\
                    &     (0.493)&     (0.494)&     (0.490)&                     \\
How often I practice sport&       3.854&       3.808&       4.018&      -0.210         \\
                    &     (1.307)&     (1.313)&     (1.298)&                     \\
How often I surf the web&       3.036&       3.003&       3.273&      -0.270         \\
                    &     (1.522)&     (1.525)&     (1.484)&                     \\
How often I watch TV&       4.700&       4.657&       5.000&      -0.343         \\
                    &     (1.654)&     (1.692)&     (1.414)&                     \\
Contribution        &      25.103&      25.192&      24.364&       0.829         \\
                    &    (19.476)&    (19.191)&    (21.039)&                     \\
I tried to maximize joint payoff in PGG&       0.569&       0.567&       0.564&       0.004         \\
                    &     (0.496)&     (0.496)&     (0.501)&                     \\
I did what I think you wanted me to do&       0.419&       0.421&       0.436&      -0.015         \\
                    &     (0.494)&     (0.495)&     (0.501)&                     \\
Calm                &       0.615&       0.606&       0.636&      -0.031         \\
                    &     (0.487)&     (0.489)&     (0.485)&                     \\
Tense               &       0.651&       0.660&       0.600&       0.060         \\
                    &     (0.477)&     (0.474)&     (0.494)&                     \\
Upset               &       0.679&       0.679&       0.618&       0.061         \\
                    &     (0.467)&     (0.467)&     (0.490)&                     \\
Relaxed             &       0.505&       0.484&       0.545&      -0.061         \\
                    &     (0.501)&     (0.501)&     (0.503)&                     \\
Happy               &       0.682&       0.663&       0.709&      -0.046         \\
                    &     (0.466)&     (0.473)&     (0.458)&                     \\
Worried             &       0.759&       0.756&       0.745&       0.011         \\
                    &     (0.428)&     (0.430)&     (0.440)&                     \\
\bottomrule
\multicolumn{5}{l}{\footnotesize Symbols $***$, $**$, and $*$ indicate significance at the 1\%, 5\% and 10\% level, respectively.}\\
\end{tabular}
}

		\end{adjustbox}
	\end{center}
	\begin{tablenotes}
		 \small \item  \textbf{Notes}: The Table show how the observable characteristics measured at the first meeting change between students who completed at least one activity among the open-ended description of themselves, the dictator game, the elicitation of fairness views, and of the network of friends -- before the COVID-19 first wave (Sample 1), and those who completed everything in June 2020 (Sample 2). Symbols $***$, $**$, and $*$ indicate significance at the 1\%, 5\% and 10\% level, respectively.
	\end{tablenotes}
	\label{tab:attrition1_ind_char}
\end{table}

\begin{table}[htbp]
	\caption{Observable characteristics and End-line sample attrition} 
	\begin{center}
		\begin{adjustbox}{width=1\textwidth}
			{
	\def\sym#1{\ifmmode^{#1}\else\(^{#1}\)\fi}
	\begin{tabular}{l*{5}{c}}

                    &        Full&    Sample 1&    Sample 2&       Diff.         \\
\midrule
Immigrant           &       0.333&       0.324&       0.478&      -0.154         \\
                    &     (0.472)&     (0.469)&     (0.511)&                     \\
Born in Italy       &       0.913&       0.918&       0.826&       0.092         \\
                    &     (0.282)&     (0.274)&     (0.388)&                     \\
Speak only Italian at home&       0.697&       0.706&       0.565&       0.141         \\
                    &     (0.460)&     (0.456)&     (0.507)&                     \\
Female              &       0.413&       0.411&       0.435&      -0.023         \\
                    &     (0.493)&     (0.493)&     (0.507)&                     \\
How often I practice sport&       3.854&       3.839&       4.087&      -0.248         \\
                    &     (1.307)&     (1.311)&     (1.240)&                     \\
How often I surf the web&       3.036&       3.044&       2.913&       0.131         \\
                    &     (1.522)&     (1.520)&     (1.593)&                     \\
How often I watch TV&       4.700&       4.708&       4.565&       0.143         \\
                    &     (1.654)&     (1.656)&     (1.647)&                     \\
Contribution        &      25.103&      25.068&      25.652&      -0.584         \\
                    &    (19.476)&    (19.451)&    (20.299)&                     \\
I tried to maximize joint payoff in PGG&       0.569&       0.567&       0.609&      -0.042         \\
                    &     (0.496)&     (0.496)&     (0.499)&                     \\
I did what I think you wanted me to do&       0.419&       0.423&       0.348&       0.076         \\
                    &     (0.494)&     (0.495)&     (0.487)&                     \\
Calm                &       0.615&       0.610&       0.696&      -0.085         \\
                    &     (0.487)&     (0.488)&     (0.470)&                     \\
Tense               &       0.651&       0.651&       0.652&      -0.001         \\
                    &     (0.477)&     (0.477)&     (0.487)&                     \\
Upset               &       0.679&       0.670&       0.826&      -0.156         \\
                    &     (0.467)&     (0.471)&     (0.388)&                     \\
Relaxed             &       0.505&       0.493&       0.696&      -0.202         \\
                    &     (0.501)&     (0.501)&     (0.470)&                     \\
Happy               &       0.682&       0.670&       0.870&      -0.199\sym{*}  \\
                    &     (0.466)&     (0.471)&     (0.344)&                     \\
Worried             &       0.759&       0.755&       0.826&      -0.071         \\
                    &     (0.428)&     (0.431)&     (0.388)&                     \\
\bottomrule
\multicolumn{5}{l}{\footnotesize Symbols $***$, $**$, and $*$ indicate significance at the 1\%, 5\% and 10\% level, respectively.}\\
\end{tabular}
}

		\end{adjustbox}
	\end{center}
	\begin{tablenotes}
		 \small \item  \textbf{Notes}: The Table show how the observable characteristics measured at the first meeting change between students who completed at least one activity among the other measures we collected (Sample 1), and those who never completed them (Sample 2). Symbols $***$, $**$, and $*$ indicate significance at the 1\%, 5\% and 10\% level, respectively.
	\end{tablenotes}
	\label{tab:attrition2_ind_char}
\end{table}

\begin{table}[htbp]
\caption{Contribution to the Public Good - Part I} 
\begin{center}
\begin{adjustbox}{width=1\textwidth}
{
\def\sym#1{\ifmmode^{#1}\else\(^{#1}\)\fi}
\begin{tabular}{l*{6}{c}}
\toprule
                &\multicolumn{1}{c}{(1)}&\multicolumn{1}{c}{(2)}&\multicolumn{1}{c}{(3)}&\multicolumn{1}{c}{(4)}&\multicolumn{1}{c}{(5)}&\multicolumn{1}{c}{(6)}\\
                &\multicolumn{1}{c}{Full}&\multicolumn{1}{c}{Full}&\multicolumn{1}{c}{Immigrants}&\multicolumn{1}{c}{Immigrants}&\multicolumn{1}{c}{Natives}&\multicolumn{1}{c}{Natives}\\
\midrule
Contribution    &                  &                  &                  &                  &                  &                  \\
Common          &    1.152\sym{***}&    1.172\sym{***}&    0.352         &    0.352         &    0.961         &    0.636         \\
                &  (0.364)         &  (0.397)         &  (3.608)         &  (3.608)         &  (0.632)         &  (1.082)         \\
Multicultural   &    2.991\sym{***}&    3.076\sym{***}&    0.878         &    0.878         &    2.820\sym{***}&    2.938\sym{**} \\
                &  (0.773)         &  (0.793)         &  (3.148)         &  (3.148)         &  (0.954)         &  (1.209)         \\
Immigrant       &    3.475\sym{**} &    3.740\sym{**} &                  &                  &                  &                  \\
                &  (1.544)         &  (1.509)         &                  &                  &                  &                  \\
Common x Immigrant&   -0.593         &   -0.864         &                  &                  &                  &                  \\
                &  (4.861)         &  (5.093)         &                  &                  &                  &                  \\
Multicultural x Immigrant&   -2.558         &   -2.616         &                  &                  &                  &                  \\
                &  (3.868)         &  (3.950)         &                  &                  &                  &                  \\
Female          &    1.016         &    1.020         &   -1.849         &   -1.849         &    2.264\sym{*}  &    2.497\sym{**} \\
                &  (1.057)         &  (1.076)         &  (1.193)         &  (1.193)         &  (1.345)         &  (1.062)         \\
Group of 4      &   -0.151         &   -0.164         &    5.566\sym{***}&    5.566\sym{***}&   -2.596         &   -2.550         \\
                &  (1.683)         &  (1.681)         &  (1.130)         &  (1.130)         &  (2.146)         &  (2.190)         \\
Group of 5      &    0.471\sym{***}&    0.639\sym{***}&    4.148\sym{***}&    4.148\sym{***}&   -0.294         &    0.505         \\
                &  (0.131)         &  (0.079)         &  (0.393)         &  (0.393)         &  (0.369)         &  (0.349)         \\
Constant        &   22.476\sym{***}&   22.350\sym{***}&   23.985\sym{***}&   23.985\sym{***}&   23.101\sym{***}&   22.851\sym{***}\\
                &  (2.670)         &  (2.722)         &  (1.003)         &  (1.003)         &  (4.169)         &  (4.432)         \\
\midrule
N Obs.          &    3,900         &    3,900         &    1,300         &    1,300         &    2,600         &    2,600         \\
Round fixed effects&      Yes         &      Yes         &      Yes         &      Yes         &      Yes         &      Yes         \\
Clustering      &      Yes         &      Yes         &      Yes         &      Yes         &      Yes         &      Yes         \\
Random Slopes   &      Yes         &      Yes         &      Yes         &      Yes         &      Yes         &      Yes         \\
Individual correlation&      Yes         &      Yes         &      Yes         &      Yes         &      Yes         &      Yes         \\
Class nesting   &      Yes         &      Yes         &      Yes         &      Yes         &      Yes         &      Yes         \\
Group nesting   &       No         &      Yes         &       No         &      Yes         &       No         &      Yes         \\
\bottomrule
\multicolumn{7}{l}{\footnotesize Column 4: Post estimation tests for Common = [Common x Immigrant - Common]: p=0.4554.}\\
\multicolumn{7}{l}{\footnotesize Column 4: Post estimation tests for Multicultural = [Multicultural x Immigrant - Multicultural]: p=0.0002.}\\
\multicolumn{7}{l}{\footnotesize Column 6: Post estimation tests for Common = Multicultural: p=0.2525.}\\
\multicolumn{7}{l}{\footnotesize Column 8: Post estimation tests for Common = Multicultural: p=0.0000.}\\
\multicolumn{7}{l}{\footnotesize Symbols $***$, $**$, and $*$ indicate significance at the 1\%, 5\% and 10\% level, respectively.}\\
\end{tabular}
}

\end{adjustbox}
\end{center}
\begin{tablenotes}
	\small \item  \textbf{Notes}: Linear mixed-effects models, to account for the hierarchical structure of the data, i.e. observations are nested in groups, and the correlation over rounds of individual observations. In particular, the four-level model uses random intercepts and slopes for individual choices observed over rounds, and it account for dependence in the observations among the individuals belonging to the same group, class and school. Columns 1-2 looks at the full sample. Columns 3 and 4 at immigrants only. Columns 5-6 at natives only.
\end{tablenotes}
\label{tab:xtmixed_contribution}
\end{table}

\begin{table}[htbp]
\caption{Social Punishment Cost in Public Good - Part I} 
\begin{center}
\begin{adjustbox}{width=1\textwidth}
{
\def\sym#1{\ifmmode^{#1}\else\(^{#1}\)\fi}
\begin{tabular}{l*{6}{c}}
\toprule
                &\multicolumn{1}{c}{(1)}&\multicolumn{1}{c}{(2)}&\multicolumn{1}{c}{(3)}&\multicolumn{1}{c}{(4)}&\multicolumn{1}{c}{(5)}&\multicolumn{1}{c}{(6)}\\
                &\multicolumn{1}{c}{Full}&\multicolumn{1}{c}{Full}&\multicolumn{1}{c}{Immigrants}&\multicolumn{1}{c}{Immigrants}&\multicolumn{1}{c}{Natives}&\multicolumn{1}{c}{Natives}\\
\midrule
Common          &    3.881\sym{*}  &    4.011\sym{*}  &   -0.755         &   -0.755         &    3.509         &    3.795         \\
                &  (2.235)         &  (2.137)         &  (4.870)         &  (4.869)         &  (3.498)         &  (2.985)         \\
Multicultural   &    9.618\sym{***}&    9.669\sym{***}&   -3.384         &   -3.384         &    9.831\sym{***}&    9.787\sym{***}\\
                &  (0.144)         &  (0.193)         &  (8.464)         &  (8.464)         &  (1.845)         &  (1.693)         \\
Immigrant       &   12.094\sym{***}&   12.207\sym{***}&                  &                  &                  &                  \\
                &  (1.283)         &  (1.411)         &                  &                  &                  &                  \\
Common x Immigrant&   -4.976         &   -5.114         &                  &                  &                  &                  \\
                &  (3.869)         &  (4.061)         &                  &                  &                  &                  \\
Multicultural x Immigrant&  -11.409         &  -11.564         &                  &                  &                  &                  \\
                &  (7.600)         &  (7.701)         &                  &                  &                  &                  \\
 Contribution   &    0.516\sym{***}&    0.515\sym{***}&    0.483\sym{***}&    0.483\sym{***}&    0.541\sym{***}&    0.540\sym{***}\\
                &  (0.053)         &  (0.054)         &  (0.003)         &  (0.003)         &  (0.101)         &  (0.103)         \\
Punished subject contrib.&   -0.060         &   -0.056         &   -0.117         &   -0.117         &   -0.026         &   -0.018         \\
                &  (0.196)         &  (0.195)         &  (0.213)         &  (0.213)         &  (0.219)         &  (0.219)         \\
Average contrib. others&   -0.410\sym{***}&   -0.409\sym{***}&   -0.442\sym{***}&   -0.442\sym{***}&   -0.403\sym{***}&   -0.399\sym{***}\\
                &  (0.033)         &  (0.032)         &  (0.048)         &  (0.048)         &  (0.080)         &  (0.077)         \\
Lagged p. received&    0.062         &    0.059         &   -0.005         &   -0.005         &    0.089\sym{*}  &    0.082\sym{*}  \\
                &  (0.080)         &  (0.081)         &  (0.137)         &  (0.137)         &  (0.049)         &  (0.049)         \\
Female          &   -9.922\sym{**} &   -9.840\sym{**} &  -11.667\sym{***}&  -11.667\sym{***}&   -9.435\sym{*}  &   -9.280         \\
                &  (4.352)         &  (4.395)         &  (0.153)         &  (0.154)         &  (5.687)         &  (5.754)         \\
Group of 4      &   14.345\sym{***}&   14.417\sym{***}&    8.223\sym{***}&    8.223\sym{***}&   16.086\sym{***}&   16.782\sym{***}\\
                &  (2.963)         &  (2.976)         &  (1.055)         &  (1.055)         &  (5.006)         &  (4.752)         \\
Group of 5      &   25.388\sym{***}&   25.549\sym{***}&   17.391\sym{***}&   17.391\sym{***}&   28.807\sym{***}&   29.318\sym{***}\\
                &  (2.049)         &  (2.049)         &  (3.351)         &  (3.351)         &  (0.423)         &  (0.003)         \\
Constant        &   27.250\sym{***}&   27.078\sym{***}&   45.320\sym{***}&   45.320\sym{***}&   25.541\sym{***}&   24.907\sym{***}\\
                &  (0.367)         &  (0.290)         &  (2.590)         &  (2.590)         &  (3.445)         &  (3.301)         \\
\midrule
N Obs.          &      909         &      909         &      312         &      312         &      597         &      597         \\
Round fixed effects&      Yes         &      Yes         &      Yes         &      Yes         &      Yes         &      Yes         \\
Random Slopes   &      Yes         &      Yes         &      Yes         &      Yes         &      Yes         &      Yes         \\
Individual correlation&      Yes         &      Yes         &      Yes         &      Yes         &      Yes         &      Yes         \\
Class nesting   &      Yes         &      Yes         &      Yes         &      Yes         &      Yes         &      Yes         \\
Group nesting   &       No         &      Yes         &       No         &      Yes         &       No         &      Yes         \\
\bottomrule
\multicolumn{7}{l}{\footnotesize Column 2: Post estimation tests for Common = [Common x Immigrant - Common]: p=0.0000.}\\
\multicolumn{7}{l}{\footnotesize Column 4: Post estimation tests for Multicultural = [Multicultural x Immigrant - Multicultural]: p=0.0000.}\\
\multicolumn{7}{l}{\footnotesize Column 4: Post estimation tests for Common = Multicultural: p=0.4646.}\\
\multicolumn{7}{l}{\footnotesize Column 6: Post estimation tests for Common = Multicultural: p=0.0000.}\\
\multicolumn{7}{l}{\footnotesize Symbols $***$, $**$, and $*$ indicate significance at the 1\%, 5\% and 10\% level, respectively.}\\
\end{tabular}
}

\end{adjustbox}
\end{center}
	\begin{tablenotes}
		\small \item  \textbf{Notes}: Linear mixed-effects models, to account for the hierarchical structure of the data, i.e. observations are nested in groups, and the correlation over rounds of individual observations. In particular, the four-level model uses random intercepts and slopes for individual choices observed over rounds, and it account for dependence in the observations among the individuals belonging to the same group, class and school. Columns 1-2 looks at the full sample. Columns 3 and 4 at immigrants only. Columns 5-6 at natives only.
	\end{tablenotes}
\label{tab:xtmixed_socialp_cost}
\end{table}

\begin{table}[htbp]
\caption{Anti-Social Punishment in Public Good - Part II} 
\begin{center}
\begin{adjustbox}{width=1\textwidth}
{
\def\sym#1{\ifmmode^{#1}\else\(^{#1}\)\fi}
\begin{tabular}{l*{6}{c}}
\toprule
                &\multicolumn{1}{c}{(1)}&\multicolumn{1}{c}{(2)}&\multicolumn{1}{c}{(3)}&\multicolumn{1}{c}{(4)}&\multicolumn{1}{c}{(5)}&\multicolumn{1}{c}{(6)}\\
                &\multicolumn{1}{c}{Full}&\multicolumn{1}{c}{Full}&\multicolumn{1}{c}{Immigrants}&\multicolumn{1}{c}{Immigrants}&\multicolumn{1}{c}{Natives}&\multicolumn{1}{c}{Natives}\\
\midrule
Common          &   -3.635\sym{***}&   -3.635\sym{***}&    7.171\sym{***}&    7.170\sym{***}&   -2.968\sym{***}&   -2.968\sym{***}\\
                &  (1.059)         &  (1.059)         &  (2.044)         &  (2.044)         &  (0.545)         &  (0.545)         \\
Multicultural   &   -5.160\sym{**} &   -5.160\sym{**} &    1.056         &    1.056         &   -6.332\sym{*}  &   -6.332\sym{*}  \\
                &  (2.323)         &  (2.324)         &  (3.529)         &  (3.528)         &  (3.716)         &  (3.716)         \\
Immigrant       &    0.853         &    0.853         &                  &                  &                  &                  \\
                &  (0.878)         &  (0.878)         &                  &                  &                  &                  \\
Common x Immigrant&    8.334         &    8.334         &                  &                  &                  &                  \\
                &  (5.714)         &  (5.714)         &                  &                  &                  &                  \\
Multicultural x Immigrant&    6.025\sym{***}&    6.025\sym{***}&                  &                  &                  &                  \\
                &  (0.549)         &  (0.549)         &                  &                  &                  &                  \\
 Contribution   &   -0.622\sym{***}&   -0.622\sym{***}&   -0.654\sym{***}&   -0.654\sym{***}&   -0.573\sym{***}&   -0.573\sym{***}\\
                &  (0.049)         &  (0.049)         &  (0.029)         &  (0.029)         &  (0.025)         &  (0.025)         \\
Punished subject contrib.&    0.212\sym{***}&    0.212\sym{***}&    0.310\sym{***}&    0.310\sym{***}&    0.195\sym{***}&    0.195\sym{***}\\
                &  (0.001)         &  (0.001)         &  (0.118)         &  (0.118)         &  (0.036)         &  (0.036)         \\
Average contrib. others&   -0.013         &   -0.013         &    0.076\sym{***}&    0.076\sym{***}&   -0.070         &   -0.070         \\
                &  (0.049)         &  (0.049)         &  (0.000)         &  (0.000)         &  (0.067)         &  (0.067)         \\
Lagged p. received&    0.030\sym{**} &    0.030\sym{**} &    0.059         &    0.059         &    0.019\sym{***}&    0.019\sym{***}\\
                &  (0.013)         &  (0.013)         &  (0.063)         &  (0.063)         &  (0.002)         &  (0.002)         \\
Female          &   -5.301\sym{***}&   -5.301\sym{***}&   -4.906         &   -4.905         &   -3.481         &   -3.481         \\
                &  (1.555)         &  (1.555)         &  (4.157)         &  (4.157)         &  (3.630)         &  (3.630)         \\
Group of 4      &    5.268\sym{**} &    5.268\sym{**} &   -0.828         &   -0.828         &   10.680\sym{***}&   10.680\sym{***}\\
                &  (2.260)         &  (2.260)         &  (4.650)         &  (4.650)         &  (0.112)         &  (0.112)         \\
Group of 5      &    8.050\sym{***}&    8.050\sym{***}&   -4.514         &   -4.513         &   13.380\sym{***}&   13.381\sym{***}\\
                &  (0.244)         &  (0.244)         &  (6.202)         &  (6.202)         &  (2.162)         &  (2.162)         \\
Constant        &   31.182\sym{***}&   31.182\sym{***}&   34.499\sym{***}&   34.500\sym{***}&   27.053\sym{***}&   27.054\sym{***}\\
                &  (2.073)         &  (2.073)         &  (4.334)         &  (4.334)         &  (0.657)         &  (0.657)         \\
\midrule
N Obs.          &      751         &      751         &      267         &      267         &      484         &      484         \\
Round fixed effects&      Yes         &      Yes         &      Yes         &      Yes         &      Yes         &      Yes         \\
Random Slopes   &      Yes         &      Yes         &      Yes         &      Yes         &      Yes         &      Yes         \\
Individual correlation&      Yes         &      Yes         &      Yes         &      Yes         &      Yes         &      Yes         \\
Class nesting   &      Yes         &      Yes         &      Yes         &      Yes         &      Yes         &      Yes         \\
Group nesting   &       No         &      Yes         &       No         &      Yes         &       No         &      Yes         \\
\bottomrule
\multicolumn{7}{l}{\footnotesize Column 2: Post estimation tests for Common = [Common x Immigrant - Common]: p=0.0463.}\\
\multicolumn{7}{l}{\footnotesize Column 2: Post estimation tests for Multicultural = [Multicultural x Immigrant - Multicultural]: p=0.0017.}\\
\multicolumn{7}{l}{\footnotesize Column 4: Post estimation tests for Common = Multicultural: p=0.0000.}\\
\multicolumn{7}{l}{\footnotesize Column 6: Post estimation tests for Common = Multicultural: p=0.2888.}\\
\multicolumn{7}{l}{\footnotesize Symbols $***$, $**$, and $*$ indicate significance at the 1\%, 5\% and 10\% level, respectively.}\\
\end{tabular}
}

\end{adjustbox}
\end{center}
	\begin{tablenotes}
		\small \item  \textbf{Notes}: Linear mixed-effects models, to account for the hierarchical structure of the data, i.e. observations are nested in groups, and the correlation over rounds of individual observations. In particular, the four-level model uses random intercepts and slopes for individual choices observed over rounds, and it account for dependence in the observations among the individuals belonging to the same group, class and school. Columns 1-2 looks at the full sample. Columns 3 and 4 at immigrants only. Columns 5-6 at natives only.
	\end{tablenotes}
\label{tab:xtmixed_antisocialp_cost}
\end{table}

\begin{table}[htbp]
	\caption{Contribution to the Public Good - Part II} 
	\begin{center}
		\begin{adjustbox}{width=1\textwidth}
			{
\def\sym#1{\ifmmode^{#1}\else\(^{#1}\)\fi}
\begin{tabular}{l*{6}{c}}
\toprule
                &\multicolumn{1}{c}{(1)}&\multicolumn{1}{c}{(2)}&\multicolumn{1}{c}{(3)}&\multicolumn{1}{c}{(4)}&\multicolumn{1}{c}{(5)}&\multicolumn{1}{c}{(6)}\\
                &\multicolumn{1}{c}{Full}&\multicolumn{1}{c}{Full}&\multicolumn{1}{c}{Immigrants}&\multicolumn{1}{c}{Immigrants}&\multicolumn{1}{c}{Natives}&\multicolumn{1}{c}{Natives}\\
\midrule
Contribution    &                  &                  &                  &                  &                  &                  \\
Common          &   -2.254         &   -2.050         &   -4.772         &   -4.772         &   -2.179         &   -2.536         \\
                &  (2.153)         &  (2.132)         &  (3.940)         &  (3.940)         &  (2.128)         &  (2.508)         \\
Multicultural   &   -0.206         &    0.075         &   -1.609         &   -1.609         &   -0.018         &    0.244         \\
                &  (2.537)         &  (2.627)         &  (2.465)         &  (2.465)         &  (2.381)         &  (2.469)         \\
Immigrant       &    2.397\sym{***}&    3.193\sym{***}&                  &                  &                  &                  \\
                &  (0.494)         &  (0.392)         &                  &                  &                  &                  \\
Common x Immigrant&   -1.904         &   -2.878         &                  &                  &                  &                  \\
                &  (2.384)         &  (3.529)         &                  &                  &                  &                  \\
Multicultural x Immigrant&   -2.489\sym{***}&   -2.748\sym{***}&                  &                  &                  &                  \\
                &  (0.060)         &  (0.179)         &                  &                  &                  &                  \\
Female          &    1.776\sym{***}&    2.170\sym{***}&   -1.949         &   -1.949         &    3.118\sym{***}&    3.926\sym{***}\\
                &  (0.319)         &  (0.673)         &  (1.982)         &  (1.982)         &  (0.998)         &  (0.333)         \\
Group of 4      &   -2.416         &   -2.334         &    2.321         &    2.320         &   -3.876         &   -3.464         \\
                &  (4.070)         &  (4.056)         &  (2.665)         &  (2.665)         &  (4.766)         &  (4.473)         \\
Group of 5      &   -1.760         &   -1.557         &    4.562\sym{***}&    4.561\sym{***}&   -3.750         &   -2.727         \\
                &  (2.828)         &  (2.560)         &  (1.367)         &  (1.367)         &  (2.592)         &  (2.562)         \\
Constant        &   23.242\sym{***}&   22.718\sym{***}&   24.301\sym{***}&   24.301\sym{***}&   23.362\sym{***}&   22.484\sym{***}\\
                &  (6.629)         &  (6.567)         &  (3.723)         &  (3.723)         &  (7.878)         &  (7.593)         \\
\midrule
N Obs.          &    3,900         &    3,900         &    1,300         &    1,300         &    2,600         &    2,600         \\
Round fixed effects&      Yes         &      Yes         &      Yes         &      Yes         &      Yes         &      Yes         \\
Clustering      &      Yes         &      Yes         &      Yes         &      Yes         &      Yes         &      Yes         \\
Random Slopes   &      Yes         &      Yes         &      Yes         &      Yes         &      Yes         &      Yes         \\
Individual correlation&      Yes         &      Yes         &      Yes         &      Yes         &      Yes         &      Yes         \\
Class nesting   &      Yes         &      Yes         &      Yes         &      Yes         &      Yes         &      Yes         \\
Group nesting   &       No         &      Yes         &       No         &      Yes         &       No         &      Yes         \\
\bottomrule
\multicolumn{7}{l}{\footnotesize Column 4: Post estimation tests for Common = [Common x Immigrant - Common]: p=0.0962.}\\
\multicolumn{7}{l}{\footnotesize Column 4: Post estimation tests for Multicultural = [Multicultural x Immigrant - Multicultural]: p=0.5679.}\\
\multicolumn{7}{l}{\footnotesize Column 6: Post estimation tests for Common = Multicultural: p=0.0320.}\\
\multicolumn{7}{l}{\footnotesize Column 8: Post estimation tests for Common = Multicultural: p=0.0000.}\\
\multicolumn{7}{l}{\footnotesize Symbols $***$, $**$, and $*$ indicate significance at the 1\%, 5\% and 10\% level, respectively.}\\
\end{tabular}
}

		\end{adjustbox}
	\end{center}
	\begin{tablenotes}
		\small \item  \textbf{Notes}: \small \item  \textbf{Notes}: Linear mixed-effects models, to account for the hierarchical structure of the data, i.e. observations are nested in groups, and the correlation over rounds of individual observations. In particular, the four-level model uses random intercepts and slopes for individual choices observed over rounds, and it account for dependence in the observations among the individuals belonging to the same group, class and school. Columns 1-2 looks at the full sample. Columns 3 and 4 at immigrants only. Columns 5-6 at natives only.
	\end{tablenotes}
	\label{tab:xtmixed_contribution2}
\end{table}

\begin{table}[htbp]
	\caption{Contribution to the Public Good - Part I + II}
	\begin{center}
		\begin{adjustbox}{width=1\textwidth}
			{
\def\sym#1{\ifmmode^{#1}\else\(^{#1}\)\fi}
\begin{tabular}{l*{6}{c}}
\toprule
                &\multicolumn{1}{c}{(1)}&\multicolumn{1}{c}{(2)}&\multicolumn{1}{c}{(3)}&\multicolumn{1}{c}{(4)}&\multicolumn{1}{c}{(5)}&\multicolumn{1}{c}{(6)}\\
                &\multicolumn{1}{c}{Full}&\multicolumn{1}{c}{Full}&\multicolumn{1}{c}{Immigrants}&\multicolumn{1}{c}{Immigrants}&\multicolumn{1}{c}{Natives}&\multicolumn{1}{c}{Natives}\\
\midrule
Contribution    &                  &                  &                  &                  &                  &                  \\
Common          &    0.768\sym{***}&    0.777\sym{***}&   -0.264         &   -0.264         &    0.502         &    0.115         \\
                &  (0.082)         &  (0.112)         &  (3.731)         &  (3.731)         &  (0.342)         &  (0.844)         \\
Multicultural   &    2.776\sym{***}&    2.852\sym{***}&    0.539         &    0.539         &    2.534\sym{***}&    2.640\sym{**} \\
                &  (0.646)         &  (0.666)         &  (3.158)         &  (3.158)         &  (0.858)         &  (1.128)         \\
Immigrant       &    3.581\sym{**} &    3.834\sym{**} &                  &                  &                  &                  \\
                &  (1.554)         &  (1.512)         &                  &                  &                  &                  \\
Common x Immigrant&   -0.920         &   -1.192         &                  &                  &                  &                  \\
                &  (5.329)         &  (5.542)         &                  &                  &                  &                  \\
Multicultural x Immigrant&   -2.763         &   -2.810         &                  &                  &                  &                  \\
                &  (4.060)         &  (4.116)         &                  &                  &                  &                  \\
Female          &    0.895         &    0.881         &   -2.007\sym{*}  &   -2.007\sym{*}  &    2.108\sym{*}  &    2.360\sym{***}\\
                &  (0.900)         &  (0.911)         &  (1.124)         &  (1.124)         &  (1.253)         &  (0.893)         \\
Group of 4      &   -0.258         &   -0.280         &    5.770\sym{***}&    5.770\sym{***}&   -2.809         &   -2.757         \\
                &  (1.648)         &  (1.659)         &  (1.096)         &  (1.096)         &  (2.173)         &  (2.267)         \\
Group of 5      &    0.443\sym{***}&    0.575\sym{***}&    4.217\sym{***}&    4.217\sym{***}&   -0.520         &    0.192         \\
                &  (0.128)         &  (0.093)         &  (0.751)         &  (0.751)         &  (0.586)         &  (0.615)         \\
Constant        &   22.790\sym{***}&   22.694\sym{***}&   24.256\sym{***}&   24.256\sym{***}&   23.542\sym{***}&   23.325\sym{***}\\
                &  (2.508)         &  (2.561)         &  (0.967)         &  (0.967)         &  (4.028)         &  (4.330)         \\
\midrule
N Obs.          &    7,800         &    7,800         &    2,600         &    2,600         &    5,200         &    5,200         \\
Round fixed effects&      Yes         &      Yes         &      Yes         &      Yes         &      Yes         &      Yes         \\
Clustering      &      Yes         &      Yes         &      Yes         &      Yes         &      Yes         &      Yes         \\
Random Slopes   &      Yes         &      Yes         &      Yes         &      Yes         &      Yes         &      Yes         \\
Individual correlation&      Yes         &      Yes         &      Yes         &      Yes         &      Yes         &      Yes         \\
Class nesting   &      Yes         &      Yes         &      Yes         &      Yes         &      Yes         &      Yes         \\
Group nesting   &       No         &      Yes         &       No         &      Yes         &       No         &      Yes         \\
\bottomrule
\multicolumn{7}{l}{\footnotesize Column 4: Post estimation tests for Common = [Common x Immigrant - Common]: p=0.6055.}\\
\multicolumn{7}{l}{\footnotesize Column 4: Post estimation tests for Multicultural = [Multicultural x Immigrant - Multicultural]: p=0.0022.}\\
\multicolumn{7}{l}{\footnotesize Column 6: Post estimation tests for Common = Multicultural: p=0.1610.}\\
\multicolumn{7}{l}{\footnotesize Column 8: Post estimation tests for Common = Multicultural: p=0.0000.}\\
\multicolumn{7}{l}{\footnotesize Symbols $***$, $**$, and $*$ indicate significance at the 1\%, 5\% and 10\% level, respectively.}\\
\end{tabular}
}

		\end{adjustbox}
	\end{center}
	\begin{tablenotes}
		\small \item \textbf{Notes}: Linear mixed-effects models, to account for the hierarchical structure of the data, i.e. observations are nested in groups, and the correlation over rounds of individual observations. In particular, the four-level model uses random intercepts and slopes for individual choices observed over rounds, and it account for dependence in the observations among the individuals belonging to the same group, class and school. Columns 1-2 looks at the full sample. Columns 3 and 4 at immigrants only. Columns 5-6 at natives only.
	\end{tablenotes}
    \label{tab:xtmixed_contribution12}
\end{table}

\begin{table}[htbp]
	\caption{Individual characteristics - Immigrants and Natives} 
	\begin{center}
		\begin{adjustbox}{width=0.9\textwidth}
			{
	\def\sym#1{\ifmmode^{#1}\else\(^{#1}\)\fi}
	\begin{tabular}{l*{6}{c}}
                    &        Full&     Natives&  Immigrants&       Diff.         \\
\midrule
Female              &       0.413&       0.408&       0.423&      -0.015         \\
                    &     (0.493)&     (0.492)&     (0.496)&                     \\
How often I practice sport&       3.854&       3.935&       3.692&       0.242         \\
                    &     (1.307)&     (1.252)&     (1.402)&                     \\
How often I surf the web&       3.036&       2.904&       3.300&      -0.396\sym{*}  \\
                    &     (1.522)&     (1.497)&     (1.543)&                     \\
How often I watch TV&       4.700&       4.812&       4.477&       0.335         \\
                    &     (1.654)&     (1.637)&     (1.672)&                     \\
Number of siblings  &       1.152&       1.000&       1.495&      -0.495\sym{***}\\
                    &     (0.905)&     (0.820)&     (0.994)&                     \\
Mother has a job    &       0.787&       0.842&       0.677&       0.165\sym{***}\\
                    &     (0.410)&     (0.365)&     (0.469)&                     \\
Mother has a degree &       0.536&       0.573&       0.462&       0.112\sym{*}  \\
                    &     (0.499)&     (0.496)&     (0.500)&                     \\
Father has a degree &       0.482&       0.481&       0.485&      -0.004         \\
                    &     (0.500)&     (0.501)&     (0.502)&                     \\
Father has a job    &       0.795&       0.823&       0.738&       0.085         \\
                    &     (0.404)&     (0.382)&     (0.441)&                     \\
I maximized my own payoff&       0.782&       0.819&       0.708&       0.112\sym{*}  \\
                    &     (0.413)&     (0.386)&     (0.457)&                     \\
I did what I think you wanted me to do&       0.419&       0.412&       0.434&      -0.023         \\
                    &     (0.494)&     (0.493)&     (0.498)&                     \\
Mistakes in control question 1&       0.846&       0.735&       1.069&      -0.335         \\
                    &     (2.339)&     (2.300)&     (2.409)&                     \\
Mistakes in control question 2&       1.044&       0.846&       1.438&      -0.592         \\
                    &     (3.171)&     (2.636)&     (4.017)&                     \\
Calm                &       0.615&       0.627&       0.592&       0.035         \\
                    &     (0.487)&     (0.485)&     (0.493)&                     \\
Tense               &       0.651&       0.662&       0.631&       0.031         \\
                    &     (0.477)&     (0.474)&     (0.484)&                     \\
Upset               &       0.679&       0.688&       0.662&       0.027         \\
                    &     (0.467)&     (0.464)&     (0.475)&                     \\
Relaxed             &       0.505&       0.500&       0.515&      -0.015         \\
                    &     (0.501)&     (0.501)&     (0.502)&                     \\
Happy               &       0.682&       0.708&       0.631&       0.077         \\
                    &     (0.466)&     (0.456)&     (0.484)&                     \\
Worried             &       0.759&       0.781&       0.715&       0.065         \\
                    &     (0.428)&     (0.415)&     (0.453)&                     \\
\end{tabular}
}
		\end{adjustbox}
	\end{center}
	\begin{tablenotes}
		 \small \item  \textbf{Notes}: The Table show how the observable characteristics differ between natives (Column 2) and immigrants (Column 3) at the baseline. Symbols $***$, $**$, and $*$ indicate significance at the 1\%, 5\% and 10\% level, respectively.
	\end{tablenotes}
	\label{tab:individual_char}
\end{table}

\begin{table}[htbp]
	\caption{Contribution to the Public Good - Mechanisms} 
	\begin{center}
		\begin{adjustbox}{width=1\textwidth}
			{
\def\sym#1{\ifmmode^{#1}\else\(^{#1}\)\fi}
\begin{tabular}{l*{6}{c}}
\toprule
                &\multicolumn{1}{c}{(1)}&\multicolumn{1}{c}{(2)}&\multicolumn{1}{c}{(3)}&\multicolumn{1}{c}{(4)}&\multicolumn{1}{c}{(5)}&\multicolumn{1}{c}{(6)}\\
                &\multicolumn{1}{c}{Contribution}&\multicolumn{1}{c}{Contribution}&\multicolumn{1}{c}{Contribution}&\multicolumn{1}{c}{Contribution}&\multicolumn{1}{c}{Contribution}&\multicolumn{1}{c}{Contribution}\\
\midrule
Contribution    &                  &                  &                  &                  &                  &                  \\
Common          &    0.672\sym{**} &    0.830\sym{***}&    0.640\sym{***}&    0.627         &   -0.224         &    0.213         \\
                &  (0.296)         &  (0.302)         &  (0.077)         &  (0.436)         &  (0.800)         &  (0.551)         \\
Multicultural   &    2.923\sym{***}&    2.842\sym{***}&    2.766\sym{***}&    2.797\sym{***}&    2.779\sym{***}&    2.838\sym{***}\\
                &  (0.662)         &  (0.662)         &  (0.710)         &  (0.708)         &  (0.662)         &  (0.686)         \\
Immigrant       &    3.771\sym{***}&    3.808\sym{***}&    3.300\sym{***}&    3.334\sym{***}&    3.060\sym{**} &    3.749\sym{**} \\
                &  (1.274)         &  (1.475)         &  (0.899)         &  (0.705)         &  (1.524)         &  (1.780)         \\
Common x Immigrant&   -0.917         &   -0.972         &   -0.694         &   -0.630         &    2.064         &    0.932         \\
                &  (5.103)         &  (5.982)         &  (5.008)         &  (5.577)         &  (4.795)         &  (5.191)         \\
Multicultural x Immigrant&   -2.796         &   -2.749         &   -2.798         &   -2.786         &   -2.635         &   -2.138         \\
                &  (4.023)         &  (4.222)         &  (3.819)         &  (4.067)         &  (6.323)         &  (5.024)         \\
Not having siblings   &    1.363         &                  &                  &    0.443         &                  &                  \\
                &  (1.211)         &                  &                  &  (0.377)         &                  &                  \\
Mother has a degree=1&                  &    0.681         &                  &    0.088         &                  &                  \\
                &                  &  (1.869)         &                  &  (2.895)         &                  &                  \\
Mother has a job=1&                  &                  &   -2.152         &   -1.918         &                  &                  \\
                &                  &                  &  (1.389)         &  (2.353)         &                  &                  \\
1st PCA (SES)            &                  &                  &                  &                  &   -0.208         &                  \\
                &                  &                  &                  &                  &  (0.728)         &                  \\
Share sent in Dictator Game&                  &                  &                  &                  &                  &    4.429         \\
                &                  &                  &                  &                  &                  &  (4.498)         \\
Female          &    0.840         &    0.918         &    0.914         &    0.901         &    1.589\sym{*}  &    0.296         \\
                &  (0.984)         &  (0.996)         &  (0.856)         &  (1.024)         &  (0.883)         &  (1.262)         \\
Group of 4      &   -0.304         &   -0.252         &   -0.346         &   -0.343         &    0.064         &    0.801         \\
                &  (1.710)         &  (1.739)         &  (1.739)         &  (1.896)         &  (1.561)         &  (1.301)         \\
Group of 5      &    0.573\sym{***}&    0.642\sym{**} &    0.581\sym{**} &    0.587         &    0.243         &    0.220         \\
                &  (0.124)         &  (0.300)         &  (0.271)         &  (0.540)         &  (0.554)         &  (0.668)         \\
Constant        &   22.411\sym{***}&   22.253\sym{***}&   24.587\sym{***}&   24.233\sym{***}&   22.920\sym{***}&   21.739\sym{***}\\
                &  (2.292)         &  (3.805)         &  (3.716)         &  (6.365)         &  (6.597)         &  (4.330)         \\
\midrule
N Obs.          &    7,800         &    7,800         &    7,800         &    7,800         &    6,460         &    6,220         \\
Round fixed effects&      Yes         &      Yes         &      Yes         &      Yes         &      Yes         &      Yes         \\
Clustering      &      Yes         &      Yes         &      Yes         &      Yes         &      Yes         &      Yes         \\
Random Slopes   &      Yes         &      Yes         &      Yes         &      Yes         &      Yes         &      Yes         \\
Individual correlation&      Yes         &      Yes         &      Yes         &      Yes         &      Yes         &      Yes         \\
Class nesting   &      Yes         &      Yes         &      Yes         &      Yes         &      Yes         &      Yes         \\
Group nesting   &      Yes         &      Yes         &      Yes         &      Yes         &      Yes         &      Yes         \\
\bottomrule
\multicolumn{7}{l}{\footnotesize Symbols $***$, $**$, and $*$ indicate significance at the 1\%, 5\% and 10\% level, respectively.}\\
\end{tabular}
}

		\end{adjustbox}
	\end{center}
	\begin{tablenotes}
		\small \item  \textbf{Notes}: Linear mixed-effects models, to account for the hierarchical structure of the data, i.e. observations are nested in groups, and the correlation over rounds of individual observations. In particular, the four-level model uses random intercepts and slopes for individual choices observed over rounds, and it account for dependence in the observations among the individuals belonging to the same group, class and school. Columns 1 include the number of siblings, Column 2 control for the level of education of the mother, in Column 3 we add the working status of the mother, Column 4 include all the proxies of the socio-economic background jointly, Column 5 include the principal component of the socio-economic backrgound, while Column 6 controls for the unconditional "generosity" as measures by the share of endowment kept in the Dictator Game.
	\end{tablenotes}
	\label{tab:xtmixed_mech}
\end{table}

\begin{table}[htbp]
	\caption{Contribution to the Public Good - Context Mechanisms} 
	\begin{center}
		\begin{adjustbox}{width=1\textwidth}
			{
\def\sym#1{\ifmmode^{#1}\else\(^{#1}\)\fi}
\begin{tabular}{l*{5}{c}}
\toprule
                &\multicolumn{1}{c}{(1)}&\multicolumn{1}{c}{(2)}&\multicolumn{1}{c}{(3)}&\multicolumn{1}{c}{(4)}&\multicolumn{1}{c}{(5)}\\
                &\multicolumn{1}{c}{Contribution}&\multicolumn{1}{c}{Contribution}&\multicolumn{1}{c}{Contribution}&\multicolumn{1}{c}{Contribution}&\multicolumn{1}{c}{Contribution}\\
\midrule
    &                  &                  &                  &                  &                  \\
Common          &    2.387         &    1.813         &    5.858         &    4.407         &    1.061         \\
                &  (4.788)         &  (2.126)         & (10.569)         &  (6.350)         &  (1.819)         \\
Multicultural   &    4.500\sym{***}&    3.387\sym{***}&   10.122\sym{**} &    5.330\sym{***}&    1.742\sym{***}\\
                &  (0.295)         &  (1.038)         &  (4.912)         &  (0.051)         &  (0.351)         \\
Common $\times$ Immigrants in the class&   -5.616         &                  &                  &                  &                  \\
                & (10.544)         &                  &                  &                  &                  \\
Multicultural $\times$ Immigrants in the class&   -5.792         &                  &                  &                  &                  \\
                &  (4.723)         &                  &                  &                  &                  \\
Common $\times$ High share of immigrants born in Italy=1&                  &   -2.472         &                  &                  &                  \\
                &                  &  (2.404)         &                  &                  &                  \\
Multicultural $\times$ High share of immigrants born in Italy=1&                  &   -1.576\sym{***}&                  &                  &                  \\
                &                  &  (0.137)         &                  &                  &                  \\
Common $\times$ Network Density&                  &                  &  -22.446         &                  &                  \\
                &                  &                  & (52.123)         &                  &                  \\
Multicultural $\times$ Network Density&                  &                  &  -31.329         &                  &                  \\
                &                  &                  & (30.869)         &                  &                  \\
Common $\times$ Centrality-in among Immigrants&                  &                  &                  &   -1.091         &                  \\
                &                  &                  &                  &  (2.350)         &                  \\
Multicultural $\times$ Centrality-in among Immigrants&                  &                  &                  &   -0.798         &                  \\
                &                  &                  &                  &  (0.620)         &                  \\
Common $\times$ Mentioned as friends immigrants&                  &                  &                  &                  &   -0.751         \\
                &                  &                  &                  &                  &  (2.969)         \\
Multicultural $\times$ Mentioned as friends immigrants&                  &                  &                  &                  &    0.444         \\
                &                  &                  &                  &                  &  (1.055)         \\
Immigrants in the class&   10.222\sym{**} &                  &                  &                  &    9.644\sym{**} \\
                &  (4.561)         &                  &                  &                  &  (3.896)         \\
High share of immigrants born in Italy=1&                  &    0.744         &                  &                  &                  \\
                &                  &  (1.044)         &                  &                  &                  \\
Network Density &                  &                  &   12.910         &                  &                  \\
                &                  &                  & (39.701)         &                  &                  \\
Centrality-in among Immigrants&                  &                  &                  &    0.541         &                  \\
                &                  &                  &                  &  (1.639)         &                  \\
Mentioned as friends immigrants&                  &                  &                  &                  &   -0.895         \\
                &                  &                  &                  &                  &  (1.668)         \\
Female          &   -0.475\sym{**} &   -0.038\sym{***}&   -0.529\sym{***}&   -0.342         &   -0.415         \\
                &  (0.216)         &  (0.006)         &  (0.144)         &  (0.506)         &  (1.043)         \\
Constant        &   20.468\sym{***}&   23.257\sym{***}&   20.931\sym{***}&   22.007\sym{***}&   21.878\sym{***}\\
                &  (3.075)         &  (2.189)         &  (7.268)         &  (3.410)         &  (1.618)         \\
\midrule
N Obs.          &    2,160         &    2,160         &    2,160         &    2,160         &    2,140         \\
Round fixed effects&      Yes         &      Yes         &      Yes         &      Yes         &      Yes         \\
Clustering      &      Yes         &      Yes         &      Yes         &      Yes         &      Yes         \\
Random Slopes   &      Yes         &      Yes         &      Yes         &      Yes         &      Yes         \\
Individual correlation&      Yes         &      Yes         &      Yes         &      Yes         &      Yes         \\
Class nesting   &      Yes         &      Yes         &      Yes         &      Yes         &      Yes         \\
Group nesting   &      Yes         &      Yes         &      Yes         &      Yes         &      Yes         \\
\bottomrule
\multicolumn{6}{l}{\footnotesize Symbols $***$, $**$, and $*$ indicate significance at the 1\%, 5\% and 10\% level, respectively.}\\
\end{tabular}
}

		\end{adjustbox}
	\end{center}
	\begin{tablenotes}
		\small \item  \textbf{Notes}: Linear mixed-effects models, to account for the hierarchical structure of the data, i.e. observations are nested in groups, and the correlation over rounds of individual observations. In particular, the four-level model uses random intercepts and slopes for individual choices observed over rounds, and it account for dependence in the observations among the individuals belonging to the same group, class and school. Columns 1 interact the treatment with the share of immigrants in the classroom, Column 2 control for the share of first generation immigrants, in Column 3 we add the network density, Column 4, 5 and 6 include the interaction with a measure of classroom-network density, the centrality of immigrants in the network, and the number of mentions as immigrant friend (while controlling for the share of immigrants in the classroom).
	\end{tablenotes}
	\label{tab:xtmixed_group_mech}
\end{table}
\clearpage

\appendix
\section*{Appendix}\label{sec:app}
\addcontentsline{toc}{section}{Appendix}

\renewcommand{\thesection}{A.\arabic{section}}
\renewcommand{\thetable}{A.\arabic{table}}
\renewcommand{\thefigure}{A.\arabic{figure}}



\section{Institutional Background} \label{appA}
In this appendix, we provide additional detail on the institutional and demographic setting of our study, focusing on immigration trends in Italy and the organization of the Italian school system.
\subsection{Immigration in Italy} 
The enrollment of students without Italian citizenship (non-Italian) has steadily increased in Italy since the early 1990s. However, the overall share of non-Italian students remained below 1.5\% until the academic year 1999/2000. Over the following decade, this share grew significantly, reaching 7.52\% in 2009/2010—rising from 119,679 to 673,800 students. By 2019/2020, the academic year during which our experiment took place, non-Italian students accounted for 10.3\% of the total student population—an increase of less than three percentage points compared to ten years earlier. Notably, around 62.2\% of students with a migrant background were born in Italy \citep{miur2021}.\footnote{Italian citizenship is acquired primarily through the principle of \emph{jus sanguinis} (right of blood), meaning that citizenship is granted to those born to Italian parents or adopted by Italian citizens. There are limited cases in which citizenship is acquired by \emph{jus soli} (right of soil)—for instance, when a child is born in Italy to stateless parents or parents whose nationality is unknown or cannot be transmitted. Foreign nationals may apply for citizenship after at least ten years of legal residence in Italy, provided they meet specific requirements, including stable income, a clean criminal record, and no threats to national security. Additionally, Italian citizenship can be obtained through marriage: a non-Italian spouse may apply for citizenship after two years of legal residence in Italy post-marriage, assuming there has been no dissolution or separation of the union.} 

In the region of Emilia-Romagna—where this study was conducted—the share of enrolled students without Italian citizenship is 17.1\%, the highest among all Italian regions. In the municipality of Bologna, this figure is 16.4\% among middle school students, 62.6\% of whom were born in Italy. The majority of students of migrant origin in Bologna come from Europe (38\%), including 18.5\% from European Union member states, followed by students from Asia (29.0\%) and Africa (28.7\%) \citep{miur2021}.\footnote{The most represented countries of origin among non-Italian students in Bologna include: Pakistan (8.7\%), Bangladesh (6.1\%), Moldova (5.0\%), the Philippines (4.1\%), Morocco (3.6\%), China (2.4\%), Romania (2.4\%), Albania (1.7\%), India (1.1\%), and Egypt (0.9\%) \citep{miur2021}.} In this study, we identify immigrant students based on their parents' citizenship status. This definition encompasses both first-generation immigrants (students born outside of Italy) and second-generation immigrants (students born in Italy to non-Italian parents). As in many European countries—unlike the United States—Italy adopts the \emph{jus sanguinis} principle, which does not automatically confer citizenship to children born on Italian soil unless certain conditions are met. Non-Italian students born in Italy may apply for citizenship upon reaching the age of 18, provided they can demonstrate uninterrupted legal residency since birth. 

Thus, throughout the paper, we refer to any student whose parents were not born in Italy as an immigrant. Table~\ref{tab:country} presents the distribution of citizenship among immigrant students in our sample. When considering the place of birth, 75.4\% of immigrant students were born in Italy (thus within the European Union), 11.5\% were born elsewhere in Europe, 8.5\% in Asia, 1.5\% in Africa, and 0.8\% in Central or South America. 

Table~\ref{tab:macro_char} describes the socio-economic characteristics of the municipality of Bologna and the two city districts involved in our study. On average, foreign residents constitute about 15\% of Bologna’s total population. The two districts we focus on—“Borgo Panigale - Reno” and “Savena”—mirror this trend, with foreign resident shares of 16\% and 14\%, respectively. In terms of average income and educational attainment, both districts fall slightly below the citywide average, particularly in “Borgo Panigale - Reno”.

\subsection{The Educational System in Italy}

In Italy, education is publicly funded and provided free of charge to all children. Schooling is compulsory from ages 6 to 16 and is structured into three main stages: five years of primary school (\emph{scuola primaria}), three years of middle school (\emph{scuola secondaria di primo grado}), and five years of high school (\emph{scuola secondaria di secondo grado}). Upon completing middle school, students must choose one of three high school tracks: academic (\emph{liceo}), technical, or vocational. 

Italian schools enjoy a high degree of autonomy: each institution defines its own curriculum, expands its educational offerings, and organizes teaching methods, including school hours and the composition of student groups. At all school levels—primary, middle, and high—students follow all subjects within the same class and with the same cohort of peers. In both primary and middle school, it is common for students to have the same teacher for a given subject throughout the full educational cycle. For example, a student’s mathematics and science teacher typically remains the same for all three years of middle school.

In recent years, the Italian Ministry of Education has faced increasing challenges in managing the integration and placement of students with non-Italian citizenship. In response, since 2006 the Ministry has issued several sets of guidelines aimed at promoting inclusion, enhancing intercultural dialogue, and addressing the growing ethnic and cultural diversity in classrooms.\footnote{These guidelines include: \emph{Linee guida per l'accoglienza e l'integrazione degli alunni stranieri} (2006); \emph{La via italiana per la scuola interculturale} (2007); \emph{Linee guida per l'accoglienza e l'integrazione degli alunni stranieri} (2014); \emph{Diversi da chi?} (2015); and \emph{Orientamenti Interculturali. Idee e proposte per l’integrazione di alunne e alunni provenienti da contesti migratori} (2022).}

According to ministerial guidelines introduced in the 2010/2011 academic year, the proportion of students with non-Italian citizenship and limited Italian language proficiency should not, as a rule, exceed 30\% of enrolled students per class.\footnote{This threshold may be exceeded if students with non-Italian citizenship already possess adequate language skills, or reduced in cases where many students lack basic Italian proficiency. Importantly, schools are not permitted to refuse enrollment based on surpassing this threshold.} Nationally, Emilia-Romagna—the region where our study was conducted—has the highest percentage of classes that meet or exceed this 30\% limit (16.2\%) \citep{miur2021}.

Table~\ref{tab:school_char} reports a few descriptive statistics of our study population across the two participating schools. In both institutions, the average share of immigrant students per class exceeds the national threshold, standing at approximately 33\%. Similarly, the average proportion of immigrant students per experimental group is also around 33\%.
\clearpage
\section{Tables and Figures}
\addcontentsline{toc}{section}{Appendix Tables and Figures}

\begin{table}[H]
	\caption{Summary Statistics - Immigrants origins} 
	\begin{center}
		\begin{adjustbox}{width=0.8\textwidth}
			{
	\def\sym#1{\ifmmode^{#1}\else\(^{#1}\)\fi}
	\begin{tabular}{l*{6}{c}}

                    &  Where born& Mother born& Father born&Grandmother born&Grandfather born\\
\midrule
Afghanistan         &       0.023&       0.031&       0.069&       0.062&       0.046\\
Albania             &       0.008&       0.085&       0.062&       0.077&       0.069\\
Algeria             &       0.000&       0.008&       0.015&       0.008&       0.008\\
Argentina           &       0.000&       0.000&       0.000&       0.000&       0.008\\
Bangladesh          &       0.008&       0.085&       0.077&       0.077&       0.077\\
Cabo Verde          &       0.000&       0.008&       0.008&       0.008&       0.008\\
China               &       0.000&       0.054&       0.054&       0.054&       0.054\\
Colombia            &       0.000&       0.008&       0.000&       0.000&       0.000\\
Cuba                &       0.000&       0.008&       0.008&       0.008&       0.008\\
Dominican Republic  &       0.008&       0.008&       0.008&       0.008&       0.008\\
Eritrea             &       0.000&       0.038&       0.015&       0.038&       0.015\\
Ethiopia            &       0.000&       0.000&       0.023&       0.000&       0.023\\
France              &       0.000&       0.000&       0.000&       0.008&       0.000\\
Germany             &       0.000&       0.008&       0.015&       0.000&       0.000\\
Ghana               &       0.000&       0.008&       0.015&       0.000&       0.008\\
Gibraltar           &       0.000&       0.000&       0.000&       0.008&       0.000\\
Greece              &       0.000&       0.000&       0.000&       0.008&       0.000\\
India               &       0.000&       0.015&       0.023&       0.015&       0.015\\
Italy               &       0.754&       0.000&       0.000&       0.023&       0.038\\
Lebanon             &       0.000&       0.000&       0.008&       0.000&       0.008\\
Moldova             &       0.085&       0.138&       0.123&       0.123&       0.131\\
Monaco              &       0.000&       0.000&       0.000&       0.008&       0.000\\
Morocco             &       0.015&       0.146&       0.131&       0.146&       0.131\\
Mozambique          &       0.000&       0.000&       0.000&       0.000&       0.008\\
Nigeria             &       0.000&       0.038&       0.031&       0.038&       0.038\\
Oman                &       0.000&       0.000&       0.000&       0.000&       0.008\\
Pakistan            &       0.046&       0.046&       0.046&       0.038&       0.038\\
Peru                &       0.008&       0.015&       0.023&       0.008&       0.015\\
Philippines         &       0.008&       0.038&       0.038&       0.038&       0.038\\
Republic of North Macedonia&       0.000&       0.008&       0.000&       0.000&       0.000\\
Romania             &       0.000&       0.085&       0.077&       0.085&       0.069\\
Russian Federation  &       0.000&       0.000&       0.000&       0.008&       0.000\\
Senegal             &       0.000&       0.008&       0.008&       0.008&       0.008\\
Serbia              &       0.008&       0.015&       0.031&       0.015&       0.038\\
Slovakia            &       0.000&       0.008&       0.000&       0.000&       0.000\\
Sudan               &       0.000&       0.000&       0.008&       0.000&       0.008\\
Tunisia             &       0.000&       0.015&       0.015&       0.015&       0.015\\
Turkey              &       0.000&       0.008&       0.008&       0.008&       0.008\\
Ukraine             &       0.015&       0.054&       0.046&       0.038&       0.038\\
Venezuela           &       0.000&       0.000&       0.000&       0.008&       0.000\\
\midrule
Europe Union        &       0.754&       0.100&       0.092&       0.138&       0.108\\
Other Europe        &       0.115&       0.308&       0.269&       0.269&       0.285\\
Asia                &       0.085&       0.269&       0.315&       0.285&       0.285\\
Africa              &       0.015&       0.269&       0.269&       0.262&       0.269\\
Central America     &       0.008&       0.015&       0.015&       0.015&       0.015\\
South America       &       0.008&       0.023&       0.023&       0.015&       0.023\\
\bottomrule
\end{tabular}
}

		\end{adjustbox}
	\end{center}
	\begin{tablenotes}
		\tiny \item \textbf{Notes}: The table reports the nationality distribution of the immigrant students, considering the place where the subject is born (Column 1), the place where the mother (Column 2), the father (Column 3), the grandmother (Column 4), and the grandfather (Column 5) are born. The different nationalities are grouped into ``European Union" which groups together Austria, Croatia, France, Germany, Greece, Italy, Monaco, Poland, Romania, Slovakia, Spain, Switzerland, United Kingdom; ``Other Europe" grouping together Albania, Belarus, Moldova, Republic of North Macedonia, Russian Federation, Serbia, Turkey, Ukraine; ``Asia" groups together Afghanistan, Bangladesh, China, India, Lebanon, Oman, Pakistan, Philippines; ``Africa" groups together Cabo Verde, Egypt, Eritrea, Ethiopia, Ghana, Mozambique, Niger, Nigeria, Senegal, Sudan, Algeria, Morocco, Tunisia; ``Central America" groups together Cuba, Dominican Republic; ``South America" groups together American Samoa, Argentina, Brazil, Chile, Colombia, Peru, Venezuela.
	\end{tablenotes}
	\label{tab:country}
\end{table}
\clearpage

\begin{table}[htbp]
	\caption{Contribution to the Public Good - Mechanisms} 
	\begin{center}
		\begin{adjustbox}{width=1\textwidth}
			
		\end{adjustbox}
	\end{center}
	\begin{tablenotes}
		\small \item  \textbf{Notes}: Linear mixed-effects models, to account for the hierarchical structure of the data, i.e. observations are nested in groups, and the correlation over rounds of individual observations. In particular, the four-level model uses random intercepts and slopes for individual choices observed over rounds, and it account for dependence in the observations among the individuals belonging to the same group, class and school. Columns 1 include the number of siblings, Column 2 control for the level of education of the mother, in Column 3 we add the working status of the mother, Column 4 include all the proxies of the socio-economic background jointly, Column 5 include the principal component of the socio-economic backrgound, while Column 6 controls for the unconditional "generosity" as measures by the share of endowment kept in the Dictator Game.
	\end{tablenotes}
	\label{tab:xtmixed_mech}
\end{table}

\begin{table}[htbp]
	\caption{Summary Statistics - Bologna and Districts} 
	\begin{center}
		\begin{adjustbox}{width=1\textwidth}
			
{
	\def\sym#1{\ifmmode^{#1}\else\(^{#1}\)\fi}
	\begin{tabular}{l*{10}{c}}

                    &     Bologna&Borgo Panigale - Reno&      Navile&Porto Saragozza&San Donato - San Vitale&Santo Stefano&      Savena\\
\midrule
Average income 2017 &       25839&       22097&       21277&       29060&       22365&       35794&       24474\\
\addlinespace
Population 2019     &      391984&       61359&       69525&       69595&       66320&       64510&       60142\\
\addlinespace
Foreign residents 2019&       60698&       10018&       15132&        8287&       11414&        7276&        8402\\
\addlinespace
Share of pop with High School&        0.54&        0.44&        0.47&        0.64&        0.54&        0.68&        0.52\\
\addlinespace
Share of home-owners (households)&        0.61&        0.67&        0.60&        0.57&        0.58&        0.61&        0.69\\
\addlinespace
Share Foreign residents 2019&        0.15&        0.16&        0.22&        0.12&        0.17&        0.11&        0.14\\
\addlinespace
Share of immigrants from EU&        0.23&        0.25&        0.19&        0.24&        0.21&        0.25&        0.26\\
\addlinespace
Share of immigrants from other EU&        0.19&        0.23&        0.16&        0.19&        0.19&        0.19&        0.23\\
\addlinespace
Share of immigrants from west Asia&       0.013&       0.011&      0.0061&       0.016&       0.013&       0.021&       0.016\\
\addlinespace
Share of immigrants from Asia&        0.36&        0.32&        0.41&        0.36&        0.35&        0.35&        0.32\\
\addlinespace
Share of immigrants from North Africa&       0.093&       0.086&        0.12&       0.076&        0.12&       0.049&       0.071\\
\addlinespace
Share of immigrants from Africa&       0.065&       0.065&       0.074&       0.053&       0.074&       0.062&       0.049\\
\addlinespace
Share of immigrants from North America&      0.0048&     0.00090&     0.00099&      0.0091&      0.0032&       0.018&      0.0030\\
\addlinespace
Share of immigrants from Central America&      0.0080&      0.0077&      0.0065&       0.011&      0.0091&      0.0093&      0.0061\\
\addlinespace
Share of immigrants from South America&       0.038&       0.033&       0.029&       0.043&       0.037&       0.050&       0.047\\
\bottomrule

\end{tabular}
}

		\end{adjustbox}
	\end{center}
	\begin{tablenotes}
		\small \item  \textbf{Notes}: The data are provided by the Statistical Office of the Municipality of Bologna. Income is the average income in 2017 expressed in Euro. The population statistics and the related percentage shares are all taken from the 2019 census.
	\end{tablenotes}
	\label{tab:macro_char}
\end{table}

\begin{table}[htbp]
	\caption{Summary Statistics by School} 
	\begin{center}
		\begin{adjustbox}{width=0.85\textwidth}
			{
	\def\sym#1{\ifmmode^{#1}\else\(^{#1}\)\fi}
	\begin{tabular}{l*{6}{c}}

                    &    School 1&    School 2&       Diff.         \\
\midrule
Common              &       0.328&       0.362&      -0.034         \\
                    &     (0.471)&     (0.483)&                     \\
Multicultural       &       0.332&       0.362&      -0.030         \\
                    &     (0.472)&     (0.483)&                     \\
Neutral             &       0.339&       0.276&       0.064         \\
                    &     (0.474)&     (0.449)&                     \\
Class size          &      18.650&      19.517&      -0.868\sym{***}\\
                    &     (2.595)&     (1.853)&                     \\
Percent Immigrants per class&       0.336&       0.328&       0.008         \\
                    &     (0.116)&     (0.068)&                     \\
Percent immigrants per group&       0.336&       0.328&       0.008         \\
                    &     (0.271)&     (0.235)&                     \\
Immigrants per class&       6.372&       6.457&      -0.085         \\
                    &     (2.572)&     (1.701)&                     \\
Immigrants in School&      92&      38&      54         \\
                    &     &     &                     \\
School district     &       Borgo Panigale - Reno&       Savena&               \\
                    &     &     &                     \\
\midrule
Observations        &         274&         116&         390         \\
\bottomrule

\end{tabular}
}

		\end{adjustbox}
	\end{center}
	\begin{tablenotes}
		\small \item  \textbf{Notes}: The table shows the summary statistics by school. \emph{Common}, \emph{Multicultural} and \emph{Neutral} measures the share of students assigned to each treatment within each school. \emph{Class size} measures the average number of students present in the class during the experiment. \emph{Percent Immigrants per class} and  \emph{Percent immigrants per group} show the average share of immigrants students over the total number of student in each class and in each group, respectively.\emph{Immigrants per class} and \emph{Immigrants per school} show the average absolute number of students in the class and the total number of students in the school. \emph{School District} is the name of district of the municipality of Bologna where the school is located. 
	\end{tablenotes}
	\label{tab:school_char}
\end{table}

\begin{table}[htbp]
	\caption{Feelings by Treatment and Immigration Status} 
	\begin{center}
		\begin{adjustbox}{width=1\textwidth}
			{
\def\sym#1{\ifmmode^{#1}\else\(^{#1}\)\fi}
\begin{tabular}{l*{7}{c}}
\toprule
                &\multicolumn{1}{c}{(1)}&\multicolumn{1}{c}{(2)}&\multicolumn{1}{c}{(3)}&\multicolumn{1}{c}{(4)}&\multicolumn{1}{c}{(5)}&\multicolumn{1}{c}{(6)}&\multicolumn{1}{c}{(7)}\\
                &\multicolumn{1}{c}{Calm}&\multicolumn{1}{c}{Tense}&\multicolumn{1}{c}{Upset}&\multicolumn{1}{c}{Relaxed}&\multicolumn{1}{c}{Happy}&\multicolumn{1}{c}{Worried}&\multicolumn{1}{c}{Demand}\\
\midrule
Common          &    0.149         &    0.200         &   -0.084         &    0.034         &   -0.200         &   -0.147         &   -0.162\sym{*} \\
                &  (0.084)         &  (0.149)         &  (0.180)         &  (0.015)         &  (0.167)         &  (0.142)         &  (0.01)         \\
Multicultural   &   -0.109         &    0.066         &   -0.063         &    0.013         &   -0.104         &   -0.160         &   -0.023         \\
                &  (0.082)         &  (0.122)         &  (0.031)         &  (0.024)         &  (0.046)         &  (0.103)         &  (0.050)         \\
Immigrant       &   -0.091         &    0.196         &   -0.017         &    0.143         &   -0.391         &    0.011         &   -0.025         \\
                &  (0.240)         &  (0.182)         &  (0.066)         &  (0.230)         &  (0.229)         &  (0.061)         &  (0.068)         \\
Common x Immigrant&   -0.128         &   -0.573         &   -0.278         &   -0.293         &    0.244         &   -0.573         &    0.104         \\
                &  (0.222)         &  (0.411)         &  (0.115)         &  (0.284)         &  (0.084)         &  (0.244)         &  (0.035)         \\
Multicultural x Immigrant&   -0.051         &   -0.183         &   -0.202         &   -0.184         &    0.260         &   -0.233         &    0.015         \\
                &  (0.134)         &  (0.249)         &  (0.207)         &  (0.386)         &  (0.062)         &  (0.278)         &  (0.144)         \\
Female          &   -0.002         &    0.162         &    0.210         &   -0.053         &   -0.038         &    0.039         &   -0.018         \\
                &  (0.088)         &  (0.027)         &  (0.093)         &  (0.077)         &  (0.115)         &  (0.056)         &  (0.095)         \\
Constant        &    2.701\sym{***}&    2.780\sym{***}&    3.044\sym{***}&    2.458\sym{***}&    3.016\sym{**} &    3.367\sym{**} &    0.484\sym{*}  \\
                &  (0.021)         &  (0.008)         &  (0.041)         &  (0.025)         &  (0.097)         &  (0.086)         &  (0.052)         \\
\midrule
N Obs.          &      389         &      389         &      389         &      389         &      389         &      389         &      389         \\
Clustering      &      Yes         &      Yes         &      Yes         &      Yes         &      Yes         &      Yes         &      Yes         \\
Random Slopes   &      Yes         &      Yes         &      Yes         &      Yes         &      Yes         &      Yes         &      Yes         \\
Class nesting   &      Yes         &      Yes         &      Yes         &      Yes         &      Yes         &      Yes         &      Yes         \\
\bottomrule
\multicolumn{8}{l}{\footnotesize OLS regression. Symbols $***$, $**$, and $*$ indicate significance at the 1\%, 5\% and 10\% level, respectively.}\\
\end{tabular}
}

		\end{adjustbox}
	\end{center}
    \begin{tablenotes}
		\small \item  \textbf{Notes}: The feelings are reported as answer to the question ``I feel ..." as in the short 6-items Spielberger State-Trait Anxiety Inventory \citep{marteau1992development}.
	\end{tablenotes}
	\label{tab:feelings}
\end{table}

\begin{table}[htbp]
	\caption{Mistakes by Treatment and Immigration Status} 
	\begin{center}
		\begin{adjustbox}{width=1\textwidth}
			{
\def\sym#1{\ifmmode^{#1}\else\(^{#1}\)\fi}
\begin{tabular}{l*{3}{c}}
\toprule
                &\multicolumn{1}{c}{(1)}&\multicolumn{1}{c}{(2)}&\multicolumn{1}{c}{(3)}\\
                &\multicolumn{1}{c}{Mistakes in control question 1}&\multicolumn{1}{c}{Mistakes in control question 2}&\multicolumn{1}{c}{Contribution}\\
\midrule
         &                  &                  &                  \\
Common          &   -0.422         &   -0.259         &    0.664\sym{***}\\
                &  (0.151)         &  (0.205)         &  (0.006)         \\
Multicultural   &   -0.121         &   -0.135         &    2.853\sym{***}\\
                &  (0.069)         &  (0.262)         &  (0.413)         \\
Immigrant       &    0.188         &    0.588         &    3.921\sym{***}\\
                &  (0.052)         &  (1.260)         &  (1.218)         \\
Common x Immigrant&    0.033         &   -0.352         &   -1.164         \\
                &  (0.459)         &  (1.613)         &  (4.945)         \\
Multicultural x Immigrant&    0.881         &    0.706\sym{**} &   -2.402         \\
                &  (0.558)         &  (0.039)         &  (3.972)         \\
Female          &   -0.372         &   -0.202         &    0.709         \\
                &  (0.182)         &  (0.386)         &  (0.822)         \\
Mistakes in control question 1&                  &                  &   -0.375\sym{***}\\
                &                  &                  &  (0.037)         \\
Mistakes in control question 2&                  &                  &   -0.128         \\
                &                  &                  &  (0.223)         \\
Constant        &    1.032\sym{**} &    1.036         &   23.178\sym{***}\\
                &  (0.063)         &  (0.237)         &  (2.027)         \\
\midrule
N Obs.          &      390         &      390         &    7,800         \\
Round fixed effects&                  &                  &                  \\
Clustering      &      Yes         &      Yes         &      Yes         \\
Random Slopes   &      Yes         &      Yes         &      Yes         \\
Individual correlation&       No         &       No         &      Yes         \\
Class nesting   &      Yes         &      Yes         &      Yes         \\
Group nesting   &       No         &       No         &      Yes         \\
\bottomrule
\multicolumn{4}{l}{\footnotesize Symbols $***$, $**$, and $*$ indicate significance at the 1\%, 5\% and 10\% level, respectively.}\\
\end{tabular}
}

		\end{adjustbox}
	\end{center}
    \begin{tablenotes}
		\small \item  \textbf{Notes}: The mistake output variables count the number of times the subjects has reported a wrong answer to the control question 1 and 2 we asked befor the main experiment to check if the instructions were clear. Column 3 estimate equation \ref{t:eqn1} directly controlling for the level of understanding by including the number of mistakes in the control questions.
	\end{tablenotes}
	\label{tab:mistakes}
\end{table}

\begin{table}[htbp]
	\caption{OLS: Contribution - Part I} 
	\begin{center}
		\begin{adjustbox}{width=1\textwidth}
			{
\def\sym#1{\ifmmode^{#1}\else\(^{#1}\)\fi}
\begin{tabular}{l*{4}{c}}
\toprule
                &\multicolumn{1}{c}{(1)}&\multicolumn{1}{c}{(2)}&\multicolumn{1}{c}{(3)}&\multicolumn{1}{c}{(4)}\\
                &\multicolumn{1}{c}{Full}&\multicolumn{1}{c}{Full}&\multicolumn{1}{c}{Immigrants}&\multicolumn{1}{c}{Natives}\\
\midrule
Common          &    1.017         &    1.328         &   -0.290         &    1.124         \\
                &  (0.748)         &  (0.925)         &  (1.289)         &  (0.925)         \\
Multicultural   &    2.092\sym{***}&    3.103\sym{***}&    0.116         &    2.933\sym{***}\\
                &  (0.747)         &  (0.913)         &  (1.408)         &  (0.914)         \\
Immigrant       &    2.245\sym{***}&    3.484\sym{***}&                  &                  \\
                &  (0.659)         &  (1.102)         &                  &                  \\
Common x Immigrant&                  &   -0.744         &                  &                  \\
                &                  &  (1.574)         &                  &                  \\
Multicultural x Immigrant&                  &   -3.135\sym{*}  &                  &                  \\
                &                  &  (1.617)         &                  &                  \\
Female          &    1.505\sym{**} &    1.561\sym{**} &   -0.678         &    2.883\sym{***}\\
                &  (0.621)         &  (0.620)         &  (1.202)         &  (0.748)         \\
Group of 4      &                  &                  &                  &                  \\
                &                  &                  &                  &                  \\
Group of 5      &                  &                  &                  &                  \\
                &                  &                  &                  &                  \\
Constant        &   19.547\sym{***}&   19.052\sym{***}&   23.853\sym{***}&   18.588\sym{***}\\
                &  (0.640)         &  (0.722)         &  (0.985)         &  (0.730)         \\
\midrule
N Obs.          &    3,900         &    3,900         &    1,300         &    2,600         \\
Round Fixed effects&      Yes         &      Yes         &      Yes         &      Yes         \\
Classroom fixed effect&      Yes         &      Yes         &      Yes         &      Yes         \\
Robust standard errors&      Yes         &      Yes         &      Yes         &      Yes         \\
\bottomrule
\multicolumn{5}{l}{\footnotesize Column 2: Post estimation tests for Common = [Common x Immigrant - Common]: p=0.2654.}\\
\multicolumn{5}{l}{\footnotesize Column 2: Post estimation tests for Multicultural = [Multicultural x Immigrant - Multicultural]: p=0.0022.}\\
\multicolumn{5}{l}{\footnotesize Column 3: Post estimation tests for Common = Multicultural: p=0.7719.}\\
\multicolumn{5}{l}{\footnotesize Column 4: Post estimation tests for Common = Multicultural: p=0.0390.}\\
\multicolumn{5}{l}{\footnotesize OLS regression, symbols $***$, $**$, and $*$ indicate significance at the 1\%, 5\% and 10\% level, respectively.}\\
\end{tabular}
}

		\end{adjustbox}
	\end{center}
	\label{tab:contribution}
\end{table}

\begin{table}[htbp]
	\caption{OLS: Contribution - Part I + Part II} 
	\begin{center}
		\begin{adjustbox}{width=0.9\textwidth}
			{
\def\sym#1{\ifmmode^{#1}\else\(^{#1}\)\fi}
\begin{tabular}{l*{4}{c}}
\toprule
                &\multicolumn{1}{c}{(1)}&\multicolumn{1}{c}{(2)}&\multicolumn{1}{c}{(3)}&\multicolumn{1}{c}{(4)}\\
                &\multicolumn{1}{c}{Full}&\multicolumn{1}{c}{Full}&\multicolumn{1}{c}{Immigrants}&\multicolumn{1}{c}{Natives}\\
\midrule
Common          &   -0.668         &   -0.198         &   -2.044\sym{**} &   -0.241         \\
                &  (0.513)         &  (0.634)         &  (0.893)         &  (0.632)         \\
Multicultural   &    0.664         &    1.515\sym{**} &    0.055         &    1.570\sym{**} \\
                &  (0.517)         &  (0.632)         &  (0.971)         &  (0.630)         \\
Immigrant       &    1.550\sym{***}&    2.779\sym{***}&                  &                  \\
                &  (0.452)         &  (0.780)         &                  &                  \\
Common x Immigrant&                  &   -1.215         &                  &                  \\
                &                  &  (1.084)         &                  &                  \\
Multicultural x Immigrant&                  &   -2.577\sym{**} &                  &                  \\
                &                  &  (1.121)         &                  &                  \\
Female          &    1.811\sym{***}&    1.837\sym{***}&   -1.275         &    3.242\sym{***}\\
                &  (0.426)         &  (0.426)         &  (0.831)         &  (0.516)         \\
Group of 4      &                  &                  &                  &                  \\
                &                  &                  &                  &                  \\
Group of 5      &                  &                  &                  &                  \\
                &                  &                  &                  &                  \\
Constant        &   17.682\sym{***}&   17.203\sym{***}&   21.425\sym{***}&   16.508\sym{***}\\
                &  (0.444)         &  (0.502)         &  (0.712)         &  (0.504)         \\
\midrule
N Obs.          &    7,800         &    7,800         &    2,600         &    5,200         \\
Round Fixed effects&      Yes         &      Yes         &      Yes         &      Yes         \\
Classroom fixed effect&      Yes         &      Yes         &      Yes         &      Yes         \\
Robust standard errors&      Yes         &      Yes         &      Yes         &      Yes         \\
\bottomrule
\multicolumn{5}{l}{\footnotesize Column 2: Post estimation tests for Common = [Common x Immigrant - Common]: p=0.6963.}\\
\multicolumn{5}{l}{\footnotesize Column 2: Post estimation tests for Multicultural = [Multicultural x Immigrant - Multicultural]: p=0.0081.}\\
\multicolumn{5}{l}{\footnotesize Column 3: Post estimation tests for Common = Multicultural: p=0.0251.}\\
\multicolumn{5}{l}{\footnotesize Column 4: Post estimation tests for Common = Multicultural: p=0.0022.}\\
\multicolumn{5}{l}{\footnotesize OLS regression, symbols $***$, $**$, and $*$ indicate significance at the 1\%, 5\% and 10\% level, respectively.}\\
\end{tabular}
}

		\end{adjustbox}
	\end{center}
	\label{tab:contribution12}
\end{table}

\clearpage
\section{Classroom Social Networks}\label{sec:network}
\addcontentsline{toc}{section}{Classroom Social Networks}

This Appendix offers a detailed overview of friendship patterns and classroom networks. Each student was asked to nominate up to five classmates as friends, identified by unique IDs using the methodology applied by \cite{landini2016friendship} and \cite{chen2016group}. We first analyze the distribution of mentions by friends, distinguishing between immigrant and native statuses, using the same definition for immigrant status as in the main paper (both parents not born in Italy). Next, we construct directed classroom networks and calculate centrality and density measures, which are also employed in the regression analysis.

\subsection{Friendship}\label{friendship}

We measure friendship as mentions by classmates. Table \ref{tab:friends} below provides a few summary statistics of the share of mentions in the classroom for each member of the classroom, divided by immigrant status.

\begin{table} [H]   \caption{Friend Status Statistics} \label{tab:friends} 
\centering \begin{tabular}{lccccc}

    \toprule
    Own status & Friend status & Mean & Median & SD & N \\
    \midrule
    Immigrant & Native & 69.5 & 75 & 28.3 & 94 \\
    Native    & Native & 75.4 & 80 & 22.1 & 216 \\
    \bottomrule
\end{tabular}
\end{table}

The table shows that Natives attract a significantly larger share of mentions compared to Immigrants, both from Natives and Immigrants (on average, 75.4\% and 69.5\%, respectively). The Natives have a higher likelihood of mentioning Natives as friends relative to
Immigrants, but the difference is small and not statistically significant significant (Wilcoxon rank sum test, p-value=0.108).

\subsection{Classroom Networks}\label{classroom-networks}

\begin{figure}[]
  \centering
  \begin{tabular}{cccc}
    \includegraphics[width=32mm]{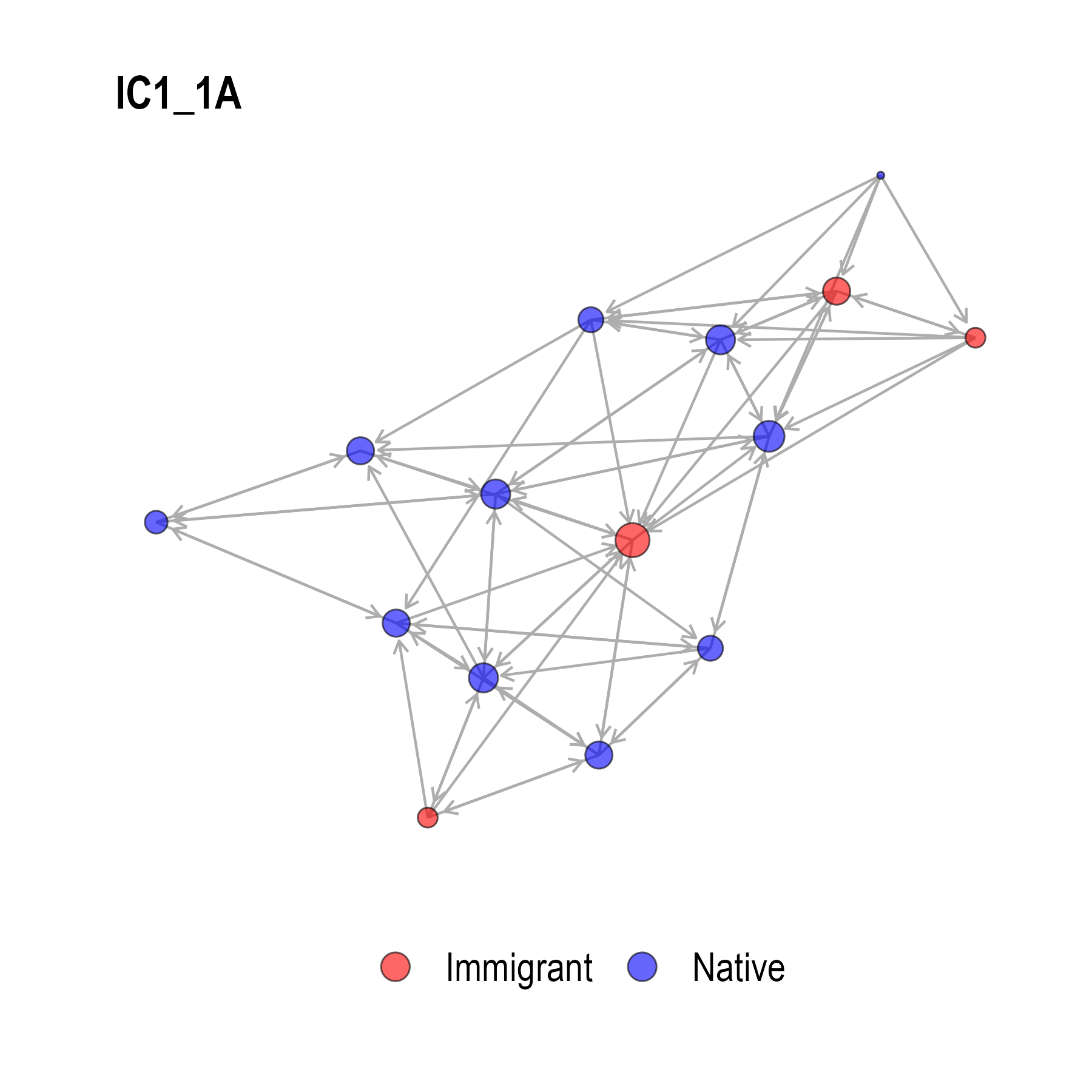} &\includegraphics[width=32mm]{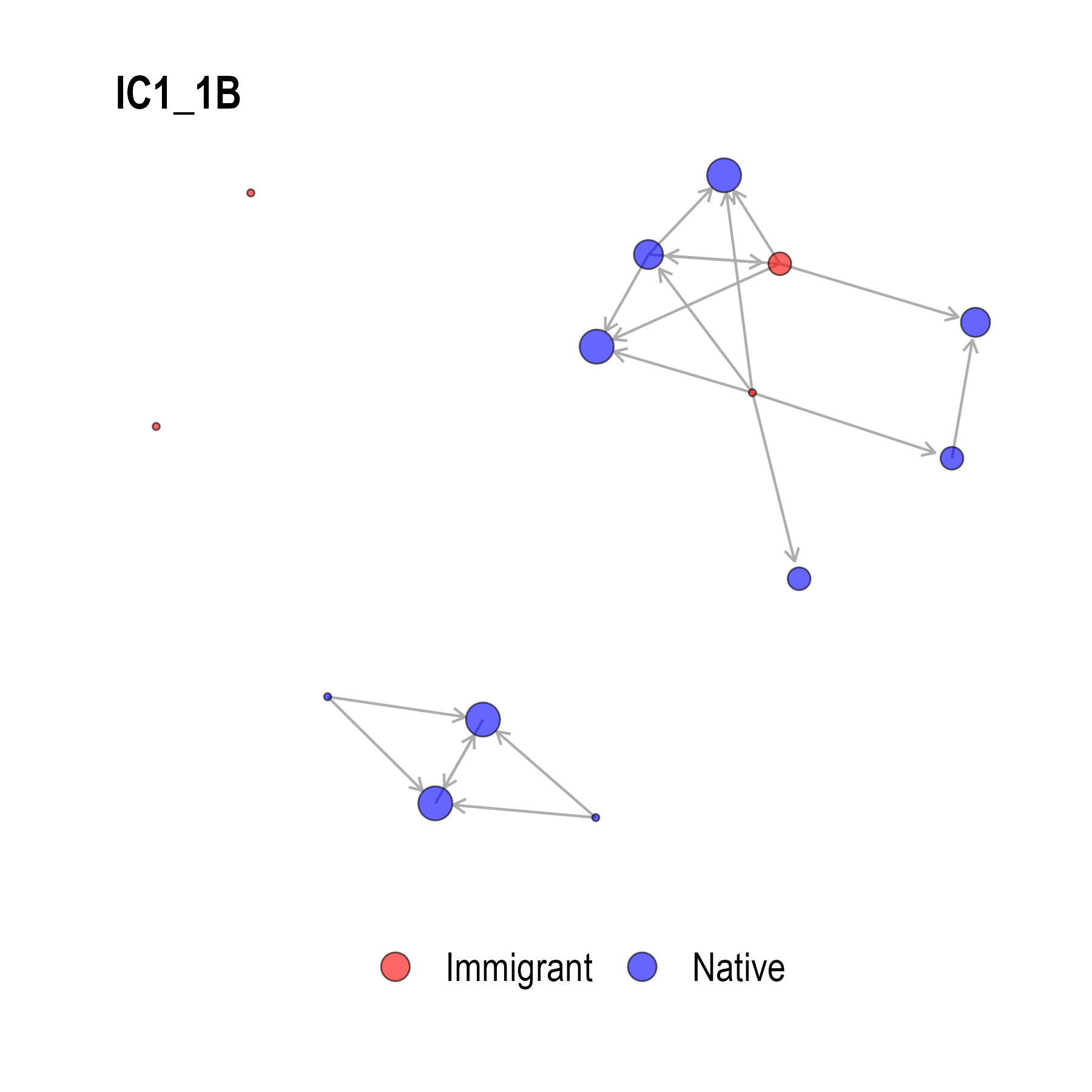}&   \includegraphics[width=32mm]{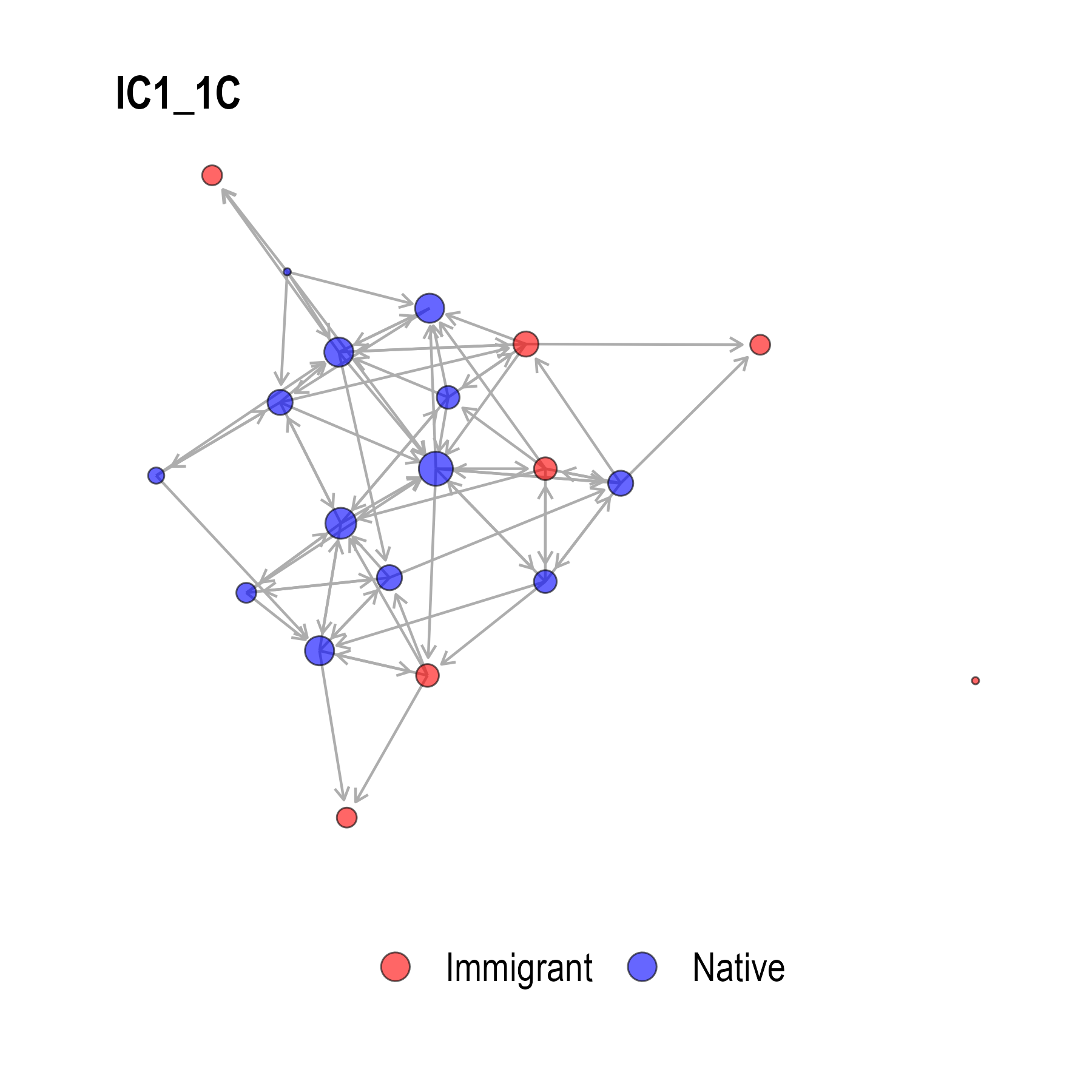} &   \includegraphics[width=32mm]{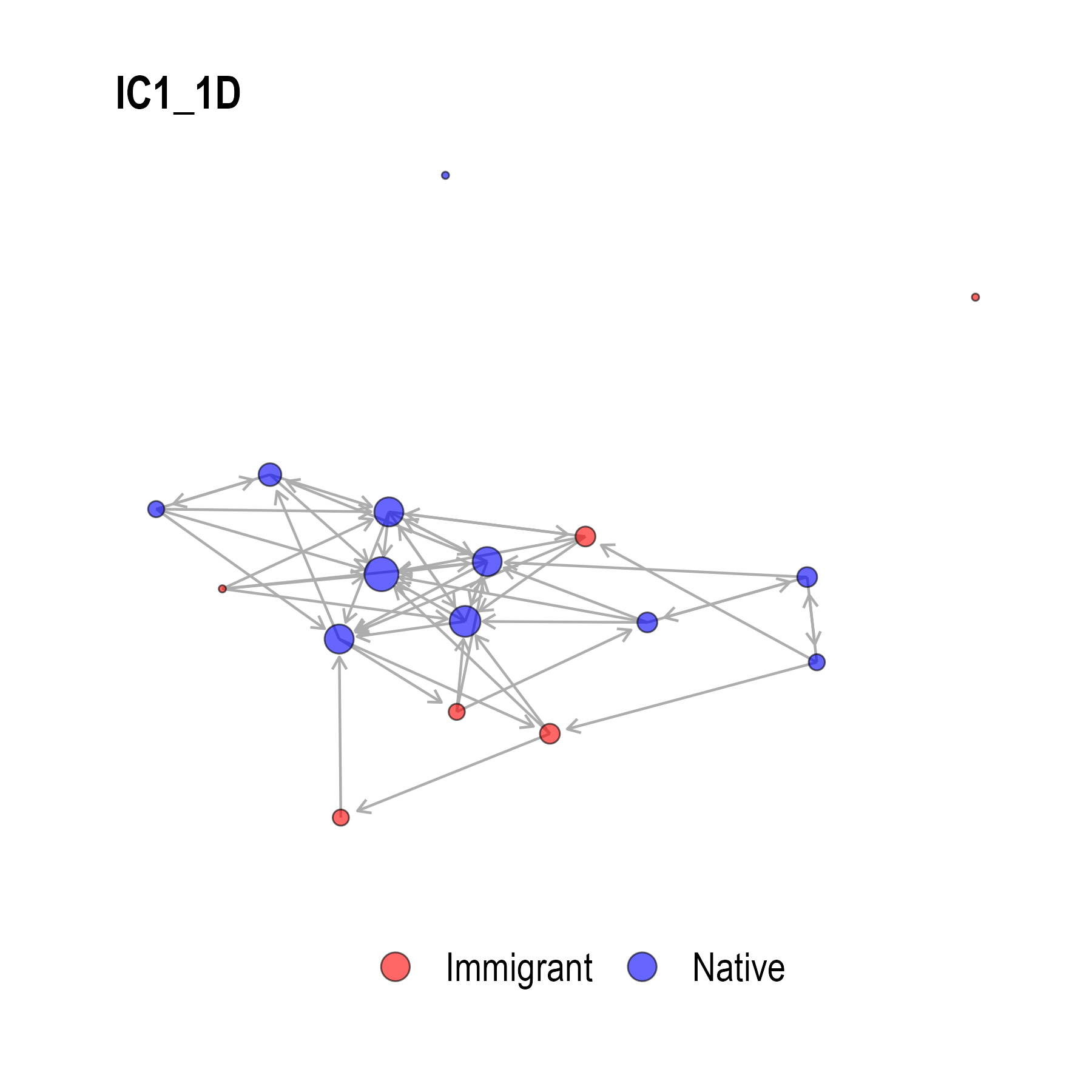} \\

    \includegraphics[width=32mm]{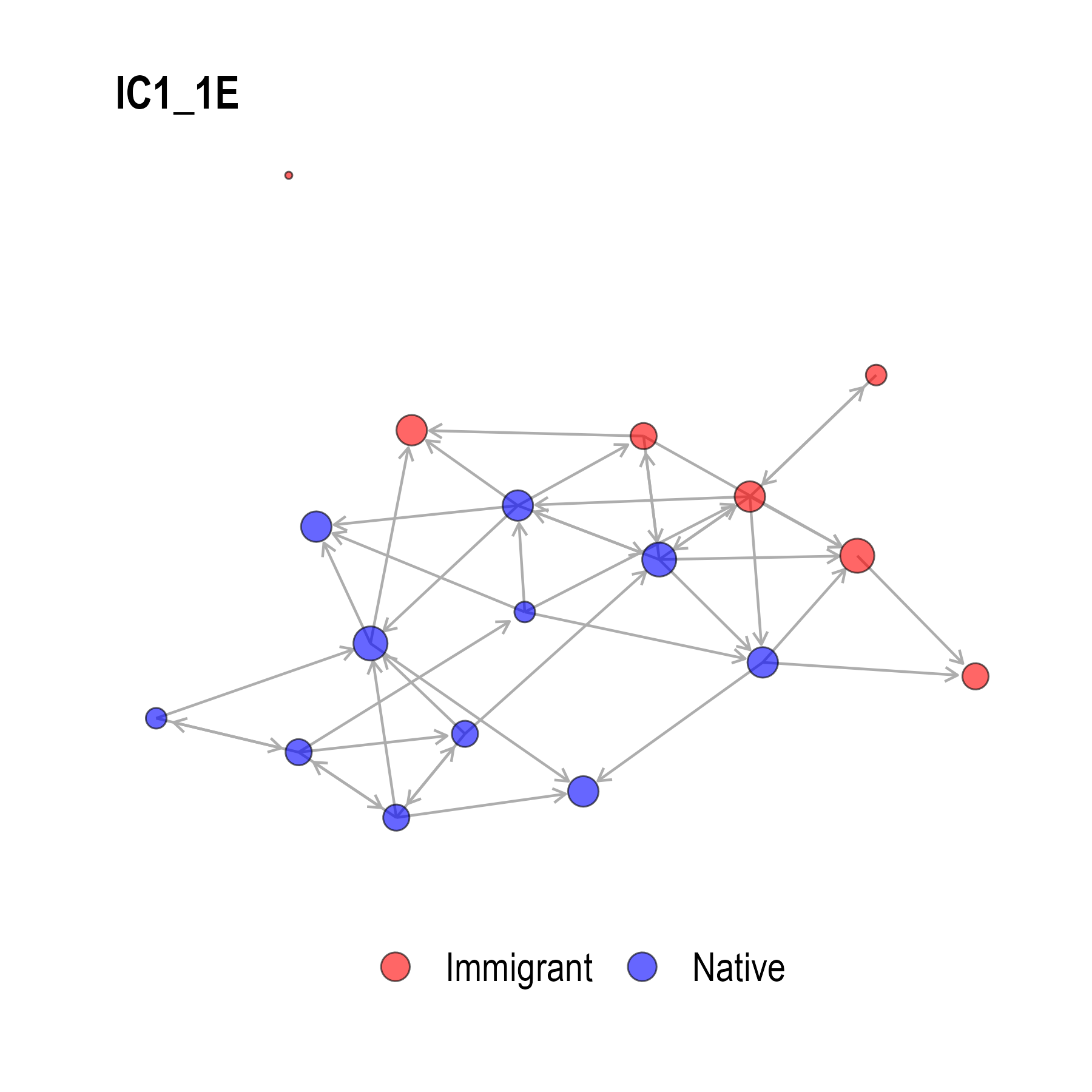} & \includegraphics[width=32mm]{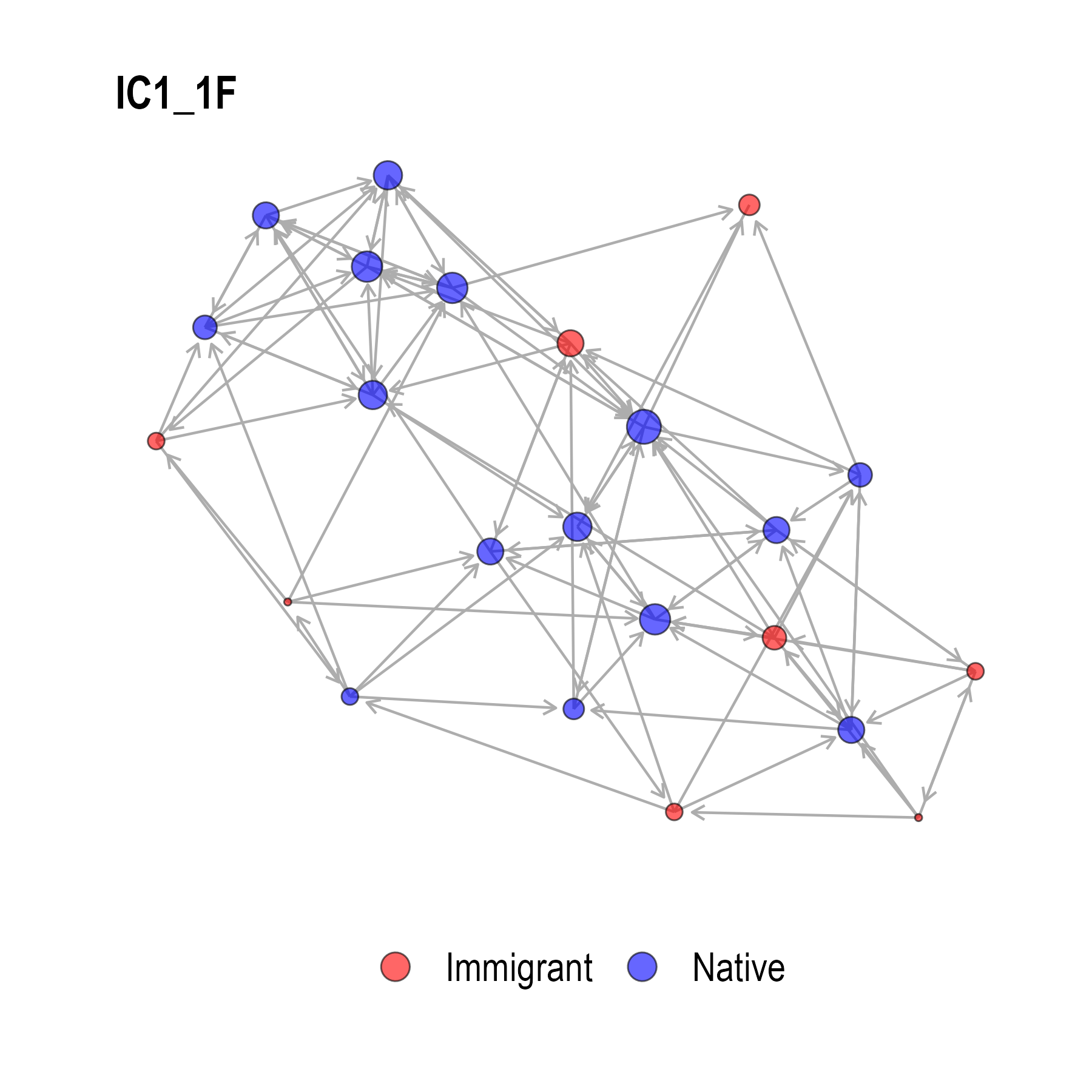} &\includegraphics[width=32mm]{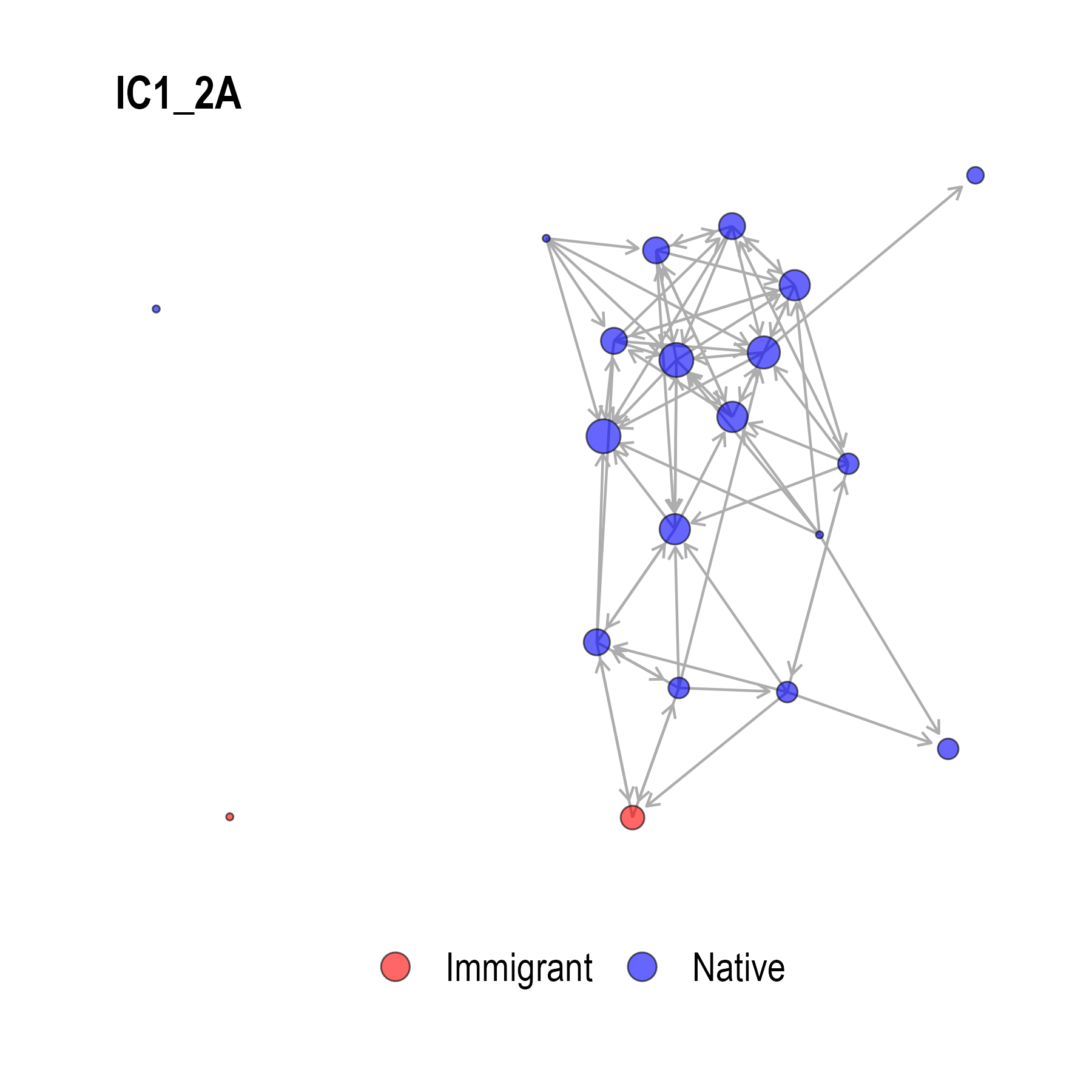}&   \includegraphics[width=32mm]{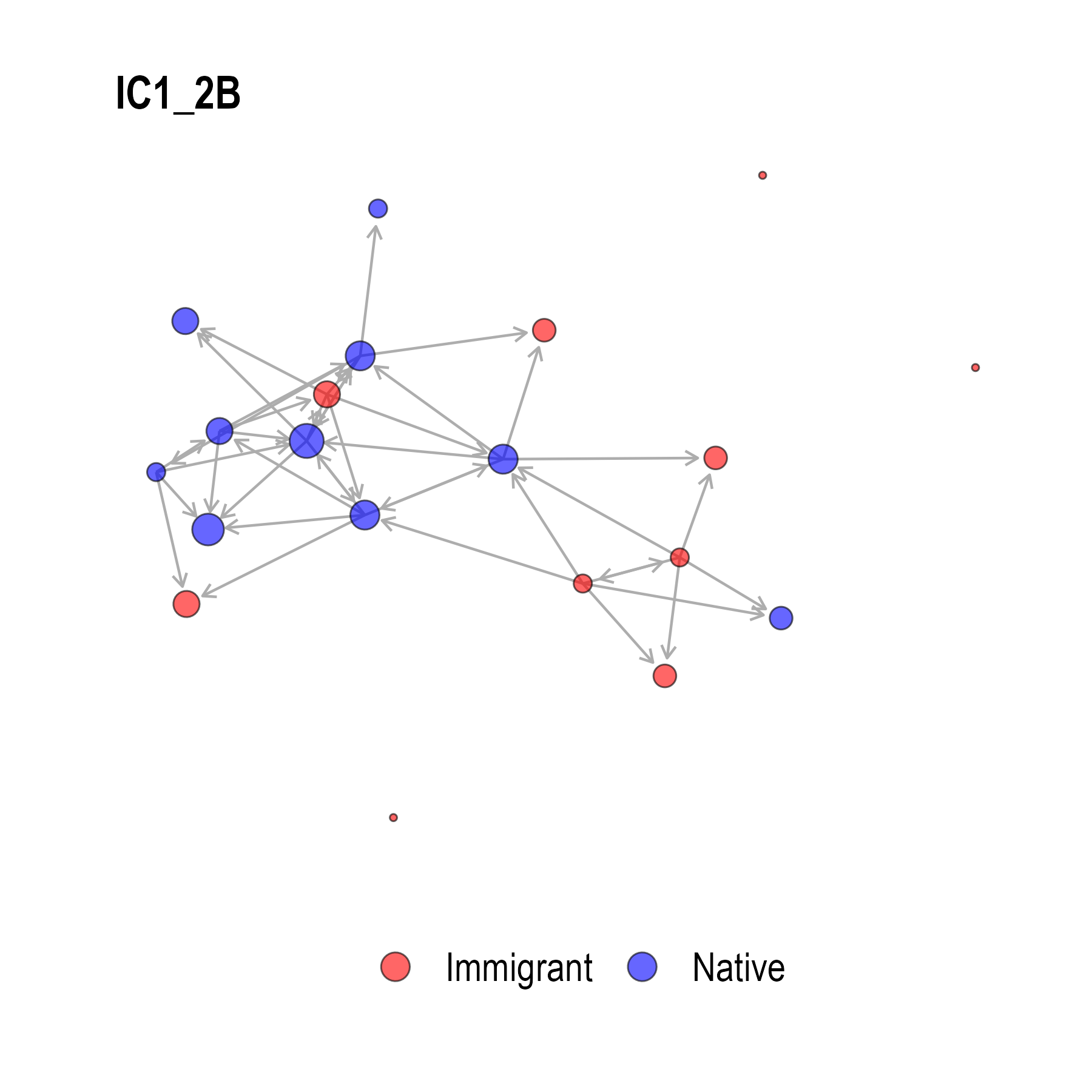}  \\

    \includegraphics[width=32mm]{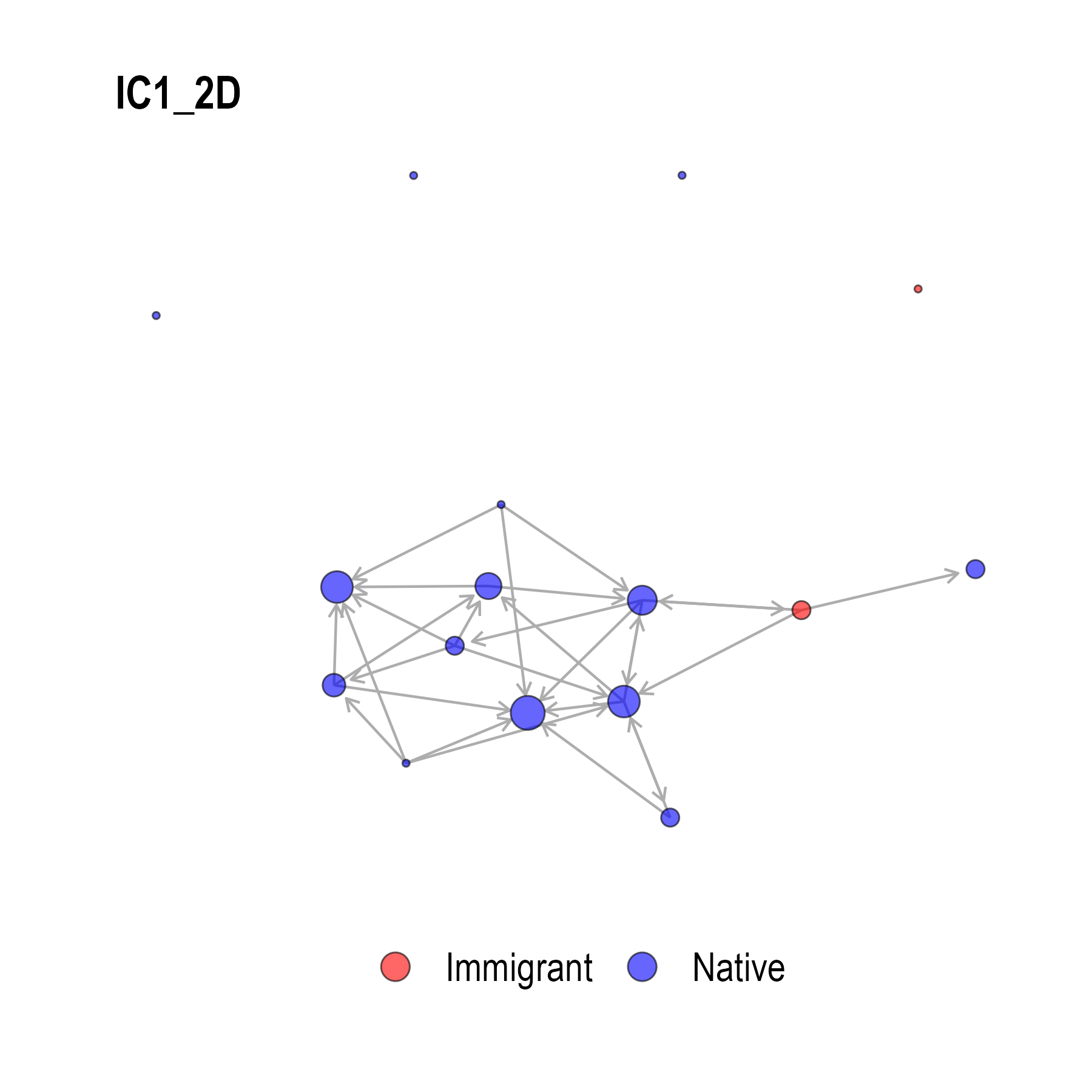} & \includegraphics[width=32mm]{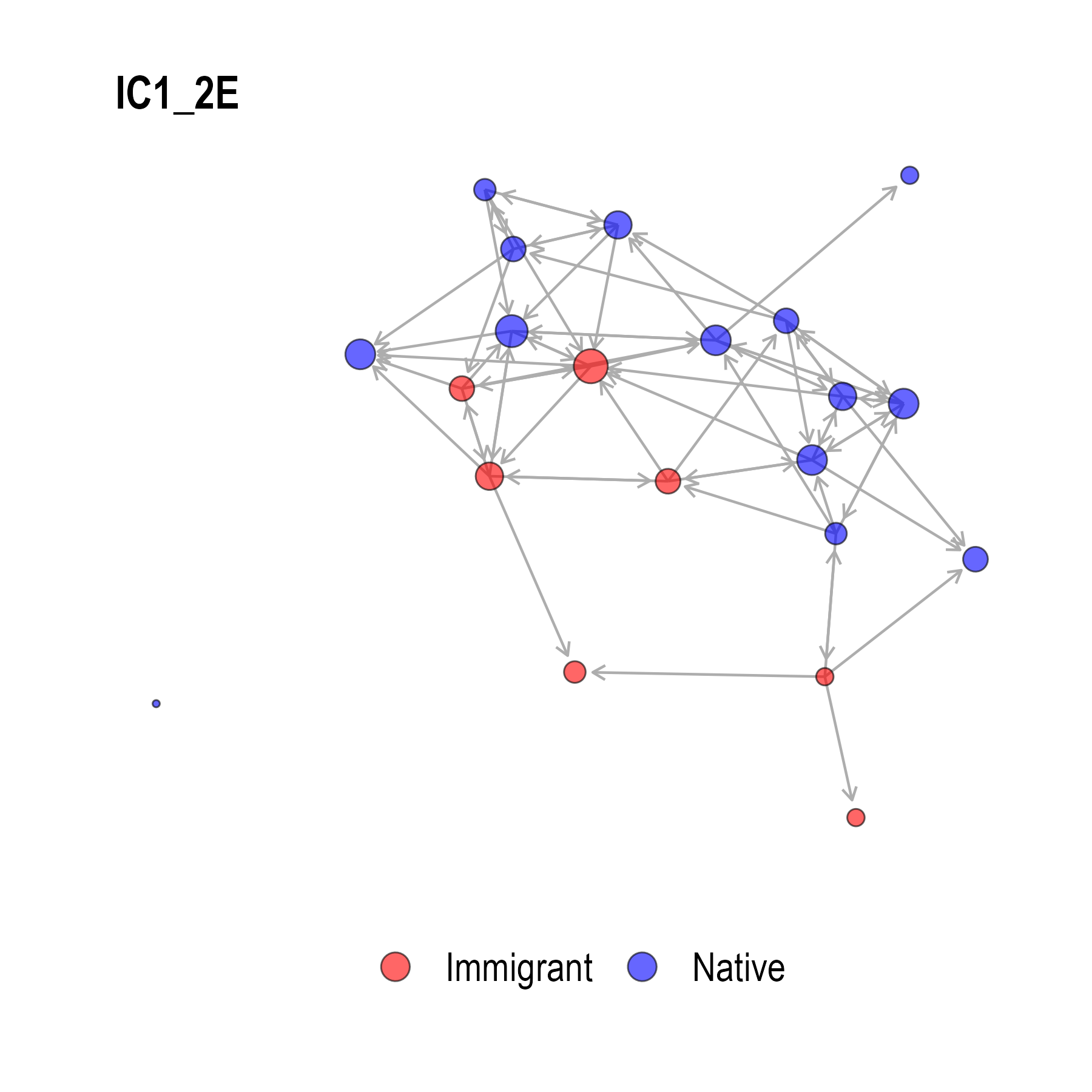} &\includegraphics[width=32mm]{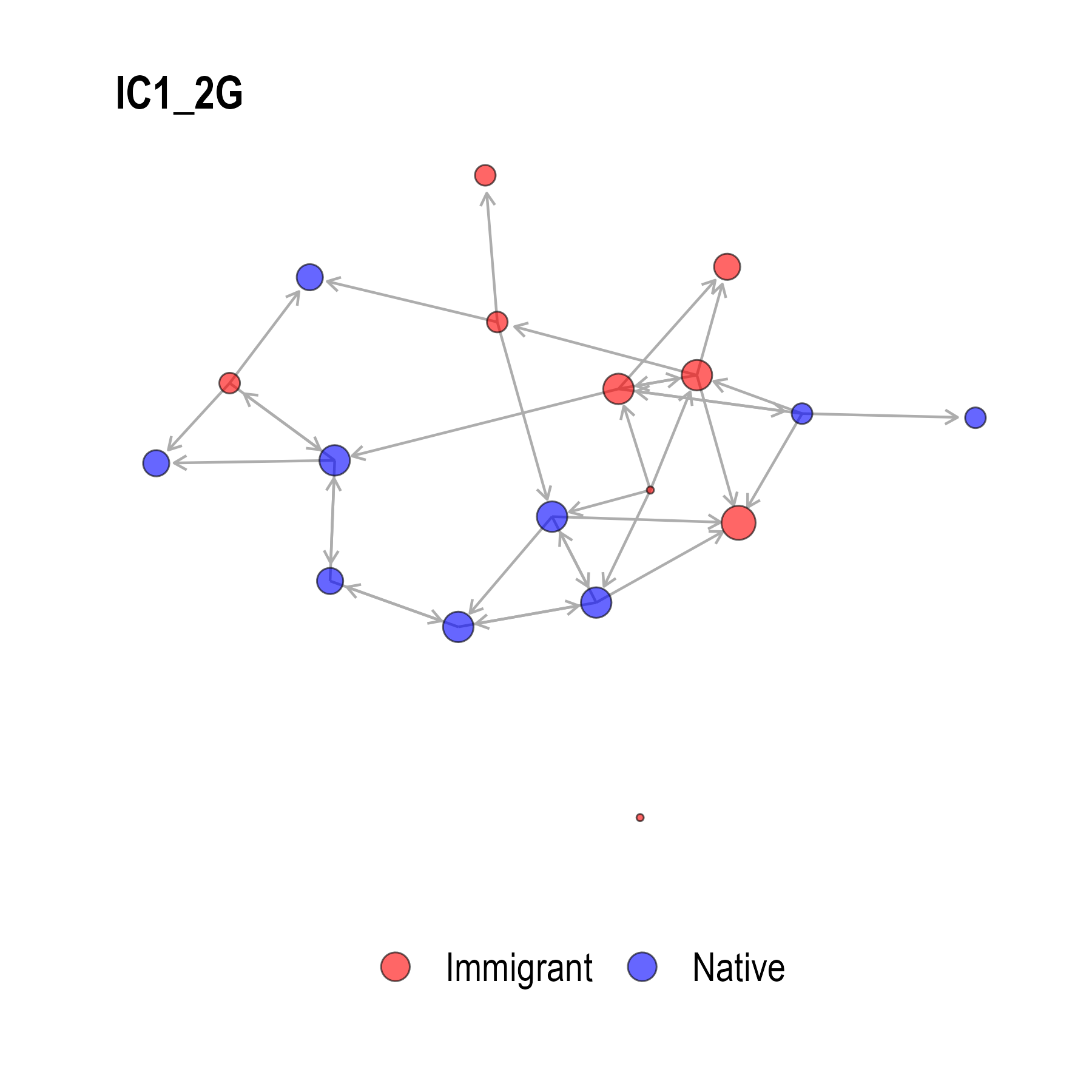}&   \includegraphics[width=32mm]{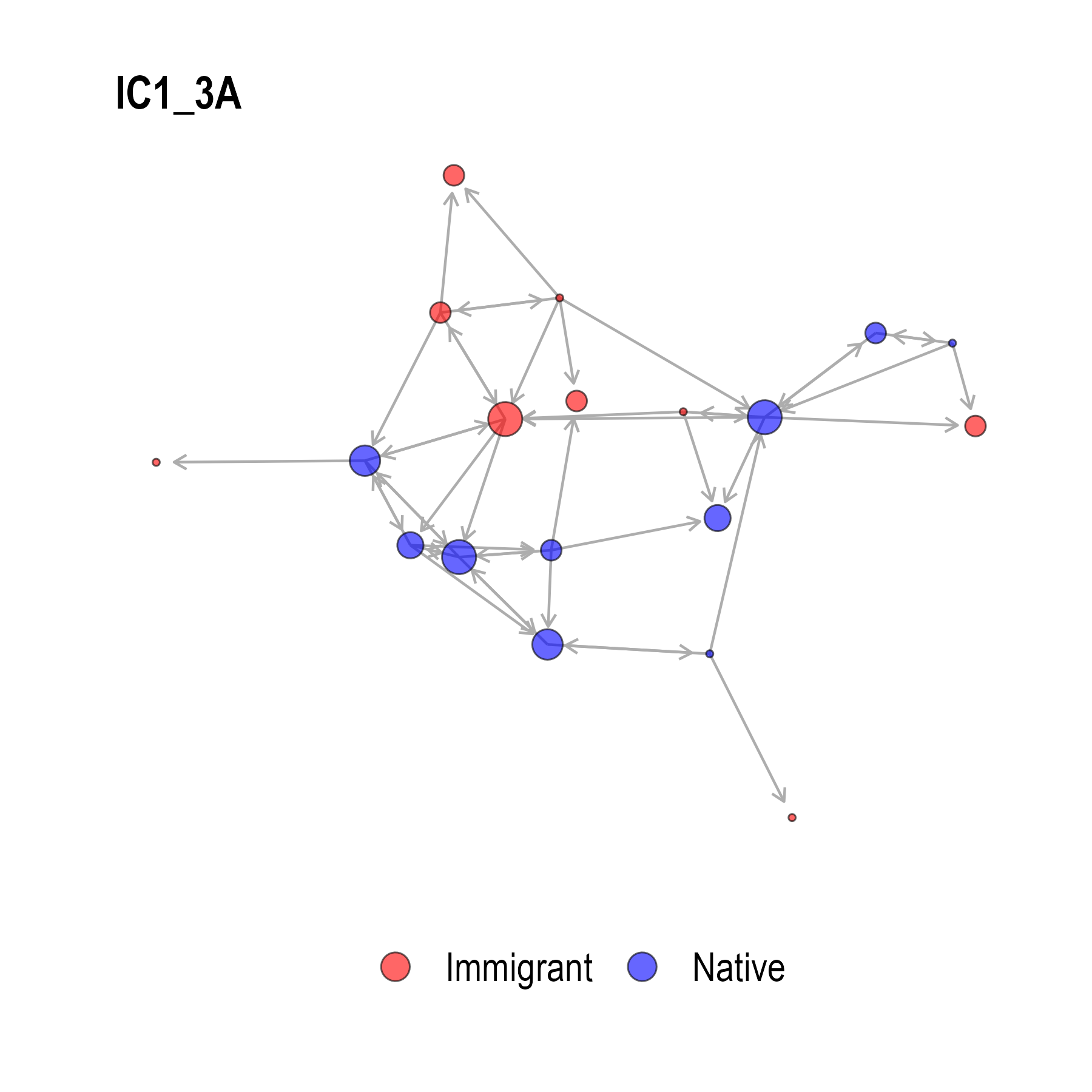}  \\

    \includegraphics[width=32mm]{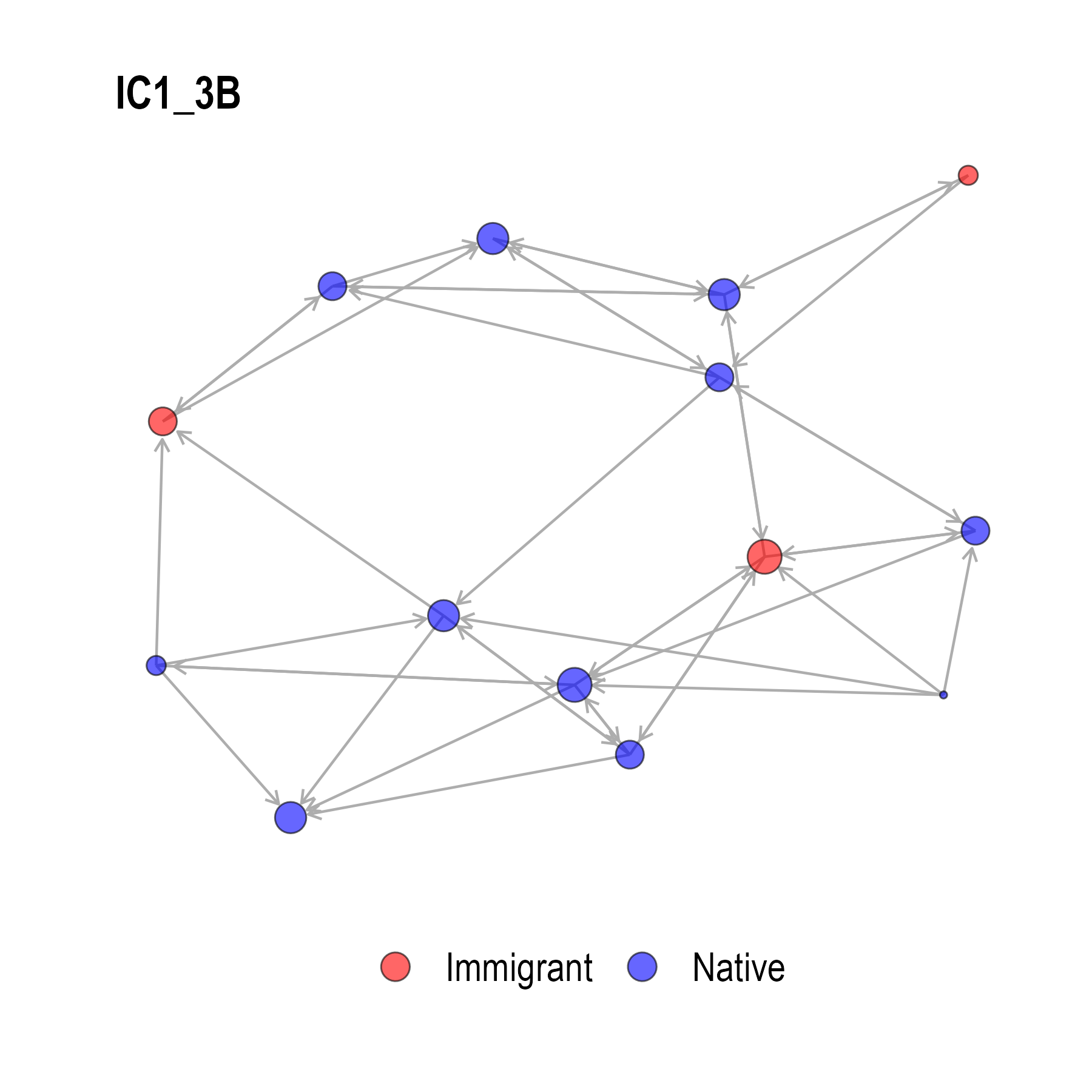} & \includegraphics[width=32mm]{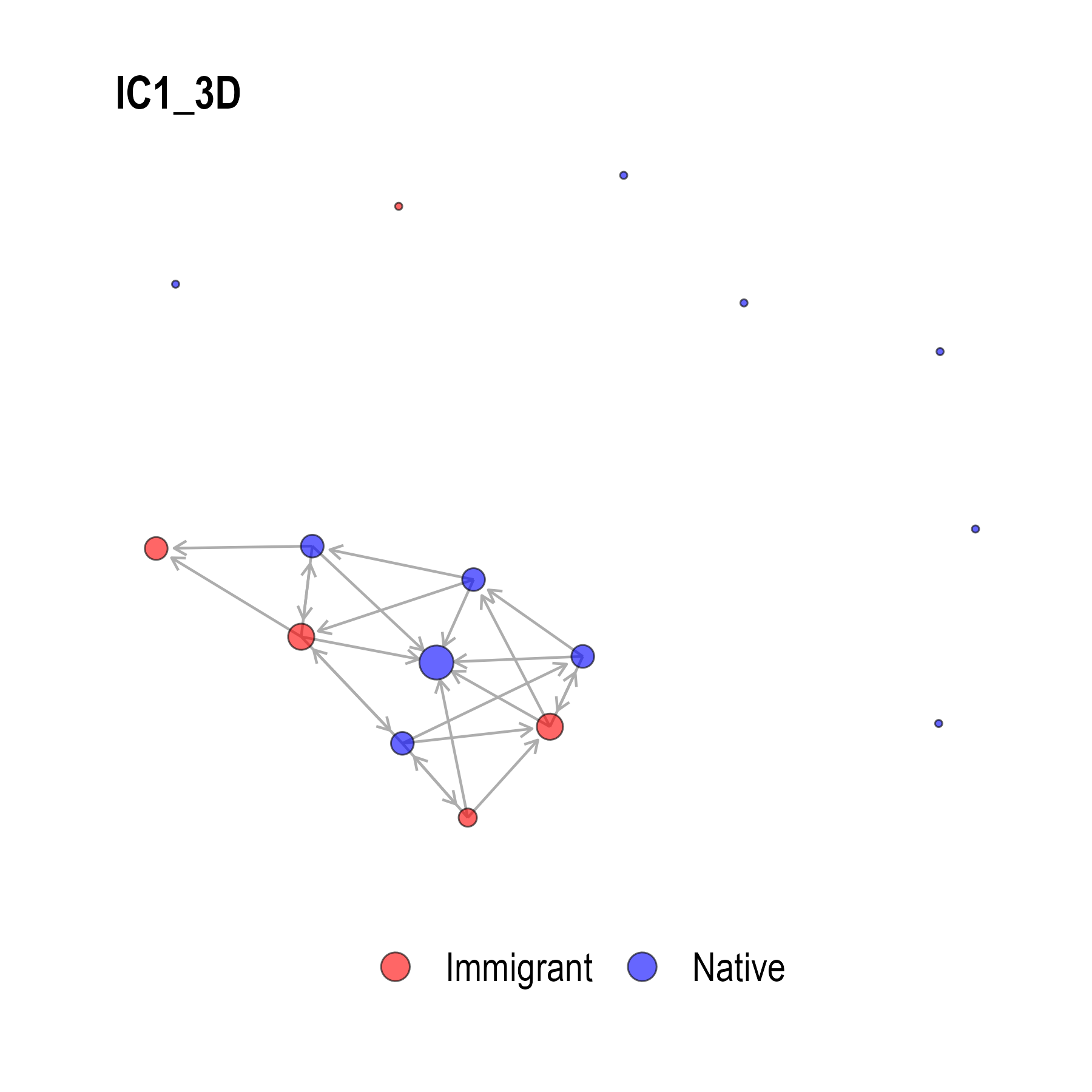} &\includegraphics[width=32mm]{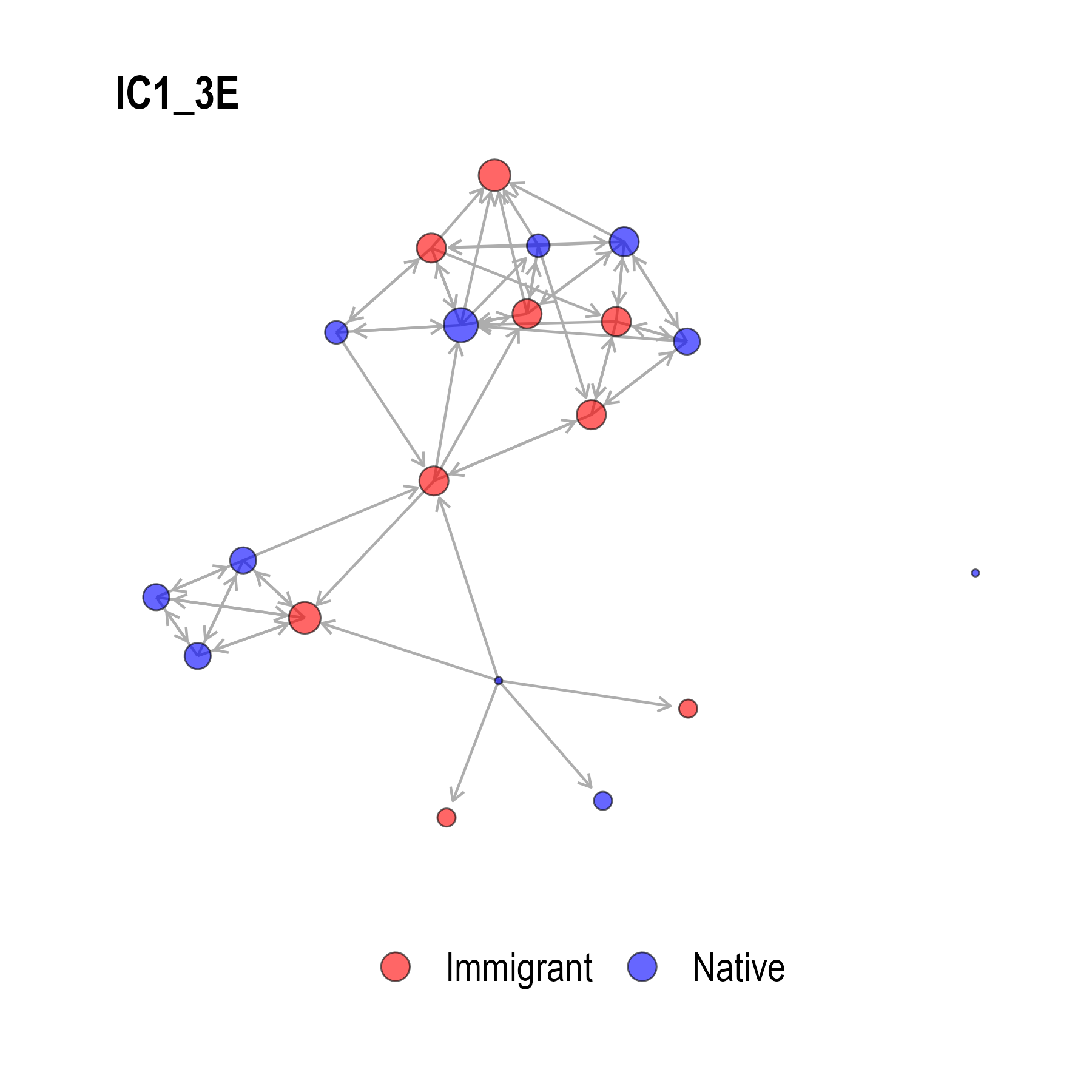}&   \includegraphics[width=32mm]{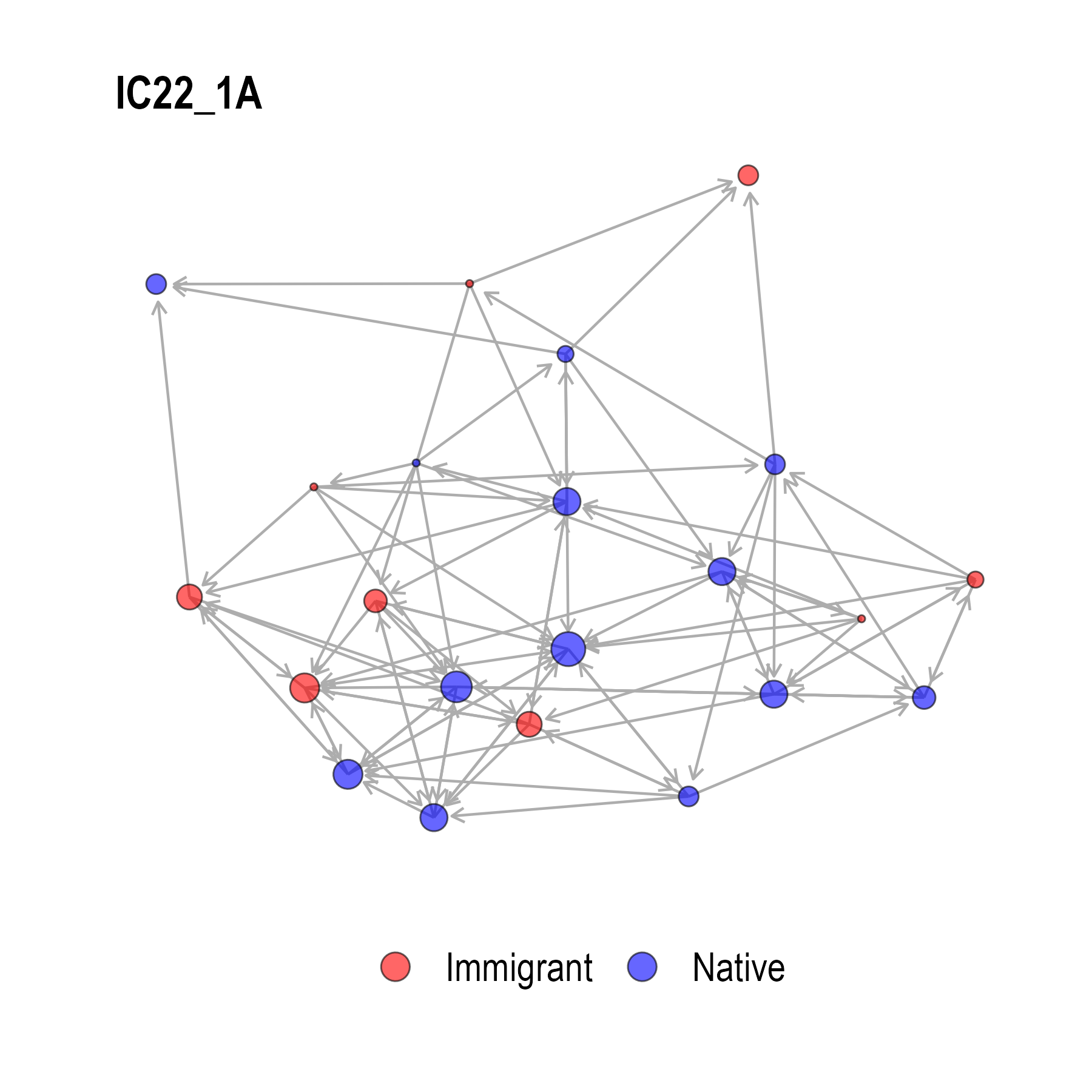}  \\

    \includegraphics[width=32mm]{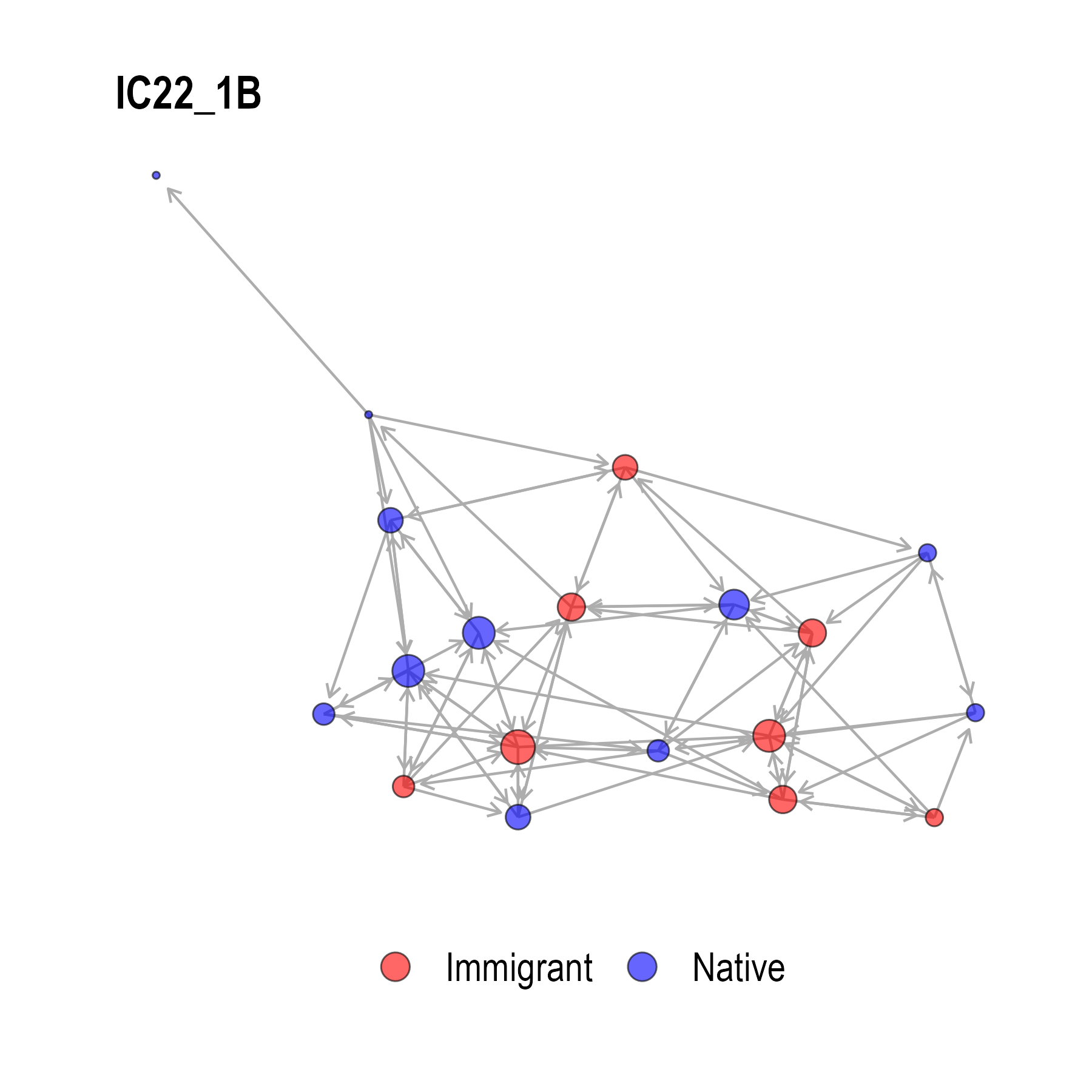} & \includegraphics[width=32mm]{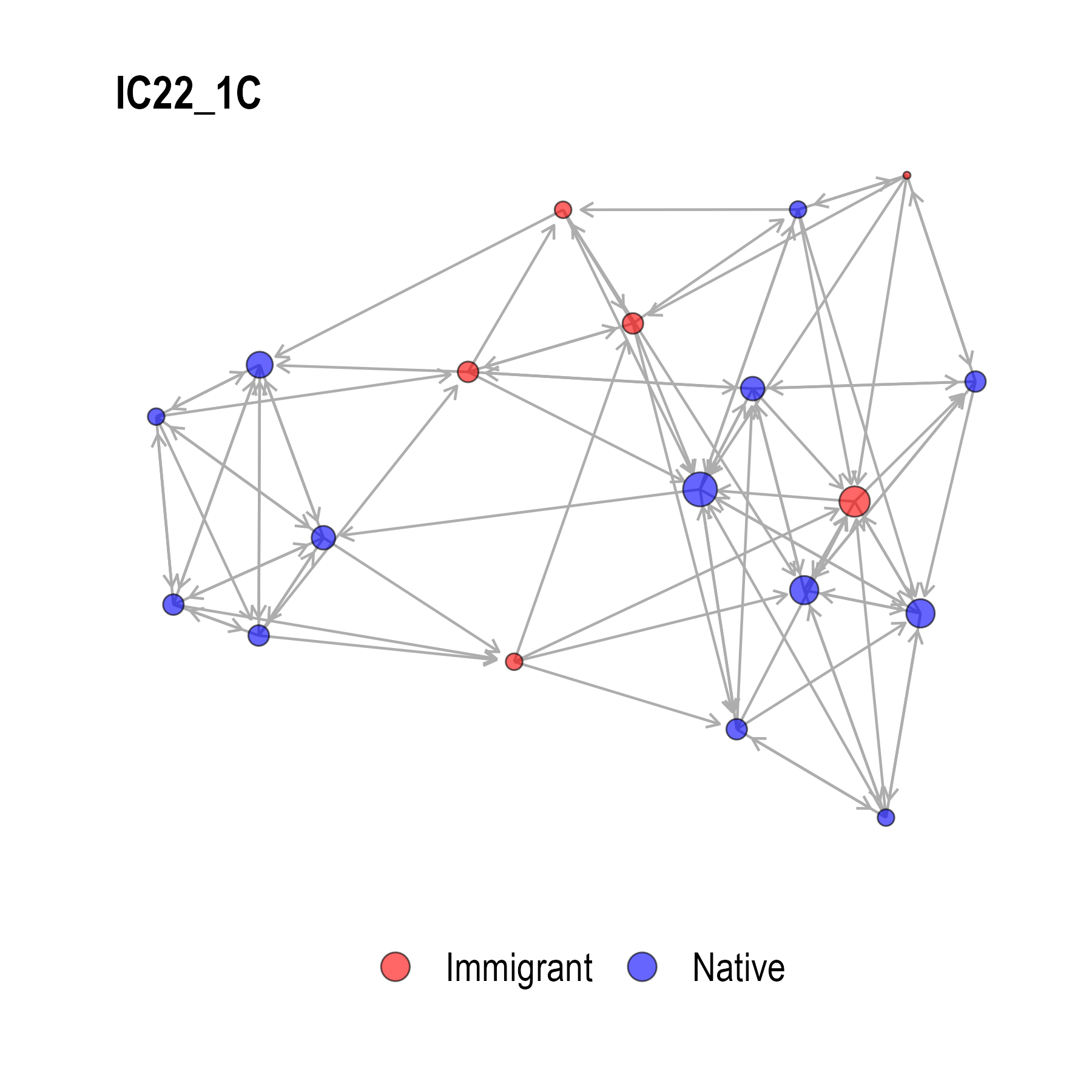} &\includegraphics[width=32mm]{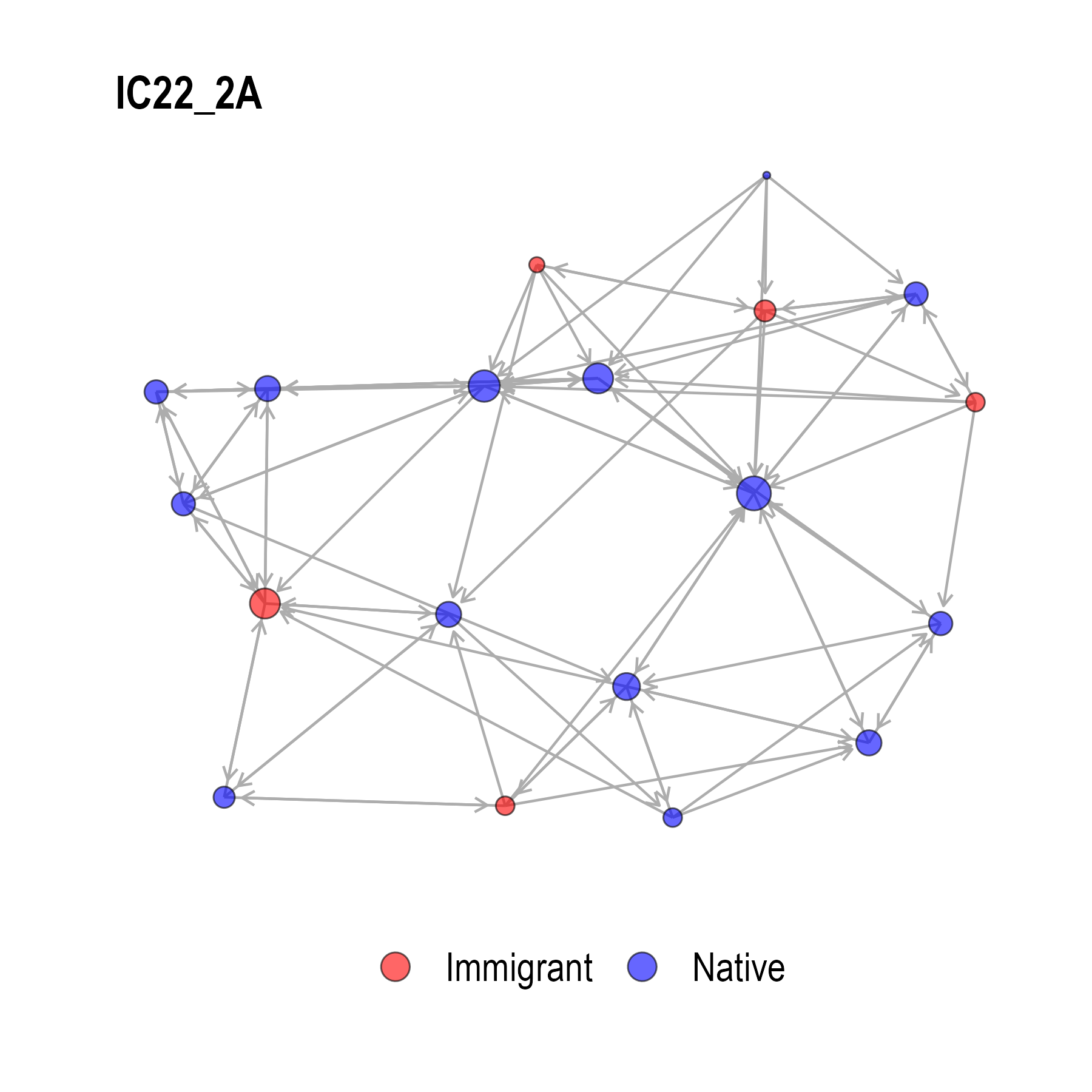}&   \includegraphics[width=32mm]{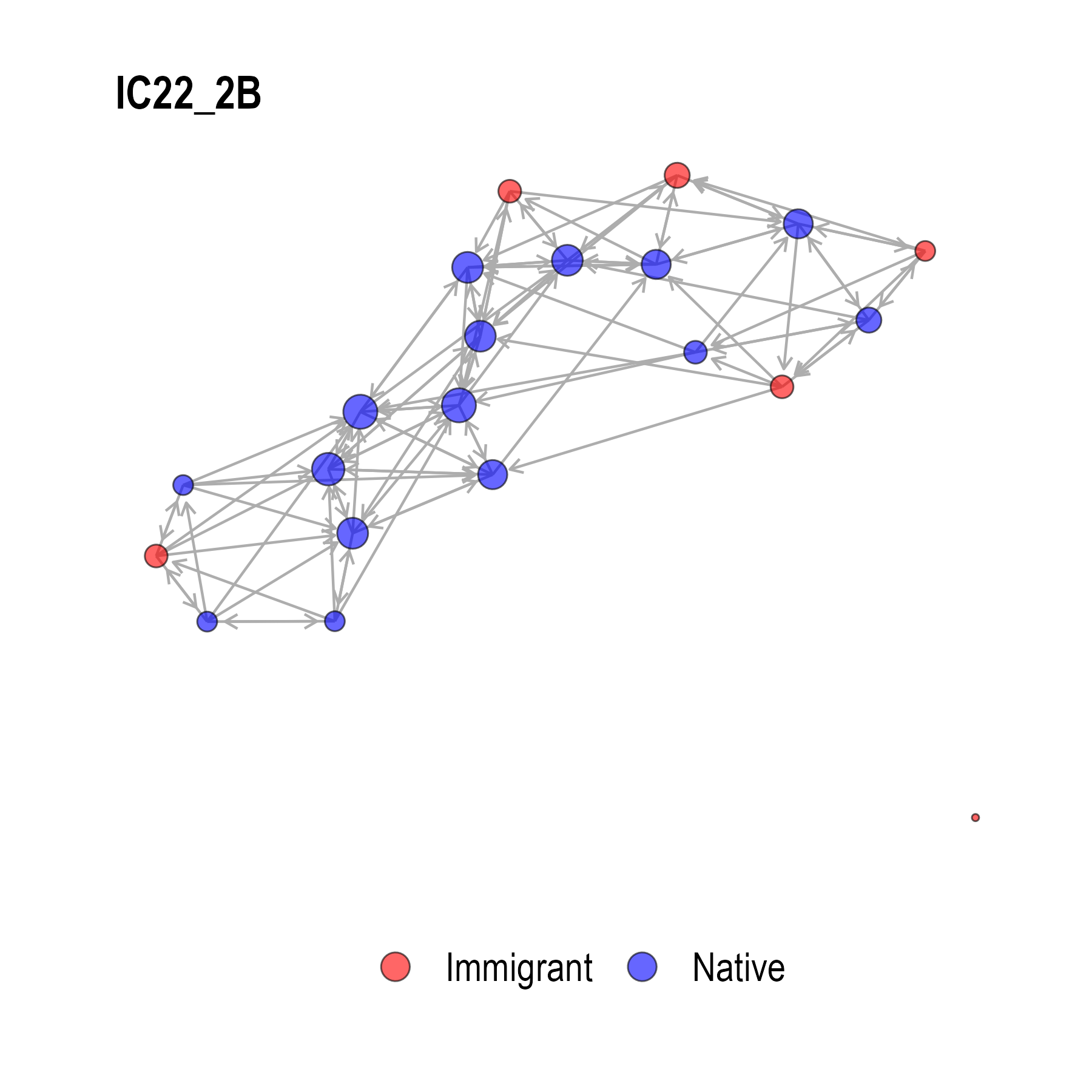} \\
    
    \includegraphics[width=32mm]{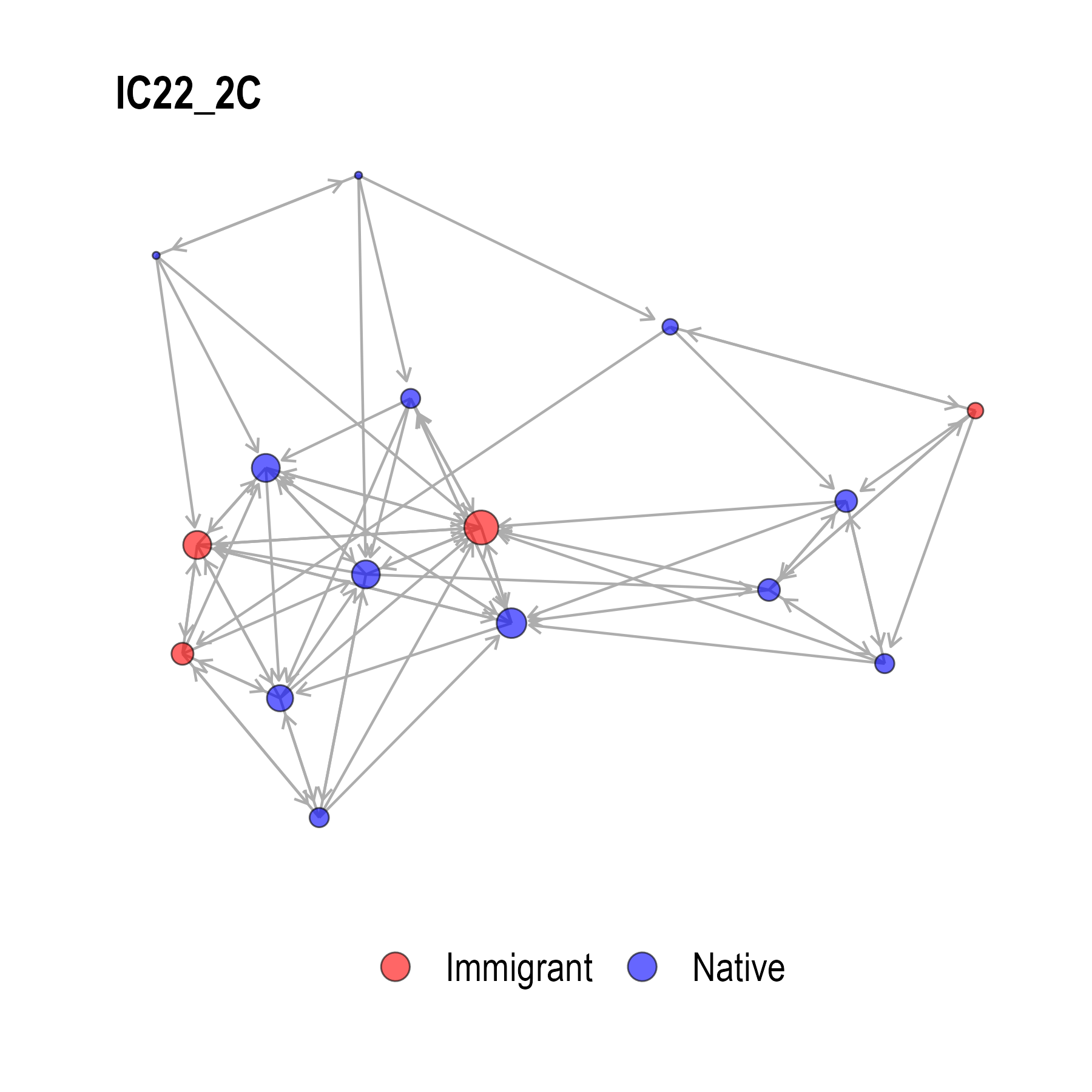} & & & \\
  \end{tabular}
  \caption{Classroom Social Networks\label{fig:networks}}
  \end{figure}

In this section, we examine the structure of classroom social networks in greater detail. Only students who participated in the study are included. Each classroom network is represented as a directed graph,
where nodes correspond to students and edges indicate friendship nominations. Since nominations are limited to classmates, all connections occur within the classroom. The visualizations below illustrate these networks (Figure \ref{fig:networks}): arrows show the direction of each friendship nomination, with reciprocal ties indicated by arrows in both directions.
Node color reflects the immigrant status of each student, and node size indicates the number of incoming nominations (i.e., how many classmates
mentioned the student as a friend).

Overall, the structure of classroom networks seems to differ from classroom to classroom. In general, classmates seem to be connected to each other either directly or indirectly, but the structure of connections differs from classroom to classroom. No clear clustering in terms of immigrant status seems to emerge from the graphical representation, even though Natives seem to occupy more central positions in many classrooms.

To gain a better understanding of network structures, a few conventional network measures, adopted also in the regression analysis, are illustrated in the next section.

\subsubsection{Density}\label{density}

\begin{figure}[H]
\begin{center} \includegraphics[width=0.6\textwidth,keepaspectratio]{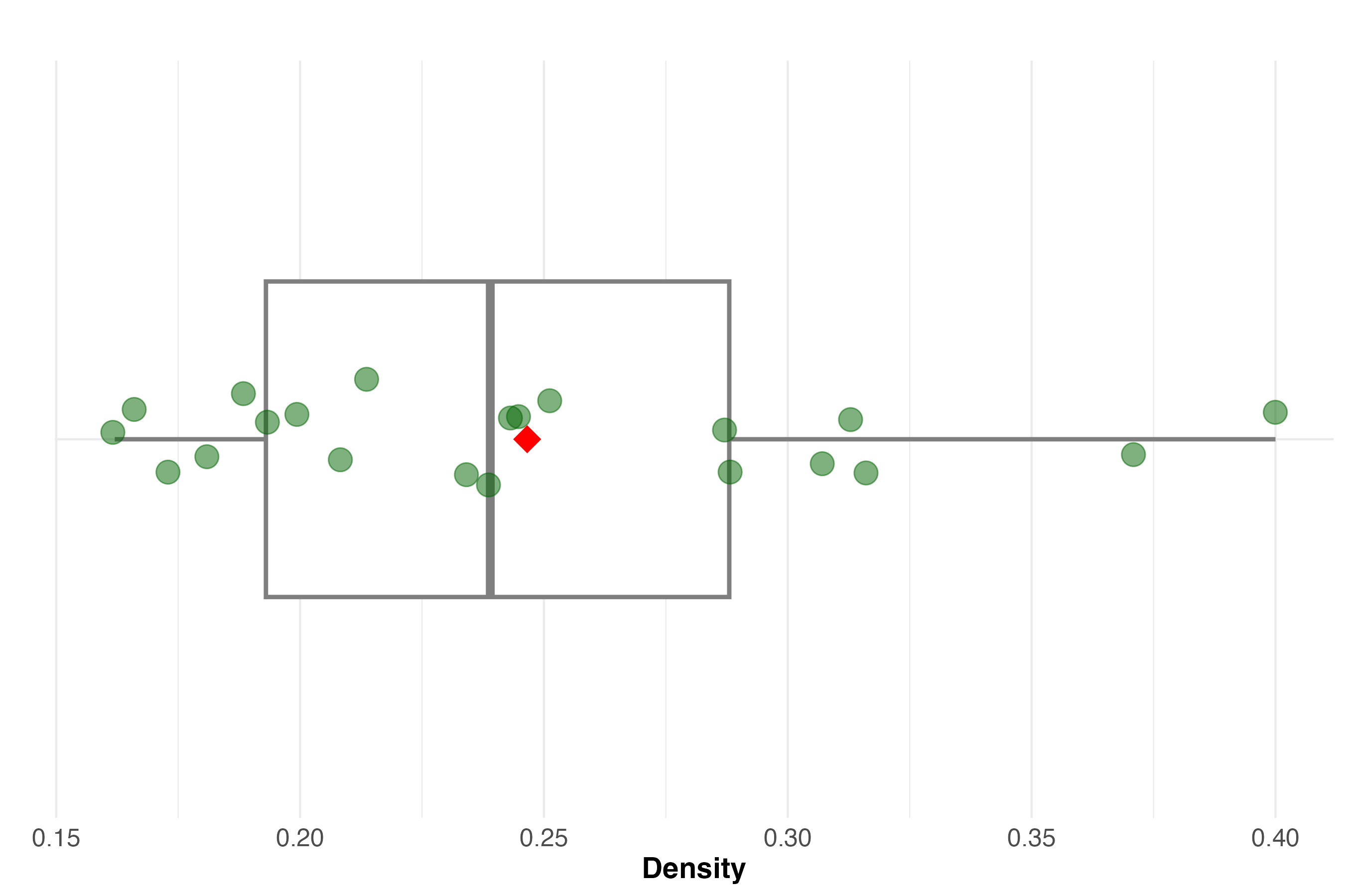}
\end{center}
\caption{Classroom Network Density \label{fig:density}}
\end{figure}

Network density quantifies the overall connectedness within a classroom network. It is calculated as the ratio of actual connections to all possible connections among students. A density of 1 means every student is connected to every other student, while values near 0 reflect sparse networks with few ties. Figure \ref{fig:density} displays the distribution of density values for each classroom (dots), highlights quartile ranges with a boxplot, and marks the average density with a diamond symbol.

Overall, classroom networks exhibit low to moderate density, with many classrooms having density values below 0.25. This indicates that while some students are well-connected, many others have fewer connections,
leading to a generally sparse network structure across classrooms.

\subsubsection{Degree Centrality: In-ties}\label{degree-centrality-in-ties}

The centrality measure we examine is degree centrality, which reflects the number of connections each student receives. In our directed classroom networks, we focus on incoming ties---representing how many classmates mention a student as a friend.

Figure \ref{fig:centrality} below summarizes the distribution of degree centrality for each classroom, separately for immigrants and natives.

\begin{figure}[H]
{\begin{center}
    \includegraphics[width=0.7\textwidth,keepaspectratio]{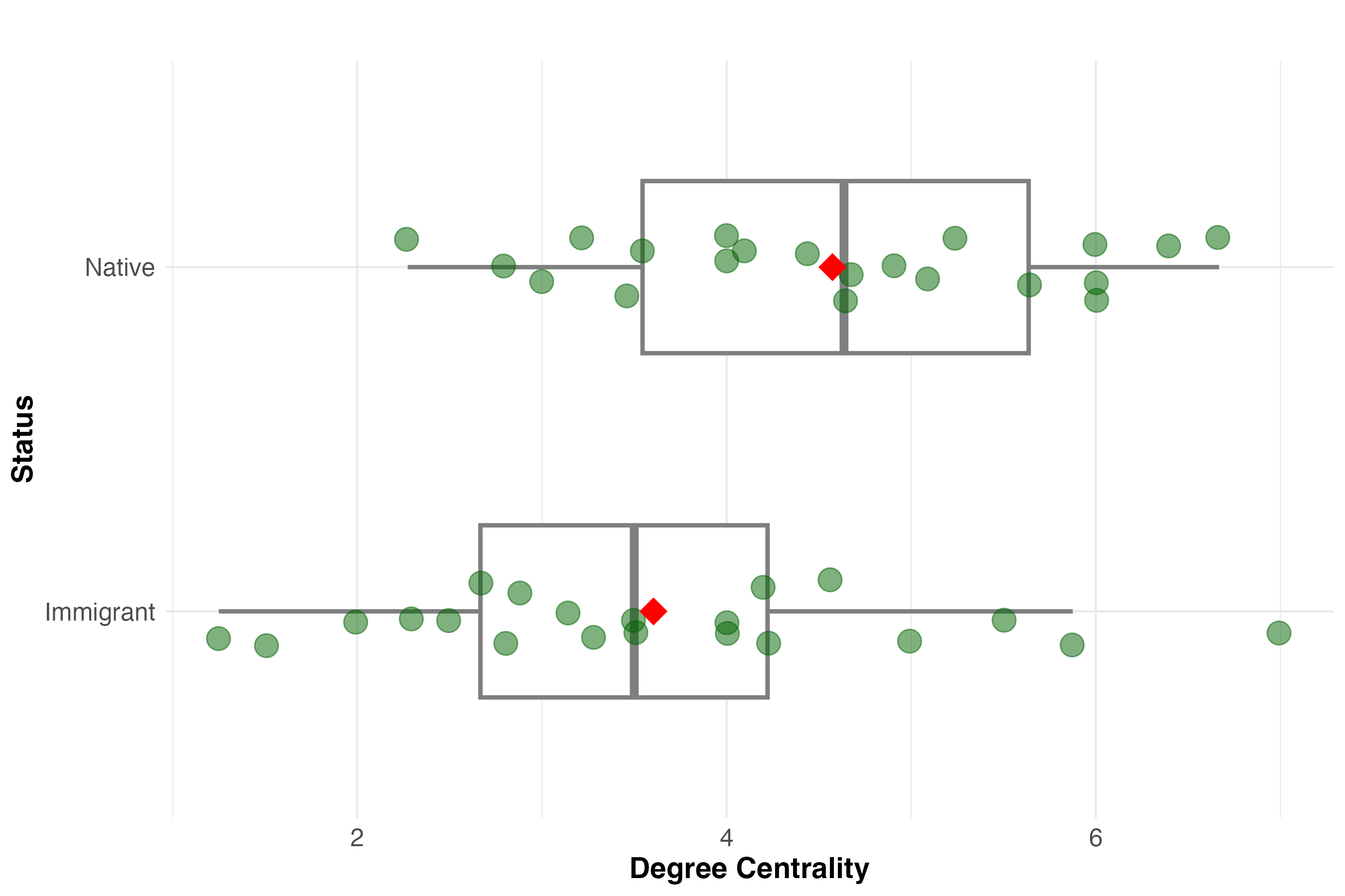}
\end{center}}
\caption{Network Degree Centrality (In-ties) by Immigrant Status \label{fig:centrality}}
\end{figure}%

As the figure shows, Natives generally have more incoming connections than Immigrants: in 17 classrooms out of 21 (80.9\%), the natives display a higher value than the immigrants. This reflects the overall higher popularity of Natives observed in the friendship nominations.

The evidence from the friendship nominations and the network measures points to a consistent pattern: classroom social structures are heterogeneous across classes, but native students generally occupy more
central and visible positions. Descriptive statistics and degree-centrality measures show that natives receive more incoming nominations in the majority of classrooms, yet the graphical layouts and class-level diagnostics do not reveal strong, uniform clustering by
immigrant status. Differences in popularity and centrality are often modest and vary with classroom composition and density. These findings suggest a recurring native advantage in social visibility within
classrooms, but the magnitude and robustness of this advantage depend on local classroom structure.

\end{document}